\documentclass[12pt]{article}
\pdfoutput=1

\usepackage[all]{xy}

\usepackage{xcolor}
\usepackage{dsfont}

\usepackage{hyperref}
\usepackage{subcaption}
\usepackage[utf8]{inputenc}
\usepackage{ytableau}
\usepackage[vcentermath]{youngtab}

\usepackage{setspace}

\usepackage{amsmath, amssymb, amsthm, float, graphicx}
\numberwithin{equation}{section}
\usepackage{booktabs}

\hypersetup{colorlinks=false, linkcolor=blue, citecolor=red}

\def\tmH{H}
\def\tmd{\delta}
\def\del{\partial}

\def\beq{\begin{eqnarray}}\def\eeq{\end{eqnarray}}
\def\be{\begin{equation}}\def\ee{\end{equation}}

\def\g{\gamma}
\def\r{\rho}
\def\s{\sigma}
\def\m{\mu}
\def\n{\nu}
\def\a{\alpha}
\def\e{\epsilon}

\def\b{\beta}
\def\d{\delta}
\def\c{\chi}

\def\vf{\varphi}

\def\D{\Delta}
\def\G{\Gamma}
\def\l{\lambda}

\def\la{\langle}
\def\ra{\rangle}

\def\Ocal{{\mathcal{O}}}
\def\Scal{{\mathcal{S}}}
\def\Jcal{{\mathcal{J}}}
\def\Bcal{{\mathcal{B}}}
\def\Hcal{{\mathcal{H}}}

\def\Lcal{{\mathcal{L}}}

\def\G{\Gamma}

\def\tr{{\rm tr~}}

\def\thetab{\bar{\theta}}
\def\psib{\bar{\psi}}

\newcommand{\tmop}[1]{\ensuremath{\operatorname{#1}}}

\begin{document}

	\vspace*{-.6in} \thispagestyle{empty}
\begin{flushright}
CPHT-RR001.012022
\end{flushright}

\vspace{.2in} {\large
	\begin{center}
		\bf  Random Field $\phi^3$ Model and Parisi-Sourlas Supersymmetry
	\end{center}
}
\vspace{.2in}
\begin{center}
	{\bf 
		Apratim Kaviraj$^{a,b,c}$,  \ Emilio Trevisani$^{b,d}$
	} 
	\\
	\vspace{.2in} 
	{\it $^{a}$Institut de Physique Th\'{e}orique Philippe Meyer, \\ $^{b}$
		Laboratoire de Physique de l’Ecole normale sup\'erieure, ENS, \\
		Universit\'e PSL, CNRS, Sorbonne Universit\'e, Universit\'e de Paris, F-75005 Paris, France}\\
	{\it$^{c}$ DESY Hamburg, Theory Group, Notkestra\ss e 85, D-22607 Hamburg, Germany}\\
	{\it $^{{d}}$    CPHT, CNRS, Ecole Polytechnique, IP Paris, F-91128 Palaiseau, France}\\
\end{center}

\vspace{.2in}

\begin{abstract}
We use the RG framework set up in \cite{paper2} to explore the $\phi^3$ theory with a random field interaction.
According to the Parisi-Sourlas conjecture this theory admits a fixed point with emergent supersymmetry which is related to the pure Lee-Yang CFT in two less dimensions. 
We study the model using replica trick and Cardy variables in $d=8-\e$ where the RG flow is perturbative. 
Allowed perturbations are singlets under the  $S_n$ symmetry that permutes the $n$ replicas.
These are decomposed into operators with different scaling dimensions: the lowest dimensional part, `leader’, controls the RG flow in the IR; the other operators, `followers’, can be neglected.
The leaders are classified into: susy-writable, susy-null and non-susy-writable according to their mixing properties. 
We construct low lying leaders and compute the anomalous dimensions of a number of them. 
We argue that there is no operator that can destabilize the SUSY RG flow in $d\le 8$. This agrees with the well known numerical result for critical exponents of Branched Polymers 
(which are in the same universality class as the random field $\phi^3$ model)
 that match the ones of the pure Lee-Yang fixed point according to dimensional reduction in all $2\le d\le 8$. Hence this is a second strong check of the RG framework that was previously shown to correctly predict loss of dimensional reduction in random field Ising model.
\end{abstract}
\vspace{.3in}
\hspace{0.7cm} 
{March 2022}
\newpage
\tableofcontents



\section{Introduction}\label{sec:intro}

Physical systems with randomly distributed impurities 
can be modelled through   
disordered quantum field theories. Random Field (RF) models constitute a class of such theories where a disorder field is coupled to a local order parameter. 
We are interested in the simplest RF models described by the following continuous action
\be\label{RFgen}
\mathcal{S}[\phi,h]=\int d^dx \Big[ \frac{1}{2}(\partial_\mu \phi)^2+V(\phi)+h(x)\phi(x) \Big]\,,
\ee
where the disorder field $h$ is coupled to a single scalar field $\phi$ for a given potential $V$.
This continuous description conveniently captures the physics of the theory near the phase transition, where the microscopic details of the model are inconsequential. 
Typically the disorder field is chosen to be Gaussian, with a zero mean. Physical observables are obtained by taking correlation functions $\langle A(\phi) \rangle_h$ for a given configuration of $h$ 
and then computing their average $\overline{\langle A(\phi)\rangle_h}$ over the disorder. The disordered coupling is relevant if $\D_{\phi}< \frac{d}{2}$. This is called Harris criterion. Since $\D_{\phi}=\frac{d-2}{2}+\eta$ (with $\eta \ll 1$) close to the usual upper critical dimension, the disorder is strongly relevant and changes the universality class of the critical system. 

A notable example of RF theories is the Random Field Ising Model (RFIM), 
which can be defined on the lattice by adding to the usual Ising model a term which couples the  disorder field  to a local Ising spin.
 This corresponds to $V(\phi)\propto \phi^4$ in the action above. In this paper we focus on another important example: Random Field $\phi^3$ model (denoted RF $\phi^3$ model), where we set $V(\phi)\propto \phi^3$.  We discuss it in a moment. { The analysis for both $\phi^4$ and $\phi^3$ potentials appeared  in a shorter paper \cite{papersummary}}.

In  \cite{Aharony:1976jx} it was first suggested that RF theories of the type \eqref{RFgen} have a phase transition with an interesting feature: its critical point is related to that of the same system without disorder in $d-2$ dimension.
 This was explained by a conjecture due to Parisi and Sourlas in 1979 \cite{Parisi:1979ka}, in the context of RFIM. The conjecture has two essential parts:
\begin{enumerate}
	\item \textit{Emergence of supersymmetry:} The IR fixed point of the RF theory is equivalent to the fixed point of a nonunitary supersymmetric theory (with supercharges that do not transform in a spinor representation). 
	This is the Parisi-Sourlas CFT. 
	\item \textit{Dimensional reduction:} The correlation functions of the SUSY CFT restricted to a $d-2$ hyperplane are equal to those of a $d-2$ dimensional CFT with the same action as the disordered model but without the disorder.
\end{enumerate}
The above implies that critical exponents of a RF model should be the same as those of a  CFT$_{d-2}$. However this is not always true as found in the RFIM case from a number of numerical studies. Indeed the  model shows dimensional reduction from $5\to 3$ \cite{Picco2,Picco3} but not for $4\to 2$ \cite{Picco1} or $3\to 1$.\footnote{The case $3\to 1$ cannot work as there is no phase transition for Ising model in 1d.} This means that one of the two parts of Parisi-Sourlas mechanism should stop working for RFIM at some intermediate dimension. In \cite{paper1} part 2 of the conjecture was explored in the framework of axiomatic CFTs. It was shown that there was no issue with dimensional reduction in the Parisi-Sourlas SUSY CFT.

Part 1 of the conjecture was explored in \cite{paper2}. A perturbative RG framework was set up to study the model. It was shown that the emergent SUSY fixed point ot the RFIM is in fact unstable below a critical dimension. The key points of the analysis were the following:
%
\begin{itemize}
\item 
One represents the RF model by using a replica  Lagrangian. In this formulation, the disorder field is integrated out and one is left with a pure quantum field theory of $n$ coupled fields $\phi_{i=1, \dots n}$ in the limit $n\to 0$. 
\item The fields $\phi_{i}$ do not have a well defined scaling even in the Gaussian theory (for $V=0$). This is problematic for RG purposes. It is thus convenient to perform a linear map in field space \cite{CARDY1985123}, dubbed Cardy transform, which gives rise to a set of fields with well defined scaling dimensions. 
\item 
Near the upper critical dimension $d_{uc}$ the model is described by a Gaussian action perturbed by a weakly relevant interaction. This gives rise to an IR fixed point equivalent to the Parisi-Sourlas SUSY CFT.
\item There is an infinite number of other perturbations, that are irrelevant close to $d=d_{uc}$. The fixed point would become  unstable if any of these perturbations becomes marginal at $d<d_{uc}$. 
\item The perturbations are all $S_n$-singlets i.e. invariant under the $S_n$ symmetry that permutes the $n$ replicas. 
These singlets (like the fields $\phi_{i}$ themselves) do not have well defined scaling dimensions. Indeed, after Cardy transform, they can be written as a sum of operators of different dimensions.
The  lowest dimensional part of  the singlet is called ``leader operator", the other parts are named ``followers".
\item The leader operators fully control the behaviour of the $S_n$-singlet perturbations in the IR. It is therefore enough to study the anomalous dimensions of leaders (in a theory ---dubbed $\mathcal{L}_L$--- where all followers are set to zero) to know if the $S_n$-singlet perturbations are relevant or irrelevant. 
\item The leader operators are further classified into 3 categories: susy-writable, susy-null and non-susy-writable.  
\item The anomalous dimensions of a number of low dimension leader operators in each category were computed in \cite{paper2}. This showed that a few $S_n$-singlet perturbations do become marginal at some critical dimensions estimated as $d_c=4.2-4.7 $ thus rendering the fixed point unstable at $d<d_c$. 
\end{itemize}

We review the framework in detail in section \ref{sec:RG}. 
The result of the above RG framework nicely explains when and why dimensional reduction works for the RFIM. 
If the framework is correct it should also be applicable to other known cases giving results consistent with the numerical predictions. 
In this paper we test the above framework on the RF $\phi^3$ model.

Let us briefly review some features of this model.
First let us consider the pure version of the RF $\phi^3$ theory. This is described by the scalar $\phi^3$ theory which has upper critical dimension $d_{uc}=6$ . This theory has a microscopical description realized by considering the Ising model in an imaginary magnetic field. The critical properties of the model are captured by the well-known Lee-Yang universality class \cite{PhysRev.87.404, PhysRev.87.410, PhysRevLett.40.1610, PhysRevLett.54.1354}. 

Once we add the disorder, by the Harris criterion, the fixed point must be in a different universality class. 
Interestingly the universality class of the RF $\phi^3$ fixed point describes the critical properties of branched polymers in a solution. 
Branched polymers are defined as polymers ---long chains of single units called monomers--- which branch into other polymers. Depending on the interactions between the parts of the chain they can appear in extended or collapsed configurations. One can study the properties of the polymers near the transition between these two phases. 
 It is also possible to define the branched polymers on the lattice as connected clusters of sites called `Lattice Animals'. 
The number of their possible configurations and their average size, scale with some critical exponents that are known from Monte-Carlo for all dimensions $2\leq d < 8$ \cite{Gaunt_1980,Redner_1979}. 
 Lubensky and Isaacson \cite{PhysRevA.20.2130} proposed that branched polymers can also be described in field theory with the $n\to 0$ limit of a Replica Lagrangian with cubic interaction  that allows perturbative computations in $d=8-\e$  (see appendix \ref{app:BPreview} for a review). Parisi and Sourlas \cite{PhysRevLett.46.871} showed that this Lagrangian also describes the  RF $\phi^3$ model. Through the Parisi-Sourlas conjecture this is related to Lee-Yang fixed point without disorder in $d-2$ dimensions. Critical exponents in the lower dimensional theory are known from both perturbative computations (in $6-\e$ dimensions) and Monte-Carlo. They agree with the branched polymer critical exponents, showing that both parts of Parisi-Sourlas conjecture indeed work for this model. 

The fact that Parisi-Sourlas conjecture works in this model 
is supported by the result of Brydges and Imbrie \cite{zbMATH02068689} who found a manifestly supersymmetric model of Branched polymers, where the dimensional reduction can be rigorously shown (see \cite{CardyLecture} for a review). 
However that analysis does not explain why a generic non-supersymmetric model should also have a SUSY fixed point, namely why SUSY breaking perturbations are irrelevant (see  App. A.4 of  \cite{paper2} for more details on this point). 
Therefore one important goal of this paper is to explain using  the RG analysis of \cite{paper2} how RF $\phi^3$ fixed point flows to a stable SUSY fixed point that admits dimensional reduction in all allowed dimensions $d <8$.

The rest of the paper is organized as follows: In section \ref{sec:RG} we review the RG framwork of \cite{paper2} focusing on the RF $\phi^3$ theory. We introduce `leader' and `follower' operators and show how the SUSY fixed point arises. In section \ref{sec:anomalous_dim} we classify all the leader operators into three categories. In each category we compute the anomalous dimensions of the ones that have low classical dimensions and break SUSY. In section \ref{stability} we use the results to argue that the SUSY fixed point is stable.  We conclude in section \ref{sec:discussion}. There are four appendices that show various technical details and computations. 


\section{The RG flow of random field models}
\label{sec:RG}
\begin{figure}[h]
	\centering \includegraphics[width=250pt]{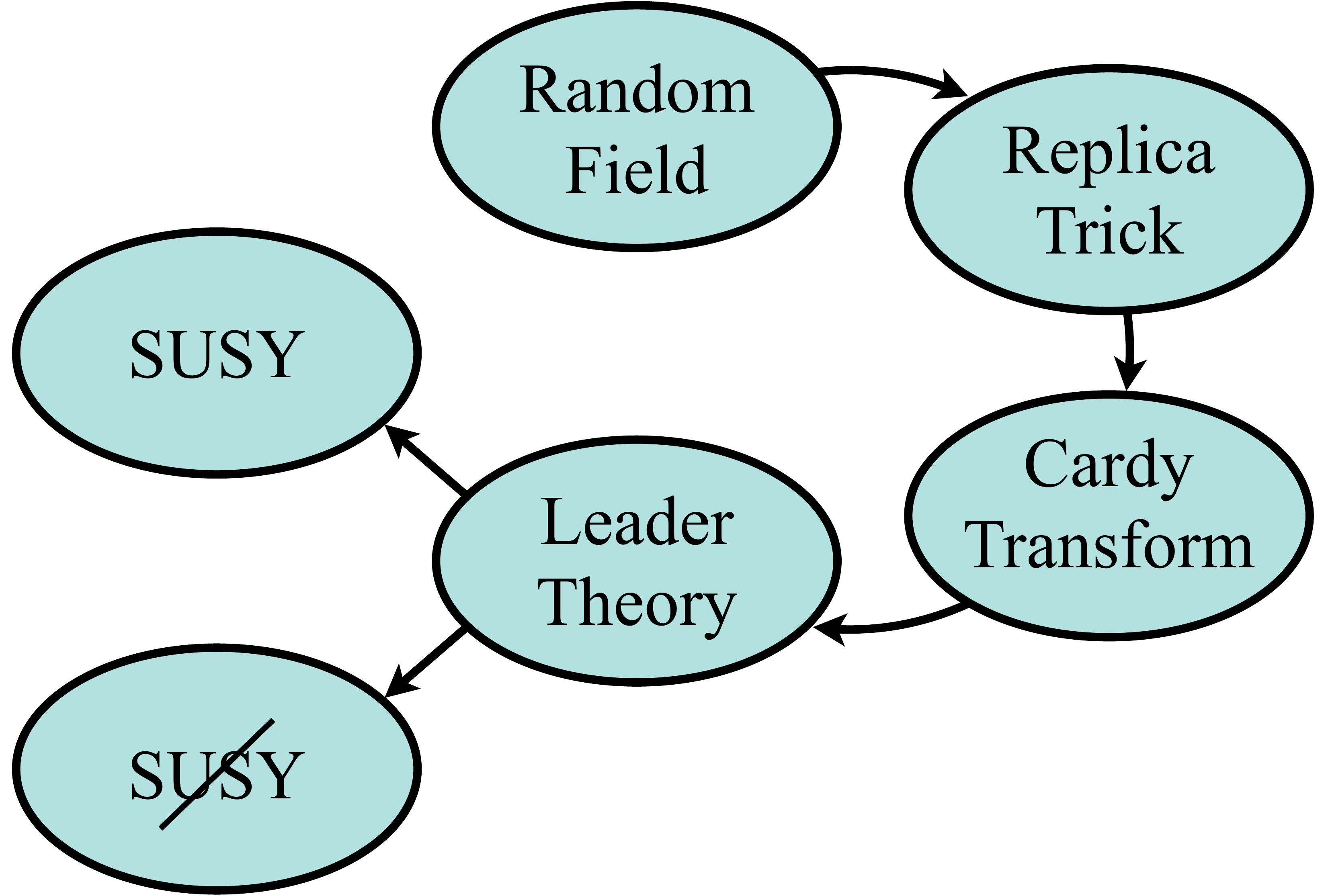}
	\
	\caption{
	A flowchart summarizing the RG framework of \cite{paper2} for a random field model. 
	\label{flowchart}
	}
\end{figure}

In this section we investigate how SUSY emerges  in the RG flow of a RF model. In order to study this question we follow the logic of \cite{paper2}, summarized in Fig.\ref{flowchart}.
We start the next subsection by reviewing how to compute quenched disorder correlation functions through the method of replicas \cite{Wegner:2016ahw}. We then define a transformation in field space (first introduced by Cardy \cite{CARDY1985123}) which makes the scaling properties of the UV Gaussian fixed point more transparent. In section \ref{RG_Cardy} we explain how to do RG in these new variables and we show that the IR properties of the model are captured by the ``leader theory'', which is specified by a simpler Lagrangian $\Lcal_{L}$. 
 Thanks to this formulation we can easily classify the spectrum of the theory and see which perturbations are relevant and whether they give rise to a SUSY IR fixed point or not. 
  In subsection  \ref{SUSY_small_eps} we will focus on the case in which  $\e=8-d$ is small and we will  easily show that  $\Lcal_{L}$  can be mapped to a SUSY model with Parisi-Sourlas supersymmetry. 
Finally in subsection \ref{dimensional_reduction} we explain that the resulting SUSY model can be dimensionally reduced to a non-SUSY theory in $d-2$ dimensions. 
 In principle, when $\e$ is of order one, new relevant SUSY breaking operators could destabilize the SUSY fixed point, as it happens for the RFIM. This scenario for RF $\phi^3$ will be considered and ruled out in  sections  \ref{sec:anomalous_dim} and \ref{stability}.



\subsection{Method of replicas and Cardy transform}\label{replicas}

Quenched averaged correlators are defined by first averaging over the field $\phi$ and in the end over the random magnetic field $h$
\begin{equation}
  \label{formula:1copy} \overline{\langle A (\phi) \rangle} = \int
  \mathcal{D}h\,\mathcal{P} (h) \hspace{0.17em} \frac{1}{Z_h}  \int \mathcal{D}
  \phi\, A (\phi) e^{- \Scal [\phi, h]} ,
\end{equation}
where overbar denotes the disorder average with distribution $ \mathcal{P}(h)$,  $\mathcal{S} [\phi, h]$ is the random field action introduced in {\eqref{RFgen}} and $A (\phi)$ is a function of
the field $\phi$, e.g.~$A (\phi)=\phi(x_1) \dots \phi(x_n)$. In this paper we will focus on a gaussian disorder distribution $ \mathcal{P} (h) \propto \exp({- \frac{1}{2 \tmH}  \int d^d x\,h (x)^2})$.
Because of the presence of the denominator $Z_h$, equation \eqref{formula:1copy} is notoriously complicated to compute.
The method of replicas is a prescription to compute these correlation functions by eliminating the factor $Z_h^{-1}$.

The idea is to multiply and divide the integrand of {\eqref{formula:1copy}} by  $Z_h^{n - 1}$. The resulting equation is independent of $n$, thus we can take the limit $n\to 0$, where the denominator $Z_h^{n }\to 1$. 
The $Z_h^{n - 1}$ in the numerator is instead considered as a product of partition functions for fields $\phi_2, \dots \phi_n$ which combine with the original functional integral where $\phi$ is renamed $\phi_1$. The net result is that we get rid of the denominator $Z_h$ at the price of having $n$ replica fields $\phi_1, \dots, \phi_n$.  The advantage of this trick is that the average over $h$ is now just a Gaussian integral which can be performed giving
\begin{equation}
  \label{eq:Sr1} \overline{\langle A (\phi) \rangle} = \lim_{n \rightarrow 0} 
  \int \mathcal{D} \vec{\phi}\, A (\phi_1) e^{- \Scal_n [\vec{\phi}]}
  \equiv \langle A (\phi_1) \rangle,
\end{equation}
\begin{equation}
  \label{Sr} \Scal_n [\vec{\phi}] = \int d^d x \left\{ \sum_{i = 1}^n \left[
  \frac{1}{2} (\partial_{\mu} \phi_i)^2 + V (\phi_i) \right] - \frac{\tmH}{2}
  \left( \sum_{i = 1}^n \phi_i \right)^2 \right\} .
\end{equation}

We can also generalize this construction to compute the disorder-average of a product of several
correlators as follows
\begin{equation}
  \overline{\la A_1 (\phi) \ra \cdots  \langle A_N (\phi) \rangle} =
  \lim_{n \to 0} \int \mathcal{D} \vec{\phi} \,A_1 (\phi_{i_1}) \dots A_N (\phi_{i_N})
  e^{- \Scal_n [\vec{\phi}]} = \langle A_1 (\phi_{i_1}) \dots A_N (\phi_{i_N})
  \rangle, \label{form2}
\end{equation}
where the result is independent of the choice of indices $i_1, \dots , i_N$ as long as they are all different. 
While the definition (\ref{formula:1copy}) eludes the standard QFT framework because of the average over disorder, the reformulation in (\ref{eq:Sr1}-\ref{form2}) is conveniently written in terms a QFT of $n$ coupled fields (the only subtlety being the limit $n \to
0$). This formulation is thus well suited to study the RG of the theory.

From the quadratic part of $\mathcal{S}_n$ we  obtain the propagator
\begin{equation}
  \mathbf{G} = \frac{\mathds{1}}{k^2} + \frac{\tmH \mathbf{M}}{k^2  (k^2 - n \tmH)}
  \hspace{0.17em} \, , \label{eq:G}
\end{equation}
where $\mathbf{M}$ is an $n \times n$ matrix whose all elements are equal to one.
We see that, in the limit $n\to 0$, the two terms in the r.h.s. of \eqref{eq:G} have different scalings in $k$. 
This in turns implies that the fields $\phi_i$ at the UV fixed point do not possess  a definite scaling dimension. The idea of Cardy {\cite{CARDY1985123}} is to disentangle the two scaling behaviors through a linear transformation in field space. 
The Cardy transform is defined as follows
\begin{equation}
  \label{eq:Cardy} \varphi = \frac{1}{2}  (\phi_1 + \rho), \qquad \omega =
  \phi_1 - \rho \hspace{0.17em} \, ,
   \qquad 
 \chi_i= \phi_i-\rho  \hspace{0.17em}, \quad (i = 2
  \ldots n) \hspace{0.17em}, \end{equation}
  with $\rho = \frac{1}{n - 1}  (\phi_2 + \ldots + \phi_n)$. By construction, the variables $\chi_i $ are not independent and satisfy the constraint $\sum_{i = 2}^n \chi_i = 0$. The quadratic Lagrangian in $n \to 0$ limit thus becomes 
\begin{equation}
\label{eq:n=0gauss}
  \Lcal^{\text{free}}= \del \varphi \del \omega - \frac{\tmH}{2} \omega^2  + \frac{1}{2}  (\del \chi_i)^2 \, ,
\end{equation}
where from now on we leave implicit the sum over $\chi_i$, where $i$ runs from $2$ to $n$. We borrowed the notation $\Lcal_L$ from \cite{paper2} and the superscript `free' denotes the $V(\phi)=0$ case.

The transformed fields $\omega$, $\varphi$, $\chi_i$ have now well-defined scaling dimensions:
\begin{equation}
  \Delta_\vf = \frac{d}{2} - 2 \hspace{0.17em}, \hspace{1cm} \Delta_\chi = \frac{d}{2} -
  1 \hspace{0.17em}, \hspace{1cm} \Delta_\omega = \frac{d}{2} \hspace{0.17em} .
  \label{dimensions}
\end{equation}
Notice that the dimension of $\vf$ is one unit below the unitarity bound, signaling that the resulting theory is actually non unitary.
By inverting the kinetic term we then obtain the free propagators, which are now scale-covariant \begin{equation}
  \langle \vf_k \vf_{- k} \rangle = \frac{H}{k^4}, \quad \langle \vf_k
  \omega_{- k} \rangle = \frac{1}{k^2}, \quad \langle \omega \omega \rangle =
  0, \quad \langle (\chi_i)_k (\chi_j)_{- k} \rangle = \frac{1}{k^2} \left(
  \delta_{i j} - \frac{1}{n - 1} \Pi_{i j} \right) \, . \label{propsL0}
\end{equation}
Here $\Pi_{i j}$ is an $(n - 1) \times (n -1)$ matrix whose components are all equal to one.
In this new basis the powers $1 / k^2$ and $1 / k^4$ of {\eqref{eq:G}} appear separated.
Notice that to get such a neat separation we had to work at $n=0$. Let us show that indeed order $n$ terms are not important to study the RG of the random field model.

To model this problem we consider a Lagrangian $\mathcal{L}$ independent of $n$ perturbed by a relevant operator $\Ocal$ with dimension $\Delta<d$, which is multiplied by a coupling $g$ of order $n$ (which we think as a small parameter),
%
%
%
%
%
\begin{equation}
  \label{On} \frac{g}{\Lambda_{\mathrm{UV}}^{d-\Delta}}  \int d^d x \hspace{0.17em}
  \Ocal(x),
\end{equation}
with $\Lambda_{\mathrm{UV}}$ the UV cutoff energy scale and $g \propto n$.
For our problem $\mathcal{L}$ is a Lagrangian obtained by setting $n=0$ (e.g. the Gaussian Lagrangian $\mathcal{L}^{\text{free}}$, where we are also allowed to turn on any number of $n=0$ perturbations discussed in the next sections) while $\Ocal$ is any relevant operator that enters at order $n$  (there are many of such operators  e.g. $\varphi^2$). 
We want to see what happens to the RG of $\mathcal{L}$ when $n$ is small and understand how we are supposed to think of a line of theories parametrized by $n$ which are smoothly connected to the $n=0$ theory (indeed this is what we are supposed to have when when we use the method of replica). This is non trivial since there are two conflicting effects: on one hand $g$ looks unimportant because $g \propto n$ is  small in the UV, on the other hand  $g$ is a relevant coupling, thus it should grow and become important in the IR. Let us explain how we should understand these fighting effects.
The coupling $g$ ---which in the UV was order-$n$--- grows along the RG and becomes order-1 at a scale
\begin{equation}
\label{ordernIRscale}
  \Lambda_{\mathrm{IR}} \sim n^{\frac{1}{d-\Delta}} \Lambda_{\mathrm{UV}}  \, .
\end{equation}
At scales $\Lambda \lesssim \Lambda_{\text{IR}}$  the flow is no longer perturbative and we have no means to predict the fate of the theory (which may flow to a gapped phase or to new fixed point).
Formula \eqref{ordernIRscale} thus tells us that any fix point of the $n = 0$ RG is ultimately  unstable with respect to order $n$ perturbations.
So in practice no matter how small $n$ is, the coupling $g$ will eventually become large and destabilize the $n=0$ RG. One could then think that it is not possible to find a smooth interpolation between the $n>0$ theories with and the $n=0$ one. On the contrary, in the following we explain how this can be done.

Indeed, as $n$ approaches zero, the scale \eqref{ordernIRscale} (at which $g\sim 1$)  becomes smaller and smaller. 
In practice the $n\ll1$ instabilities are only triggered at parametrically large distances. 
This indicates that we should only focus on the regime $\Lambda\gtrsim  \Lambda_{\text{IR}}$. Because of the IR cutoff at these scales we can only reach approximate scale invariance. However by lowering  $n$ we are able to make the approximate scale invariant region larger and larger, finally reaching a true fixed point at $n=0$ (which is the one of $\mathcal{L}$ theory).


Very importantly in the regime $\Lambda\gtrsim  \Lambda_{\text{IR}}$ all \eqref{On} contributions are perturbative. They appear as a Taylor series in $n$ which can be dropped  when taking the limit  $n\to 0$. It is thus possible to drop them from the start.
In summary from now on we are allowed to consider the theory at strictly $n=0$ and drop all perturbations that which are proportional to $n$. This is going to be assumed in the rest of the paper.

\subsection{RG in Cardy variables} \label{RG_Cardy}
In the previous section we showed how the gaussian Lagrangian at $n=0$ takes a nice form in Cardy variables.
The next step is to consider the RG of the theory. 

Independently of the form of the potential, the replica action \eqref{eq:Sr1} is $S_n$ symmetric. In order to study the RG of \eqref{eq:Sr1}  we should consider all possible perturbations compatible with $S_n$, and possible extra symmetries if they are not broken by the form of the potential. E.g. in \cite{paper2} a quartic potential was considered, which allowed for an extra $\mathbb{Z}_2$ symmetry. In this work we will consider a cubic potential $V(\phi)=\phi^3$  in \eqref{eq:Sr1}, thus the $\mathbb{Z}_2$ symmetry will not be present.

$S_n$ symmetric preturbations are easily written in terms of the replica fields, e.g.
\begin{equation}
\begin{array}{ll}
 \sigma_k \equiv \sum \phi_i^k \, ,
 \qquad  
 &\sigma_{k (\mu)} \equiv \sum \phi_i^{k - 1} \partial_{\mu} \phi_i \, ,
 \\
 \sigma_{k (\mu \nu)} \equiv \sum \phi_i^{k - 1} \partial_{\mu}
  \partial_{\nu} \phi_i \, ,
 \qquad  
& \sigma_{k (\mu) (\nu)} \equiv \sum \phi_i^{k - 2} \partial_{\mu} \phi_i
  \partial_{\nu} \phi_i
  \, , 
\end{array}
 \label{skder}
 \end{equation}
and so on.
On the other hand in order to use the results of the previous section, we need to rewrite the $S_n$-singlet operators in Cardy variables. To do so we simply use \eqref{eq:Cardy} and we set $n=0$. The lowest dimensional singlets in Cardy basis are the following quadratic operators\footnote{In principle one can also consider the perturbation $ \sigma_1= \omega$. This however does not play any role, since it can be reabsorbed in the Gaussian Lagrangian \eqref{eq:n=0gauss} by a shift of fields, $\omega \to \omega + \mbox{const}$. Indeed, as we will explain in App. \ref{class}, this perturbation is equivalent to the perturbation $\hat \phi$ in $\hat \phi^3$ theory, which is also discarded because it can be reabsorbed by a shift of $\hat \phi$.}
\begin{equation}
\label{quadraticL0}
\begin{array}{rcl}
  \sigma_{2 (\mu) (\mu)}&=& \del \varphi \del \omega  + \frac{1}{2}   (\del \chi_i)^2 
      \, ,
  \\
  \sigma_1^2&=& \omega^2 
     \, ,
   \\
    \sigma_2&=&2 \vf \omega +\chi_i^2
    \, .
  \end{array}
\end{equation}
We see that $\sigma_{2 (\mu) (\mu)}$ and $ \sigma_1^2$ generate the Gaussian Lagrangian $\Lcal^{\text{free}}$ of \eqref{eq:n=0gauss}.\footnote{It is important to see that the two terms belong to different $S_n$-singlets, so the relative coefficient is allowed to change along an  $S_n$-preserving RG flow. This is not a problem since the IR fixed point is independent of such relative rescaling, as it is explained in section 3 of \cite{paper2}.} 
Since they generate the kinetic terms, they are by definition always marginal.
 The perturbation $\sigma_2$ instead has scaling dimensions $d-2$, thus it is always strongly relevant. It takes the role of a mass term, which we need to tune to zero in order to reach an IR fixed point. The next perturbation is $\sigma _3$ which appears due to the potential $V(\phi)=\frac{g}{6} \phi^3$ of the perturbed UV action \eqref{Sr}. In Cardy variables it takes the form
\begin{equation}\label{sigma3}
\sigma _3=\big[ 3 \varphi ^2 \omega +3 \varphi  \chi_i ^2 \big]_{\frac{3 d}{2}-4} +\big[\chi_i ^3\big]_{\frac{3 d}{2}-3}- \frac{3}{2} \big[ \chi_i ^2 \omega\big]_{\frac{3 d}{2}-2}+\frac{ 1}{4} \big[\omega ^3\big]_{\frac{3 d}{2}} \, .
\end{equation}
Here the subscript of the square brackets indicate the bare dimensions of the composite operators computed by summing the dimensions of the fields $\varphi, \chi_i, \omega$ in \eqref{dimensions}. We thus conclude that the singlet $\sigma _3$ can be written as a sum of operators of different dimensions, which is actually  a recurrent feature of the model. While looking very exotic, this feature should be expected. Indeed, as we stressed in the previous section, the fields $\phi_i$ which are used to define the $S_n$ singlets do not have a well-defined scaling dimension.
Thus we do not expect to find that combinations of $\phi_i$ give operators with definite dimensions. The quadratic operators in \eqref{quadraticL0} should be considered as exceptions to this rule. 
 In general a given $S_n$ singlet perturbation $\Ocal$ is written in Cardy basis as a sum of operators of different dimensions
\begin{equation}
\Ocal=\Ocal_L+\Ocal_{F_1}+\Ocal_{F_2}+\dots \, ,
\end{equation}
where we distinguish its lowest dimensional piece $\Ocal_L$ which we call \emph{leader} from all the higher dimensional terms $\Ocal_{F_k}$ which we dub  \emph{followers}, where the follower $\Ocal_{F_k}$ has dimension $k$-units bigger than $\Ocal_L$. As it should be clear from what follows, this difference in dimension does not renormalize, so the difference between the dimensions of a leader and its followers is the same in the UV and in the IR.

It is easy to see that leaders play a much more important role than followers to characterize the IR properties of the  RG. This can be shown in two steps.
First, due to $S_n$ symmetry, the form of the multiplet must be preserved when integrating out degrees of freedom.
E.g. if we introduce a momentum cutoff $\Lambda$ and we integrate out the momentum shell $[\Lambda',\Lambda]$, where $\Lambda/\Lambda'\equiv b>1$, the form of $\Ocal$ is not allowed to change, meaning that the relative coefficients between $\Ocal_L$ and the $\Ocal_{F_k}$ remain fixed and only the overall coupling is allowed to change,
\begin{equation}
g (\Ocal_L+\Ocal_{F_1}+\Ocal_{F_2}+\dots) \underset{\textrm{integrating out}}\longrightarrow \tilde g (\Ocal_L+\Ocal_{F_1}+\Ocal_{F_2}+\dots) \, .
\end{equation}
Here $g$ is the coupling associated to $\Ocal$ at a scale $\Lambda$, while $ \tilde g$ is new coupling after integrating out the momentum shell from $[\Lambda',\Lambda]$. 
  Second, when we rescale back to initial form of the action we see that the relative coefficients get rescaled differently and follower operators are more and more suppressed as we go to the IR,
  \begin{equation}
 \tilde g  (\Ocal_L+\Ocal_{F_1}+\Ocal_{F_2}+\dots) \underset{\textrm{rescaling}}\longrightarrow g(b) (\Ocal_L+ \frac{1}{b}\Ocal_{F_1}+\frac{1}{b^2}\Ocal_{F_2}+\dots) \, .
\end{equation}
In particular in the deep IR, $b \to \infty$ and all follower operators are set to zero with respect to the leader. 
We can define an RG equation for the perturbation $\Ocal$ as the variation of $g(b)$ in the RG scale $b$. Close to the fixed point this can be linearized as follows,
\begin{equation}
\label{RGg}
 \frac{d g(g)}{d \log b}=-y_{\Ocal} g(b) \, .
\end{equation}
From this construction it should be clear that the eigenvalue $y_{\Ocal}$ can be computed in a very simple way:
it is obtained as $y_{\Ocal} = \Delta_{\Ocal_L} -d$ where  $\Delta_{\Ocal_L}$ is the the scaling dimension of the leader operator  $\Ocal_L$ computed in a simpler theory ---which we call $\Lcal_L$--- defined by dropping all follower operators.
This is a remarkable conclusion. It gives us a very simple algorithm to check if an $S_n$ perturbation is relevant or not. 
The strategy is to construct the $S_n$  perturbations, map them in Cardy variables and drop all the follower contributions. By studying the anomalous dimensions of the leader operators in the $\Lcal_L$ theory we are then able to see if $y_{\Ocal}>0$ and thus the $S_n$  perturbation falls off in the IR, or if $y_{\Ocal}<0$ and the perturbation grows.
E.g. to study the IR effect of the perturbation $\sigma _3$, we should only consider its leader piece $3 \varphi ^2 \omega +3 \varphi  \chi_i ^2$, which has bare dimension $\frac{3 d}{2}-4$ and thus gives rise to a weakly relevant perturbation in $d=8-\epsilon$ dimensions. In the next section we will detail what happens to the IR fixed point of this RG flow. Before ending this section we shall make a final remark about the $\Lcal_L$ theory.

Of course, as it is written, equation \eqref{RGg} works when there are no other degenerate (i.e. of the same scaling dimension) perturbations. 
As expected, when some $S_n$ perturbations $\Ocal_i$  have degenerate leaders $\Ocal_{i L}$, one  gets coupled equations of the form $ \frac{d g_i(g)}{d \log b}=-(y_{\Ocal})_{ij} g_j(b) $ where $y_{\Ocal}$ is a mixing matrix which must be diagonalized to obtain the eigenvalues of the correct IR scaling perturbations. However one should be careful because sometimes the chosen basis for $\Ocal_i$ is such that two distinct perturbations $\Ocal_1$, $\Ocal_2$ have the same  leader $\Ocal_{1L}=\Ocal_{2L}$ (namely $\Ocal_{1L}$ and $\Ocal_{2L}$ do not only have the same dimensions but they are really the same operator) and different followers. In this case it would not be correct to say that the $\Ocal_1$ and $\Ocal_2$ are degenerate (for a detailed exemplification of why this is the case see App. B of \cite{paper2}). 
In order to correctly diagonalize the RG one should change the basis of $\Ocal_i$ and consider as a new  perturbation their difference,
\begin{equation}
\Ocal_1-\Ocal_2=0+(\Ocal_{1 F_1}-\Ocal_{2 F_1}) +(\Ocal_{1 F_2}-\Ocal_{2 F_2})+\dots \, .
\end{equation}
The new $S_n$ perturbation $\Ocal_1-\Ocal_2$ has a leader $(\Ocal_{1 F_1}-\Ocal_{2 F_1})$ which clearly has a different scaling dimension with respect to $\Ocal_{1L}$, thus $(\Ocal_1-\Ocal_2)$ and $\Ocal_1$ are not degenerate preturbations. 
When we say that $\Lcal_L$ is a theory of leader operators, we mean that one should first diagonalize the $S_n$ perturbations as above and only then one is allowed to set the follower operators to zero. In App. \ref{class} we construct such a diagonalized basis for the low lying $S_n$ perturbations $\Ocal_i$ where each perturbation is associated to a distinct leader  $\Ocal_{i L}$. The resulting basis of $\Ocal_{i L}$ defines the possible low lying perturbations of  the theory $\Lcal_L$.

\subsection{Emergence of supersymmetry for RF $\phi^3$ at small $\epsilon=8-d$} 
\label{SUSY_small_eps}
  If we work in $d=8-\epsilon$ (for  $\epsilon \ll1$),  it is easy to see that the full list of relevant leaders in the theory $\Lcal_L$ is given by the quadratic terms in  \eqref{quadraticL0} plus the leader
 $(\sigma _3)_L$.  
The Landau-Ginzburg Lagrangian thus takes the simple form 
\begin{equation} \label{L0}
\Lcal_L=\left[ \del \varphi \del \omega - \frac{\tmH}{2} \omega^2  + \frac{1}{2}  (\del \chi_i)^2  \right]+ m^2 (2 \vf \omega +\chi_i^2) + \frac{g}{2} ( \varphi ^2 \omega + \varphi  \chi_i ^2 ) \, ,
\end{equation}
Beside the strongly relevant mass term, which must be tuned to zero to reach the IR fixed point, the Lagrangian \eqref{L0} has the single weakly relevant perturbation $(\sigma _3)_L$, which triggers a short RG flow which can be studied perturbatively. This Lagrangian will thus be the starting point for all the following computations.

Before entering the perturbative computations, we would like to show that many of the observables of the theory \eqref{L0} are captured by a SUSY Lagrangian. This in turns implies that for small $\epsilon$ the random field $\phi^3$ model has an IR fixed point with emergent SUSY. 

By looking at \eqref{L0}  one notices that $\chi_i$ only appear quadratically, thus the associated partition function is defined through a Gaussian functional integral in the fields $\chi_i$ which can be performed. The Gaussian path integral of $n-2$ bosonic fields (there are $n-1$ fields subjected to one constraint) in the limit $n \to 0$ is equal to a fermionic Gaussian path integral. This motivates the following replacement, 
\begin{equation}
  \label{fromchitopsi} \frac{1}{2}  \sum_{i = 2}^n \chi_i  [- \del^2 +f
  (\vf)] \chi_i \overset{n \to 0}{\longrightarrow} \psi [- \del^2 + f (\vf)]
  \bar{\psi} \, ,
\end{equation}
which is valid for any function $f$ of the field $\vf$, where $\psi, \psib$ are fermionic fields which transform as Lorentz scalars i.e. they are not spinors. In other words we replace a Gaussian $O(-2)$ model by a Gaussian $Sp(2)$ model, where $O(-2)$-singlets $ (\chi_i)^2$ are replaced $Sp(2)$-singlets $(\psi \psib)$. 
The new fermionic action thus takes the form 
\begin{equation} \label{Lsusy}
\Lcal_{\textrm{susy}} =\left[ \del \varphi \del \omega - \frac{\tmH}{2} \omega^2  + \del \psi  \del \psib  \right]+ 2 m^2 ( \vf \omega + \psi \psib) + \frac{g}{2} ( \varphi ^2 \omega +2 \varphi  \psi \psib ) \, .
\end{equation}
The Lagrangian $\Lcal_{\textrm{susy}} $ is invariant under a special kind of supersymmetry named after Parisi and Sourlas \cite{Parisi:1979ka}. In order to make SUSY manifest we can rewrite the action in superspace as follows
\begin{equation}
  \label{SUSYPhi} \Scal_{\textrm{superspace}} = \int d^d xd \thetab d \theta
  \left[ - \frac{1}{2} \Phi D^2 \Phi + m^2 \Phi^2 + \frac{ g}{6} \Phi^3  \right] \hspace{0.17em},
\end{equation}
where $\theta, \thetab$ are fermionic coordinates  (which also transform as Lorentz scalars), $D^2 \equiv \del^2 - H \partial_{\theta} \partial_{\thetab}$ is the super-Laplacian and the superfield $\Phi$ can be expanded in components as 
 \begin{equation}
  \Phi (x, \theta, \thetab) = \varphi (x) + \theta \psib (x) + \thetab \psi
  (x) + \theta \thetab \omega (x) \hspace{0.17em} . \label{supefield}
\end{equation}
The supersymmetry enjoyed by $\mathcal{S}_{\text{superspace}}$ is the super-Poincar\'e group $\mathbb{R}^{d|2}\rtimes \text{OSp}(d|2)$, where $\mathbb{R}^{d|2}$ denotes super-translations and $\text{OSp}(d|2)$ super-rotations.

One needs to be a bit careful with the map \eqref{fromchitopsi}. While the path-integral of the two theories is equal, the two theories are actually different since they have a different set of operators. In particular operators that are not singlets under $\text{Sp}(2)$ and $O(-2)$ cannot be mapped from one theory to the other. E.g. $\chi_i^{k}$ for $k \neq 2$ has no counterpart in the $\text{Sp}(2)$ theory. Conversely e.g. $\psi$ itself has no counterpart in the  $O(-2)$ theory. In App. C of \cite{paper2} it was shown how to explicitly write a map between the singlet operators of the two theories (this is not trivial since it is possible to map also composite operators which contain derivatives, e.g. $
\chi_i  \partial_\m \chi_i
\leftrightarrow 
\psi \partial_\m\bar \psi+\partial_\m\psi  \bar  \psi $).

Since the $\chi$- and $\psi$-theories are somewhat different, one may wander if it is correct to study the fixed point of the $\mathcal{L}_L$ by using $\Lcal_{\textrm{susy}}$. However, it is easy to see that this step is completely rigorous. Indeed the Lagrangian $\mathcal{L}_L$ had accidental $O(-2)$ symmetry, therefore along its RG flow only $O(-2)$-singlets can be produced. These can in turn be mapped to Sp$(2)$ singlets. Thus the RG flow of the two theories is restricted to live inside the subspace of operators which exists (and it is equivalent) in both theories. 

To be more pragmatic one can compute the
 beta function for $g$ with both Lagrangians \eqref{L0} and  \eqref{Lsusy}, and obtain the same result (computed in dimensional regularization):
\be
 \beta_g ={- \frac{\epsilon g}{2} -
	\frac{3 g^3 H}{4 (4 \pi)^4}} +O(g^5)\, ,
\ee
where the mass term $m$ is tuned to zero in order to reach the fixed point, which is given by 
\be
H g_{\text{${\star}$}}^2 = - \frac{2 (4 \pi)^4 }{3} \e +O(\e^2). \label{gcrit} 
\ee
We stress that the same fixed point is reached independently of the value of $H$ (indeed by rescaling $\vf \to \sqrt{H} \vf$ and $\omega \to \omega/\sqrt{H}$ we can get rid of $H$ in the kinetic terms of the actions at the price of rescaling $g \to \sqrt{H} g $). In practice we can consider $H g^2$ as a single coupling.\footnote{
An interesting observation is that by taking  $H$ negative it is possible to make the critical coupling $g_{\star}$ real.
} 
At this fixed point we can easily compute the anomalous dimension for some singlet operators and obtain the same result using both formulation \eqref{L0} and  \eqref{Lsusy}. E.g. the one-loop anomalous dimension for $\vf$ and $\omega$ is $\g_\vf=\g_{\omega}=-\frac{\e}{18}$ (see  App \ref{app:fieldanomdim}). Notice that these are equal as expected since they belong to the same SUSY multiplet in the  \eqref{Lsusy} formulation.

In this section we thus learned that for small $\epsilon=8-d$ the IR properties of the random field $\phi^3$ model are captured by the leader Lagrangian $\Lcal_L$, which can mapped using \eqref{fromchitopsi} to the explicitly SUSY Lagrangian  \eqref{Lsusy}. We can conclude that  for small $\epsilon$ the random field $\phi^3$ model has emergent supersymmetry. The question that remains to answer is what happens when $\e$ is not small.
Before entering this discussion we review a property of the SUSY Lagrangian \eqref{Lsusy} which is going to be useful in the following.

%
%

\subsection{Dimensional reduction} 
\label{dimensional_reduction}
Theories with a Parisi-Sourlas type of supersymmetry like $\mathcal{S}_{\text{superspace}}$ in equation \eqref{SUSYPhi} have a remarkable property: most of the observables of the theory can be described in terms of a model living in $\widehat d=d-2$ dimensions which does not have any supersymmetry.
More precisely correlation functions of the $d$-dimensional SUSY theory with an interaction $V(\Phi)$ when  restricted to  a $\widehat d$-dimensional spatial hyperplane are equal to correlation functions of a scalar theory with the interaction $V(\widehat\phi)$ in $\widehat d$ dimensions. This map can be established for general axiomatic CFTs just by using the superconformal symmetries \cite{paper1}.

To illustrate this map we consider a $2$-point function,
\be
\langle\Phi(x_1,0,0)\Phi(x_2,0,0)\rangle = \langle\vf(x_1)\vf(x_2)\rangle = \langle\widehat{\phi}(x_1)\widehat{\phi}(x_2)\rangle\,, \ \ (x_i\in \mathbb{R})\,,
\ee
where the correlators in the l.h.s is computed using \eqref{SUSYPhi}  while the one in the r.h.s. is computed in the theory:
\be
\label{hphi3}
\mathcal{S}=\frac{4\pi}{H}\int d^{\widehat d}x \Big[ \frac{1}{2}(\partial\widehat{\phi})^2+m^2\widehat{\phi}^2+\frac{g}{6}\widehat{\phi}^3 \Big]\,.
\ee
The ${\widehat{\phi}}^3$-theory \eqref{hphi3} is a quite well studied model that can even be found as a toy example in some QFT textbooks  (see e.g. \cite{srednicki_2007}). 
At $\widehat d=6-\e$ the one-loop beta function for $g$, after tuning $m$ to zero, is given by:
\be
\beta_g=-\frac{\e g}{2}-\frac{3g^3}{4(4\pi)^3}+O(g^5)\,.
\ee
The fixed point occurs at 
\be
g_{\star}^2=-\frac{2(4\pi)^3}{3}\e +O(\e^2)\,.
\ee
One can study anomalous dimensions of operators at this fix point and get an exact match with the ones computed using the SUSY formulation \eqref{Lsusy}, e.g. $\g_{\widehat{\phi}}=\g_{\vf}= \g_{\psi}= \g_{\omega}$.
This fact is going to be very useful in the following since it allows us to relate observables of the SUSY theory \eqref{Lsusy} to the ones of the better studied and simpler ${\widehat{\phi}}^3$-theory. 




\section{Classifying perturbations  }
\label{sec:anomalous_dim}
For infinitesimal $\epsilon=8-d$ in the previous section we proved that the IR fixed point of the RF $\phi^3$ model must be supersymmetric and should therefore undergo dimensional reduction. This however may not be the case for larger values of $\epsilon$.
 Indeed if $\epsilon$ is of order one, other operators could in principle become relevant and destabilize the RG flow, as it happens for RFIM \cite{paper2}.
 Of course, as we reviewed in the introduction, numerical simulations are compatible with dimensional reduction \cite{PhysRevLett.46.871} and therefore we do not expect a destabilization to occur. However it is interesting to see how this works out theoretically using the RG setup described in the previous sections.  With this in mind, in this section we study the one-loop anomalous dimensions of a large set of operators. This information will then be used in section \ref{stability} to check if the SUSY fixed point is indeed stable for the RF $\phi^3$ model at $\epsilon$ of order one. 

In order to achieve this goal  we follow 
the strategy outlined in the previous section.
Namely we consider the $S_n$ singlet perturbations of the Lagrangian \eqref{Sr} in the limit $n \to 0$ which are simply captured by the leader perturbations of the model $\Lcal_L$. The latter being obtained by only considering the leading pieces of the $S_n$ singlets when written in terms of Cardy variables. 
So our objective is to systematically consider low-lying $S_n$ singlets and compute the anomalous dimensions of their associated leaders, checking if at (non infinitesimal) values of $\e$  they become relevant. 

A large set of low-lying leaders is presented in App. \ref{class}. As we shall see, leaders can be classified into 3 categories: 1. Susy-writable, 2. Susy-null, and 3. Non-susy-writable operators. This classification is justified since the three types of leaders mix with themselves only in a triangular way. This can be schematically shown as below:
\begin{align}\label{mixing}
\text{susy-null} \ & \leftrightarrow \ \text{susy-null}\nonumber\\
\text{susy-writable} \ & \to \ \text{susy-writable}\,, \ \text{susy-null}\nonumber\\
\text{non-susy-writable} \ &\to \ \text{non-susy-writable}\,, \ \text{susy-writable}\,, \ \text{susy-null}\,.
\end{align}
This means a susy-null operator mixes only with other susy-null operators, a susy-writable mixes with only susy-writables and susy-nulls, while a non-susy-writable operator mixes with all 3 types. In other words a renormalized susy-null operator can be written as a linear combination of only  susy-null operators, and so on.

In the following we explain in detail how we define these three classes and we will further compute the anomalous dimensions of the low lying operators of each class. Finally in section \ref{stability} we will collect these results and we will comment on the stability of the SUSY fixed point.

\subsection{Susy-writable leaders}\label{sec:susywritable}

We refer to the operators invariant under $O(n-2)$ and which do not vanish after the substitution  $\c \to \psi$ (i.e. they become Sp$(2)$ invariant after the substitution) as susy-writable operators. They can thus be written in terms of the susy-fields $\varphi,\psi,\psib, \omega$. Since for these operators the map between $\c$ and $\psi$ is a bijection, with an abuse of language we sometimes will refer  to the operators already written in terms of $\psi, \psib$ variables as susy-writable operators. However one should keep in mind that the actual operators are the ones defined in terms of the $\chi$ fields.

Susy-writable leaders are the most frequent ones in the low lying spectrum of the $\Lcal_L$ theory. This may sound surprising since they come from $S_n$ singlets which do not have any knowledge of the emergent supersymmetry of $\Lcal_L$.
The reason why this happens is that given an $S_n$ singlet of the form  $\sum_{i = 1}^n A (\phi_i)$ (for any function $A$) its leader can be written as
\begin{align}\label{master}
\left(\sum_{i = 1}^n A (\phi_i) \right)_L& = \Big[ A'(\varphi) \omega 
+  \frac{1}{2}A''(\varphi)  \chi_i^2 \Big] \, .
\end{align}
This form is explicitly susy-writable, indeed it can be written as the highest component of the composite superfield $A(\Phi)$, namely $\partial_{\theta}\partial_{\thetab}A(\Phi)$.  Similarly, product of the leaders \eqref{master} are still  susy-writable. From this point of view it may in fact seem that only susy-writable leaders exist in the theory. This is of course false as it is easy to see.
 Indeed very often two operators built out of products of singlets of the form  \eqref{master} have the same leader. In this case  the prescription of section \ref{RG_Cardy} tells us that we should subtract them to obtain a subleading contribution. The latter is either non-susy-writable or susy-null.  Many examples of this phenomenon are presented in App. \ref{class}.  { For more discussion see App. B of \cite{papersummary}. }
 

It is important to stress that susy-writable leaders belong to a special class of susy-writable operators. 
As we exemplified below equation \eqref{master}, they can be written as the  the highest components $\Ocal_{\theta \thetab}\equiv \partial_{\theta}\partial_{\thetab} \Ocal $ of superfields $\Ocal$.
In other words, once written as susy fields, they are invariant under super-translations.  See  \cite{papersummary} App. B for a formal proof of this statement.\footnote{ It was also shown in \cite{papersummary} that any Sp$(2)$ and super-translation invariant operator is a susy-writable leader.}

On the other hand susy-writable leaders do not have to be invariant under $\text{OSp}(d|2)$ super-rotations, they are only required to be singlets of $\text{SO}(d) \times \text{Sp}(2) \subset \text{OSp}(d|2)$,\footnote{
They must be $\text{Sp}(2)$ singlets by definition of susy-writable operators.
They must be $\text{SO}(d)$ scalars since the relative $S_n$-singlets perturbations are also scalars (non-scalar perturbations would explicitly break rotation symmetry).
} as it is the case e.g. for the operators $\s_1^2$ and $\s_{2(\m)(\m)}$ in \eqref{quadraticL0}. 
Since superrotations are not preserved, along the RG flow we may also find superfields which transform in non-trivial representations $\Ocal^{a_1 \dots a_\ell}$ of $\text{OSp}(d|2)$, where  we use the notation of \cite{paper1} and define latin $\text{OSp}(d|2)$ tensor indices as $a_i=1,\dots, d,\theta, \thetab$. 
Indeed we can be obtain  singlets of $\text{SO}(d) \times \text{Sp}(2)$  by contracting the $\text{OSp}(d|2)$ indices of the superfields with the $\text{Sp}(2)$-metric, e.g. $\Ocal^{a b} g^{\text{Sp}(2)}_{a b} =-\Ocal^{\theta \thetab}+\Ocal^{\thetab \theta}$. This procedure gives rise to operators with indices set to the $\theta$ and $\thetab$ directions.
 However one should notice that the $\text{Sp}(2)$-metric cannot be contracted to graded-antisymmetric directions otherwise the result vanishes. Also operators $\Ocal$ with more that $2$ graded-symmetric indices vanish when contracted to the $\text{Sp}(2)$ metric because $\Ocal^{\theta \theta a_3 \dots a_\ell}=0$.
 In practice this means that we should only consider irreducible representations of the kind $(2,0, \dots)$, $(2,2,0, \dots)$,  $(2,2,2,0 \dots)$, etc. \footnote{Here $(a,b,\cdots)$ denotes the Young Tableaux that represents mixed symmetry OSp$(d|2)$ tensors \cite{paper1} with $a$ boxes in the first row, $b$ boxes in the second row, etc.} 

The final result is that only superfield components of the following form do contribute under the RG (recall that lower $\theta \thetab$ refer to the superfield component in $\theta, \thetab$ expansion, while upper $\theta \thetab$ refer to OSp$(d|2)$ indices)
\begin{equation}
\Scal_{\theta \thetab} \, ,
\qquad 
\Jcal^{\theta \thetab}_{\theta \thetab} \, ,
\qquad 
\Bcal^{\theta \thetab, \theta \thetab}_{\theta \thetab} \, , 
\quad 
\dots  
\end{equation} 
where $\Scal$ is a superscalar,  $\Jcal^{ab}$ transforms in the spin-$2$ representation and $\Bcal^{ab,cd}$ in the box representation $(2,2)$ where $(a,b)$ and $(c,d)$ are the graded-symmetric pairs.
 The dots take into account the contributions from other representations with more indices, e.g. $(2,2,2)$, $(2,2,2,2)$ and so on. 
In the following we will not consider the latter since all free theory primaries in these representations have very high conformal dimensions (many fields and many derivatives are needed to allow for the required antisymmetrizations). 
Notice also that these higher irreps do not appear in integer dimensions $d \leq 5$ ---e.g. the dimension of the representation $(2,2,2)$ is equal to $\frac{1}{144} (d-5) (d-1) d^2 (d+1) (d+2)$ and vanishes for $d=5$.
On the other hand they exist in non-integer dimensions. It would be interesting to study the lightest operators in these irreps and to compute their anomalous dimensions. We will not attempt this here. In the following we will focus on the lightest operators in the scalar, spin $2$ and box representations.

Luckily we already know a lot about these operators because they are related to operators in the dimensionally-reduced theory. 
Indeed, given a superfield $\mathcal{O}$ with highest component  $\mathcal{O}_{\theta\thetab}$, the following relation holds
\be\label{susyeqn}
\D_{\mathcal{O}_{\theta\thetab}}=\D_{\mathcal{O}}+2 = \D_{\widehat O}+2\, ,
\ee
where $\D_{\widehat O}$ is the dimension of the operator $\widehat O$ in the $\widehat d \equiv d-2$ dimensional theory to which ${\mathcal{O}}$ gets mapped under dimensional reduction. Therefore the susy-writable leader $\mathcal{O}_{\theta\thetab}$ is relevant in $d$ dimensions exactly when the operator $\widehat O$ is relevant in the dimensionally reduced theory.

\subsubsection*{Scalars}
Let us first consider the case of scalars.
For the $\widehat{\phi}^3$ fixed point it is known that there is only one operator which is relevant in $\widehat{d}$ dimensions, which corresponds to the mass term $\widehat{\phi}^2$.  So we must have only one susy-writable leader operator that is relevant, which corresponds to $[\Phi^2]_{\theta\thetab}$. This is indeed equal to the term $\s_2$ which was considered in the previous section (see equation \eqref{quadraticL0}) and that must be tuned to reach the fixed point. 
All other scalar operators are irrelevant in $\widehat{d}$ dimensions, thus there cannot exist other relevant susy-writable scalar leaders (remarkably we are able to make this statement without any computation).

\subsubsection*{Spin two}
We can now focus on spin two operators.
In $\widehat d=6$ the lowest spin $2$ operator is the conserved stress tensor $\widehat T$ which has dimension $\D_{\widehat T}=\widehat d=6$. Any other spin 2 operator has higher dimension. We thus conclude that in $d=8$ the leader  $\mathcal{T}_{\theta\thetab}^{\theta\thetab}$ is marginal and that any other spin two susy-writable leader is irrelevant.
One may be worried that in lower dimensions (namely $\widehat{\phi}^3$ theory in $\widehat d<6$) one would get new relevant operators. However, since the stress tensor is always marginal,  this would  imply that the new operators would have to cross it.  Since level crossing is unlikely in an interacting non-integrable model, we conclude that all  spin two operators should stay irrelevant even at $\widehat d<6$.\footnote{For unitary theories one could use the unitarity bounds to argue that that all spin-two operators above the stress tensor are irrelevant. We avoided this argument since the $\widehat{\phi}^3$ fixed point is known to be non-unitary.  }
Thus the only spin two operator that we should consider is  the component $\mathcal{T}_{\theta\thetab}^{\theta\thetab}$ of the super stress tensor itself which  corresponds to the following susy-writable leader at  $d=8$ (see App. C of \cite{paper1}) 
\begin{equation}
\label{leaderTthetathetab}
\mathcal{T}_{\theta\thetab}^{\theta\thetab}
\;
\underset{\psi \to \chi}{ \longrightarrow }
 \;
 -\frac 23 \partial \vf\partial \omega-\frac{1}{3}(\partial \chi_i)^2+2 H \omega^2= \left(-\frac 23\s_{2(\m)(\m)}+2H \s_1^2\right)_L 
  \, .
\end{equation}
One could ask what happens when we add this perturbation to the action. On one hand since $\mathcal{T}_{\theta\thetab}^{\theta\thetab}$ is always marginal one may imagine that the IR fixed point is a one-dimensional conformal manifold. However this intuition is incorrect as we explain in what follows. Indeed the super stress tensor can be defined as a variation of the action with respect to a superspace background metric $g_{ab}$.
Conversely we can say that a perturbation by $\mathcal{T}^{\theta \thetab}_{\theta \thetab}$ produces a change of $g_{\theta \thetab}$,
\begin{equation}
g_{\theta \thetab}\to g_{\theta \thetab}+ \d g_{\theta \thetab} \;  \longleftrightarrow \; \d S=- \frac{1}{2} \int  d^d x \sqrt{|g|}  \d g_{\theta \thetab} \mathcal{T}^{\theta \thetab}_{\theta \thetab} \, .
\end{equation}
Notice that the metric of  \eqref{SUSYPhi} is defined such that  $g_{\theta \thetab}=H/2$, so the perturbation $\mathcal{T}^{\theta \thetab}_{\theta \thetab}$ actually modifies only the value of $H$. This is also visible from the explicit form of the operator \eqref{leaderTthetathetab}, which when added to the quadratic Lagrangian \eqref{eq:n=0gauss} has the net effect of changing the relative coefficient of the kinetic term, namely $H$. As we explained below equation \eqref{gcrit}, changes of $H$ do not have physical consequences on the theory. Therefore $\mathcal{T}_{\theta\thetab}^{\theta\thetab}$   should be considered as a ``redundant'' perturbation (see \cite{Wegner:1976bn}) and should be discarded.

\subsubsection*{Box}
Finally we consider the box representation. 
In $\widehat{\phi}^3$ theory we can write an infinite tower of box operators  as follows (see App \ref{App:Susy_Writable} for a detailed discussion),
\begin{equation}
\widehat{B}^{(k)}_{\mu \nu , \rho \sigma} \equiv \widehat{\phi}^{k-3} \left( \widehat{\phi}_{, \mu \nu} \widehat{\phi}_{,
	\rho \sigma} \widehat{\phi} - \frac{2 \widehat{d}}{\widehat{d} - 2} \widehat{\phi}_{, \mu}
\widehat{\phi}_{, \nu} \widehat{\phi}_{, \rho \sigma}  \right)^Y \, ,
\label{Box-n}
\end{equation}
where $\widehat{B}^{(k)}$ is the lowest dimensional operators in a box representation built out of $k\geq3$ fields $\widehat{\phi}$.
Here $Y$ denotes the box Young symmetrization and subtraction of traces and the notation $A_{,\m_1 \m_2 \dots }$ stands for $\partial_{\m_1} \partial_{\m_2} \dots A$.
There is no argument that fixes the dimensions of $\widehat{B}^{(k)}$ nor were they previously computed, so we have to obtain them in $\widehat d=6-\e$ by explicit computation. 
This is done in App. \ref{app:susywrit}.
Using  \eqref{susyeqn} we thus conclude that the highest components $\mathcal{B}_{\theta\thetab}^{(k)}$   (see App \ref{App:Susy_Writable} for their explicit definition in terms of the SUSY fields) of the lightest box operators $\mathcal{B}^{(k)}$ have the following IR scaling dimension:
\begin{eqnarray}
\label{box_dim}
\D_{\mathcal{B}_{\theta\thetab}^{(k)}}&=\left(2k+6-\dfrac{k}{2}\e\right)_{\text{class}}+ \left( \dfrac{1}{6} \left(2 k^2-5 k-2\right) \epsilon \right)_{1\text{-loop}} +O(\e^2) \,, 
\end{eqnarray}
This ends our classification of the low-lying spectrum of susy-writable leaders. 
To summarize,  we explained that after the $\chi \to \psi$ map they can be classified in terms of their $\text{OSp}(d|2)$ representations. We argued that  for our problem the most important ones are the scalar, spin two and box representations.
We explained how to obtain anomalous dimensions of  susy-writable leaders using the knowledge of the dimensionally reduced theory. Without any computation we concluded that scalar and spin two leaders cannot be responsible for the destabilization of the SUSY fixed point. 
Finally we computed the dimension \eqref{box_dim} of the first  susy-writable leader in the box representation.
This is the susy-writable leader which we should worry the most about becoming relevant in $d<8$.
In section  \ref{stability} we will comment on if this perturbation can actually be responsible for destabilizing the SUSY fixed point.

\subsection{Susy-null leaders}\label{sec:susynull}

 A susy-null operator ---like a susy-writable operator--- is a composite operator built of $\vf,\omega,\chi_i$-s that is invariant under $O(n-2)$ and thus can be rewritten using the map  $\chi\to\psi,\psib$. The novelty is that after this map the resulting operator vanishes.
 E.g. $(\chi_i^2)^2$  maps to $(\psi\psib)^2=0$. 
 Since they vanish in the SUSY theory these operators have very restrictive mixing properties. Indeed susy-null perturbations cannot affect susy-writable observables. 

In App. \ref{class}  we list the susy-null leader operators with low classical dimensions in 8d. 
We find that there exists an infinite class of $S_n$ singlets which plays an important role
\be
\label{def:NkSingl}
\mathcal{N}_{k}=\frac{2}{k-3} \left(\frac{\sigma _2 \sigma _{k-2}}{k-2}-\frac{2 \sigma _1 \sigma _{k-1}}{k-1}\right) \, ,
\ee
for $k=4,5 , \dots, \infty$.
Indeed the leaders associated to $\mathcal{N}_k$ take the following simple form
\begin{equation}
(\mathcal{N}_{k})_L=\varphi^{k-4}(\chi_i^2)^2 \, ,
\end{equation}
which makes them the lowest dimensional null operators made of $k$ fields. Because of this property they cannot mix with any other operator in perturbative computations. We can then quite easily compute their one-loop dimensions, which in $d=8-\e$ are given by (see App. \ref{app:susynull})
\begin{align}
  \label{Nkgenphi3}
\D_{\mathcal{N}_k}&=\bigg(2 (k+2)-\frac{k}{2}  \epsilon\bigg)_{\textrm{class}}+\bigg(\frac{1}{18} (6 k^2-7 k-48) \epsilon\bigg)_{1\textrm{-loop}}+O(\epsilon^2)
\, .
\end{align}
We notice that all operators $\mathcal{N}_k$ for any $k\geq4$ have positive anomalous dimensions. In the next section we show the consequences of this fact on the stability of the fixed point. 

Other susy-null leaders (such as $\omega(\chi_i^2)^2$) are obtained in locality constraints  \ref{class}. However for our argument it is sufficient to consider the anomalous dimensions \eqref{Nkgenphi3}. Nevertheless, for completeness, it would be nice to also compute their anomalous dimensions.\footnote{
Indeed the main role is played by $\mathcal{N}_4$ which is the lowest dimensional leader in the susy-null class. However we will further consider $\mathcal{N}_5$ to perform a sanity check. It can be interesting to consider other low lying susy-null leaders like $\omega(\chi_i^2)^2$ (which have the same classical dimensions of $\mathcal{N}_5$) to get further sanity checks.
}
 We leave this task for future investigations.

As a last comment we stress that the operator $(\mathcal{N}_4)_L\equiv (\chi_i^2)^2$ has a positive  one-loop anomalous dimension. This is in contrast with the RFIM case of \cite{paper2} where the one-loop correction of the same operator vanishes (notice that the operator $\mathcal{N}_4$ was called $\mathcal{F}_{4}$ in \cite{paper2}).
 In that circumstance we thus needed to consider the two-loop correction which was negative and played an important role in determining the stability of the RFIM fixed point.\footnote{ 
For completeness we show the one-loop anomalous dimensions of $(\mathcal{N}_k)_L$ in the RFIM case (which we  obtained using the techniques of App. H of \cite{paper2}),
	\be\D^{\text{RFIM}}_{(\mathcal{N}_k)_L}=\bigg(k+4-\frac{k}{2}  \epsilon\bigg)_{\textrm{class}} +\bigg(\frac{(k-4) (k+3)}{6}  \epsilon\bigg)_{1\textrm{-loop}}+O(\epsilon^2)\,.
	\ee 
}

\subsection{Non-susy-writable leaders}
Finally we discuss non-susy-writable operators. In Cardy variables they are only singlets under $S_{n-1}$  which permutes the $n-1$ fields $\chi_i$, and not of $O(n-2)$. So they  cannot be mapped to $\psi, \psib$ variables, and hence break SUSY explicitly. E.g. an operator $\sum_{i=2}^n\chi_i^3$ cannot be mapped to SUSY variables.

If a  leader that is non-susy-writable becomes relevant it would clearly destabilize the SUSY fixed point. In the RG flow  non-susy-writable leaders   mix with the other two types mentioned above, although (as expected from the SUSY theory) the opposite mixing cannot occur. 

The non-susy-writable leader with the lowest classical dimension comes from a class of $S_n$ singlets which we call Feldman operators (see App \ref{class} and also App D of \cite{paper2}). They are defined as follows \cite{paper2, Feldman}:
\begin{eqnarray}
  \mathcal{F}_k & = & \sum_{i, j = 1}^n (\phi_i - \phi_j)^k = \sum_{l = 1}^{k
  - 1} (- 1)^l \binom{k}{l} \sigma_l \sigma_{k - l} \,,  \label{Fk}
\end{eqnarray}
and their leaders take the form
\begin{eqnarray}
  \left( \mathcal{F}_k\right)_L & = & \sum_{l = 2}^{k - 2} (- 1)^l \binom{k}{l} (   \chi^l_i)  \left(  \chi^{k - l}_j \right) 
  \, .
\end{eqnarray}
We will consider $k$ to be an even integer greater or equal to $6$. For all odd $k$ the operator vanishes.
Similarly, for $k=2$, $\mathcal{F}_{k=2} \to 0 $ as $n \to 0$. When $k=4$ the operator $\mathcal{F}_{4}$ just reduces to $\mathcal{N}_4$ which has a susy-null leader and it was already considered in the previous section.\footnote{In this paper we avoid the name $\mathcal{F}_{4}$ and we shall use $\mathcal{N}_{4}$  in order to keep the distinction between the family $\mathcal{F}_{k\geq 6}$ with non-susy-writable leaders and the family $\mathcal{N}_{k\geq 4}$ with susy-null leaders. However the name $\mathcal{F}_{4}$ was used in \cite{paper2}.}
The operator $\mathcal{F}_{6}$ has a non-susy-writable leader 
\begin{equation}
(\mathcal{F}_{6})_L=-10 \left(2 \c_i^3 \c_j^3 -3 \c_i^2 \c_j^4\right)  \, ,
\end{equation}
 with classical dimension $18-3\e$ in $d=8-\e$. It is the lowest of its type, and hence we want to compute its RG correction. Below is the result of IR anomalous dimensions of all  $(\mathcal{F}_k)_L$:
\be\label{Feldmangen}
\D_{(\mathcal{F}_k)_L}=\bigg(3 k-\frac{k \epsilon }{2} \bigg)_{\textrm{class}}+\bigg(\frac{1}{18} k (2 k-3) \epsilon \bigg)_{1\textrm{-loop}}+O(\epsilon^2) \,.
\ee
We show the computation in App. \ref{app:nonsusy}. We see that the one-loop correction is strictly positive. On the contrary, for the RFIM case, the first non vanishing correction comes at two loops and it is negative \cite{paper2}.\footnote{
\label{RFIM_Feld}
For convenience we quote the result below:
\be\label{FeldmangenRFIM}
\Delta^{\text{RFIM}}_{(\mathcal{F}_k)_L} = \bigg(2 k - \frac{k}{2} \e\bigg)_{\textrm{class}} + \bigg(- \frac{k (3 k - 4)}{108}
\e^2\bigg)_{2\textrm{-loops}} + O (\epsilon^3) \, .
\ee}
This difference is very important for the stability of the respective SUSY fixed point as we will explain in the next section.

Of course there are infinitely many more non-susy-writable leaders. Let us discuss another infinite family which is interesting because it contains some of the lowest dimensional operators of this type. The family can be defined by the following combination of $S_n$ singlets
\begin{equation}
\mathcal{G}_{k}\equiv  \frac{\sigma _3 \sigma _{k-3}}{3 (k-5)}+\frac{\sigma _1 \sigma _{k-1}}{k-1}-\frac{(k-4) \sigma _2 \sigma _{k-2}}{(k-5) (k-2)} \, ,
\end{equation}
for $k=6,7, \dots$.
The leaders of $\mathcal{G}_{k}$ can be written in a very compact way as a composite of the leader of $\mathcal{F}_{6}$ and powers of $\vf$,
\begin{equation}
(\mathcal{G}_{k})_L \propto \vf^{k-6}(\mathcal{F}_{6})_L \, .
\end{equation}
This implies that their classical dimension is  $2 (k+3)-\frac{k}{2}  \epsilon$  in $d=8-\e$. Since they are written as a product of the lowest dimensional  non-susy-writable leader $(\mathcal{F}_{6})_L$ times powers of the lowest dimensional field $\vf$,  the operators $(\mathcal{G}_{k})_L$ are the lowest dimensional non-susy-writable operators made of $k$ fields. Because of this property in perturbation theory they do not mix with other operators and it is easy to compute their anomalous dimensions. The result is
\begin{equation}
\label{dimF6k}
\D_{(\mathcal{G}_{k})_L}=\left( 2 (k+3)-\frac{k}{2} \epsilon \right)_{\textrm{class}}+\left( \frac{1}{18} (k (6 k-7)-120) \epsilon\right)_{1\textrm{-loop}}+O(\epsilon^2) \, .
\end{equation}
Once again we notice that the one-loop correction is always positive.\footnote{
A similar computation for the RFIM gives
\be\label{Feldmangen-F6k}
\Delta^{\text{RFIM}}_{(\mathcal{G}_{k})_L} = \bigg(k+6 -\frac{k}{2} \epsilon  \bigg)_{\textrm{cl.}} +\left( \frac{(k-6)(k+5)}{6}  \epsilon\right)_{1\textrm{-loop}}+ O (\epsilon^2) \, ,
\ee
where the anomalous dimensions  for $k>6$ are always positive.
}

For our argument we will mostly focus on the  lowest dimensional non-susy-writable leader $(\mathcal{F}_{6})_L$, however it will be useful to know the dimensions of the other leaders of the families $(\mathcal{F}_{k})_L$ and $(\mathcal{G}_{k})_L$ as a sanity check.   In particular we found it useful to introduce the family $(\mathcal{G}_{k})_L$ since $(\mathcal{G}_{7})_L$ in $d=8$ has dimension equal to $20$ and thus lies well below  $(\mathcal{F}_{8})_L$ which has dimension $24$.

\section{Stability of the SUSY fixed points}
\label{stability}
In the previous section we computed the one-loop scaling dimensions $\Delta$ of the low-lying operators of the $\mathcal{L}_L$ theory in $d=8-\e$. It is now time to take our conclusions on the stability of the SUSY fixed point when $\e$ is of order one. 
The strategy is simple and it amounts to checking if any scaling dimension  $\Delta$, as a function of $d$, can cross the marginality line $\D=d$ at some dimension $d<8$. 

One may be worried that we considered only some low lying operators, and that there exist still an infinite number of operators which we did not take into account and which may possibly cross the marginality line. If one has to check the whole spectrum of the theory, the problem would be intractable and our strategy would not have any hope to work. However it is possible to argue that there is no need to check the higher dimensional operators because level crossing is unlikely to occur in non-integrable theories.

Let us explain this point in more detail. Operator mixing occurs in perturbation theory between operators of the same symmetry, which also satisfy  extra selection rules like having the same dimension and being composite of the same number of fields. In a non-perturbative setup these extra selection rules are lost and any two operators with the same symmetry can mix. Let us consider two operators with the same symmetry and dimensions $\D_1(d)$ and $\D_2(d)$ computed from perturbation theory.
Let us assume that at certain dimension $d=\tilde{d}$  we have a crossing $\D_1(\tilde d)=\D_2(\tilde d)$. At this point we should not fully trust the functions  $\D_i(d)$ since  nonperturbative mixing  effect can modify the dimensions.  The modification is computed by diagonalizing the mixing matrix of the inner products between the associated states. The corrected  $\D_i(d)$ can either repel or become complex conjugate close to $d=\tilde d$, depending on operators having norms of same sign or opposite signs respectively (see also section 10 of \cite{paper2}).  One point of view about crossing is that when this happens one should not trust anymore the perturbative computation. However for our problem this point of view seems too pessimistic.

Indeed it is important to take into account that in the context of $\e$-expansion the computations $\D_i(d)$ (when $\e$ is of order one) tend to be much more reliable for operators with low classical dimension. In other words, by computing higher orders in perturbation theory, the dimensions of the low lying spectrum do not dramatically change, while the higher dimensional operators may have very important corrections.
For this reason, when level crossing occurs in $\e$-expansion between operators with low and high classical dimensions, the most likely scenario  is that the crossing is non-perturbatively resolved  by repulsion of the higher dimensional operator. The operators with lowest classical dimensions, roughly speaking, provide a barrier which is unlikely to be crossed by the higher dimensional ones.
With this in mind we can conclude that  in $\e$-expansion by knowing the low lying spectrum of operators in all symmetry sectors,  we can answer with reasonable confidence the question of stability of the fixed point.

In our case the mixing may occur if two operators have leaders in the same class between: non-susy-writable, susy-null and susy-writable of a given OSp$(d|2)$ representation. By the argument above it should be enough to study the operator with lowest classical dimension for each sector. However, to be extra cautious,   we further considered  (at least) one operator above it. By checking if the operators above cross the ones below we thus have a measure of whether non-perturbative mixing effect may be important and  slightly affect the result (namely if there is no crossing we should be more confident about our computation).

We are now ready to analyze the result of section \ref{sec:anomalous_dim}.
\begin{table}[t]\centering
	\begin{tabular}{@{}ccc@{}}
		\toprule
		Leaders $\mathcal{O}$\quad \qquad & Type \quad &  IR dimension:
		$\Delta_{\mathcal{O}}$\\
		\midrule
		$(\mathcal{N}_4)_L $ & Susy-null &  $12-2 \epsilon+\frac{10 \epsilon }{9}+ O (\epsilon^2)$\\
		$(\mathcal{N}_5)_L $ & Susy-null & $14-\frac{5 \epsilon }{2}+\frac{67 \epsilon }{18}+ O (\epsilon^2)$\\
		$\mathcal{B}^{(3)}_{\theta\thetab}$ & Susy-writable (box) &  $12-\frac{4 \epsilon }{3}+O(\e^2)$\\
		$\mathcal{B}^{(4)}_{\theta\thetab}$ & Susy-writable (box)& $14-\frac{\e}{3}+O(\e^2)$\\
		$(\mathcal{F}_6)_L$ & Non-susy-writable &  $18+O(\e^2)$\\
		$(\mathcal{G}_{7})_L$ & Non-susy-writable & $20+\frac{31 \e }{9}+O(\e^2)$\\
		\bottomrule
	\end{tabular}
	\caption{\label{dangerous ops}Summary of the computation of anomalous dimensions of the first 
	leader operators of each type. The computations are performed in  the $\mathcal{L}_L$ theory for  the RF $\phi^3$ model in $d=8-\e$.}
\end{table}
 We found that the lowest dimensional susy-null leader is $(\mathcal{N}_4)_L =(\chi_i^2)^2$. This operator belongs to the infinite family of susy-null leaders  $(\mathcal{N}_k)_L$ with dimensions computed  in formula \eqref{Nkgenphi3} for generic $k$. For susy-writable leaders we argued that the operators in the box representation of OSp$(d|2)$ have a better chance to play a role. The lowest dimensional one is $\mathcal{B}^{(3)}_{\theta\thetab}$ and belongs to the infinite family of operators $\mathcal{B}^{(k)}_{\theta\thetab}$ ($k\geq 3$) with dimensions that increase with $k$ as computed in \eqref{box_dim}. In the non-susy-writable sector the  lowest dimensional operator is $(\mathcal{F}_6)_L$ which comes from the infinite family $(\mathcal{F}_k)_L$ for $k=6,8,10,\dots$. Their dimensions are given in equation \eqref{Feldmangen}. We also considered  another infinite family of non-susy-writable leaders $(\mathcal{G}_{k})_L$ which contain the operator $(\mathcal{G}_{k=7})_L$ that lies right above $(\mathcal{F}_6)_L$. Their dimensions are given in equation \eqref{dimF6k}.
  
  In Table \ref{dangerous ops} we summarize  the anomalous dimensions of the lowest dimensional operators  $(\mathcal{N}_4)_L$, $\mathcal{B}^{(3)}_{\theta\thetab}$,  $(\mathcal{F}_6)_L$ of each type  and we compare them with the dimensions of one operator above.
 From the table it is easy to see that the lowest operators  are never crossed by the operators above. Moreover one can also check that they are never crossed by any higher dimensional operator of the same type. 
We thus conclude that non-perturbative mixing should not affect our analysis. 

\begin{figure}[h]
	\centering \includegraphics[width=360pt]{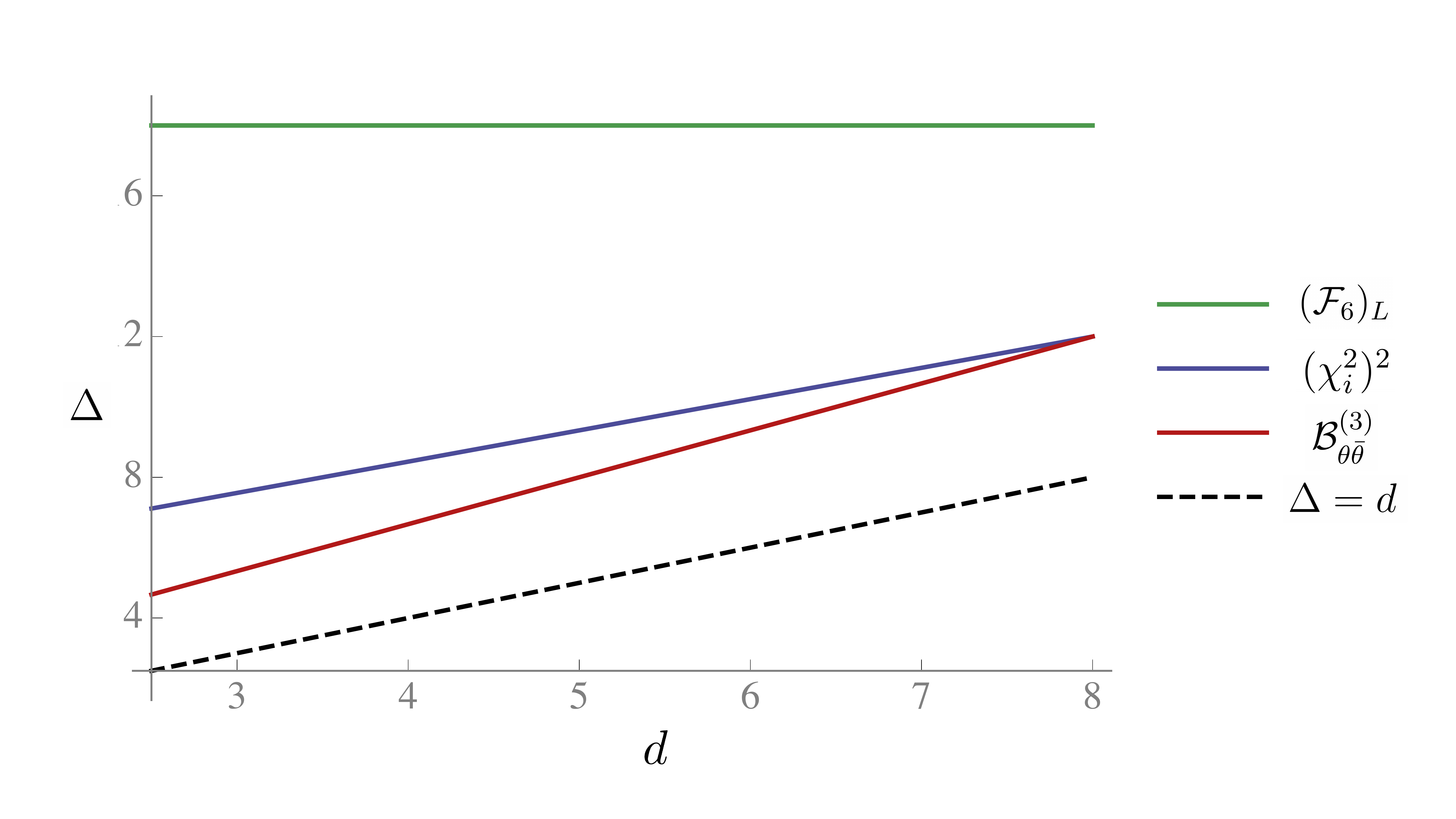}
	\
	\caption{Plot showing dimensions of the lowest dangerous leaders from the three categories vs spacetime dimension $d$.\label{stable}
	 }
\end{figure}

In what follows we can then  focus on each of the lowest operators of the three categories. We have plotted their IR dimensions as a function of the spacetime dimension $d$ in Figure \ref{stable} . 
Comparing with the marginality line $\D=d$ we see that none of the operators becomes relevant. Since we do not find any new relevant perturbation we can conclude that the  SUSY fixed point persists even at $8-d=O(1)$ and thus that the RF $\phi^3$ model always undergoes dimensional reduction.\\

Of course our conclusion has to be taken with a grain of salt. Besides the non-perturbative mixing (which should not significantly change our results), there is a  more important source of uncertainty. Indeed our conclusion relies on a one-loop computation of the anomalous dimensions. Since we are extrapolating to $\epsilon \sim O(1)$, higher loop corrections may become important. It would be very interesting to compute them and see if our results are confirmed.\footnote{We checked that a Pad\'e[1,1] approximation does not change our observation that none of the considered operators crosses marginality. However it would be nice include higher loop computations in the analysis.} We leave this task for the future.

One of our main motivations for this work was to check if the RG setup introduced in \cite{paper2} to study  the RFIM model would give consistent results also for  RF $\phi^3$.
Recall the discussion from section \ref{sec:intro} that the RF $\phi^3$ model describes the near critical behavior of branched polymers. As first established in \cite{PhysRevLett.46.871} according to numerical evidence, the critical point of branched polymers is  related to the non-disordered Lee-Yang fixed point (which describes critical point of Ising model in an imaginary magnetic field) via dimensional reduction in all $2\leq d<8$. Consequently it is expected that the critical RF $\phi^3$ model in this range of dimensions is always described by a supersymmetric fixed point. Our conclusion above is therefore consistent with this expectation (see the next section for a discussion on the subtle $d=2$ case).

\begin{figure}[h]
 \centering
     \begin{subfigure}[b]{0.3\textwidth}
         \includegraphics[height=4.1cm
         ]{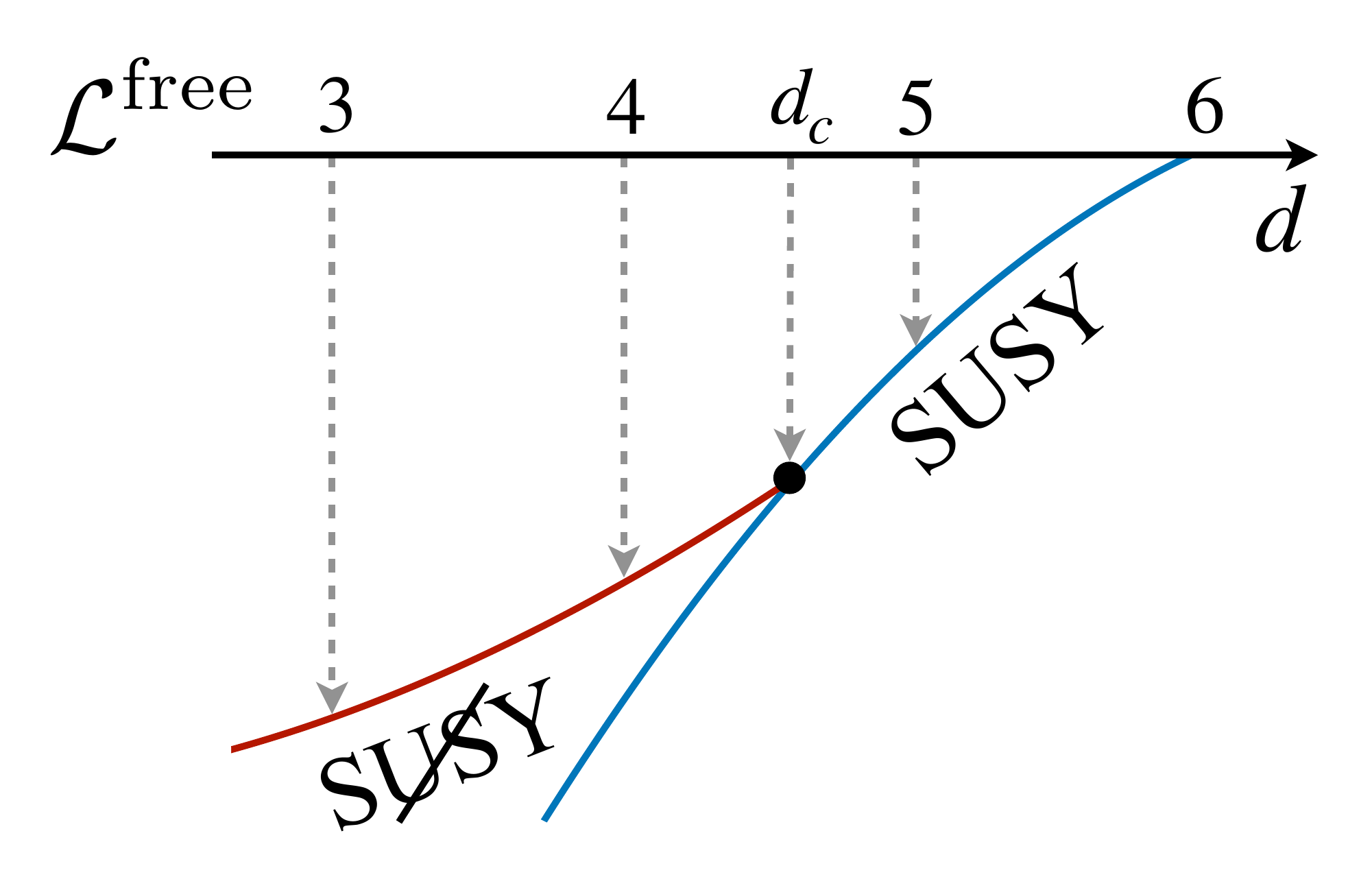}
         \caption{RFIM}
        \label{fig:RFIM_RG}
     \end{subfigure}
    \qquad 
        \qquad
        \qquad       
  \begin{subfigure}[b]{0.3\textwidth}
         \centering
         \includegraphics[
         height=4cm
         ]{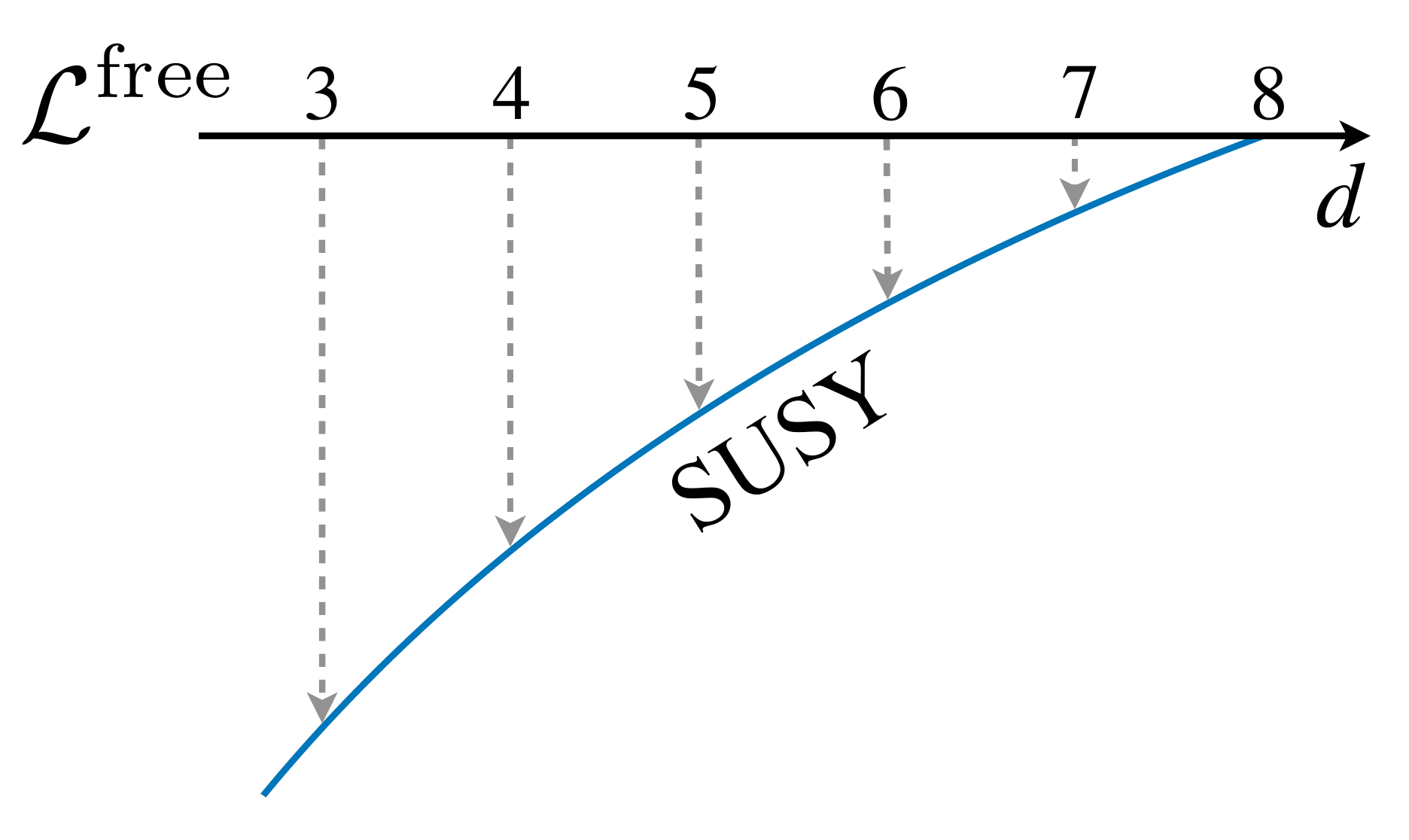}
         \caption{RF $\phi^3$}
         \label{fig:RF3_RG}
     \end{subfigure}
      \qquad\qquad
 	\caption{
	RG flow of the leader theory in the RFIM (Fig.\ref{fig:RFIM_RG})
	vs RF $\phi^3$ theory 
	(Fig.\ref{fig:RF3_RG}). In RFIM we have an unstable SUSY fixed point below $d_c\approx 4.2$ - $4.7$. For RF $\phi^3$ the SUSY fixed point  is stable. \label{phi4vsphi3}
	 }
\end{figure}

Another motivation was to compare how the single RG setup works in different ways for RFIM  and RF $\phi^3$ (see Fig.\ref{phi4vsphi3}).
Indeed in  \cite{paper2} we found that for the RFIM the SUSY fixed point becomes unstable at a critical dimension $d_c\approx 4.2$ - $4.7$ (namely for $d \leq d_c$ the fixed point is non supersymmetric). On the other hand we expected no instability for the  RF $\phi^3$. It was interesting for us to see how this would come about. This  paper provides an answer to this puzzle.
The main source of instability for the RFIM is due to $(\mathcal{N}_4)_L$ (which we called $(\mathcal{F}_4)_L$ in \cite{paper2})  and $(\mathcal{F}_6)_L$. They are found to have vanishing one-loop anomalous dimension. The first non-vanishing contribution is found at two-loops and it is negative (see footnote \ref{RFIM_Feld}). So the main difference  is that in the RFIM their one-loop correction  vanishes, leaving a leading two-loop negative correction. For the RF $\phi^3$ we did not compute the two-loop correction of $(\mathcal{N}_4)_L$ and $(\mathcal{F}_4)_L$ because their one-loop anomalous dimension is already positive and fairly large.\footnote{Still it would be interesting to compute the two-loop correction.}
One can easily understand why the one-loop corrections behave differently in the two models. Indeed these are computed in a different way: one needs to insert two interaction vertices to compute a one-loop correction in the RF $\phi^3$ while a single vertex is required in  the RFIM (this is also true for the pure versions of the models and it is simply due to the fact that the vertices have odd or even number of fields). Operators made  of only $\c_i$ fields (like the Feldman operators)  clearly do not receive  corrections from the insertion of a single vertex  since the latter always contains other fields (e.g. the vertex $\vf^k \c_i^2$ has extra powers of $\vf^k$ which cannot be contracted with the fields $\c_i$ of the Feldman operator). Therefore by construction in the RFIM all Feldman operators have zero anomalous dimension at one loop, while for the RF $\phi^3$ this need not be the case and indeed we find that it is not. 
We thus conclude that by applying the single RG framework of \cite{paper2} we obtain the expected results in two different models. This gives us more confidence that the framework of \cite{paper2}  is indeed correct.


\section{Discussion}
\label{sec:discussion}
In this paper  we carry out a perturbative RG analysis of the  Random Field $\phi^3$ model. We use the framework introduced to study the Random Field Ising Model in \cite{paper2}.

The results of  \cite{paper2} are compatible with the numerical observations \cite{Picco2, Picco3, Picco1} that supersymmetry (and thus dimensional reduction) of the RFIM fixed point are lost in $d=3,4$ and are recovered in $d=5$, where $d=6$ is the upper critical dimension.
In  \cite{paper2} this is explained by the fact that (at least) one operator becomes relevant in dimensions $d<d_c$ where $d_c\approx 4.2$ - $4.7$.
For the RF $\phi^3$ model the numerical expectation \cite{PhysRevLett.46.871}  is that the SUSY fixed point is always reached in $2\leq d <8$ (where $d=8$ is the upper critical dimension, while the subtle $d=2$ case is discussed below). The main result of this work is a check that the framework introduced in \cite{paper2} is compatible with the numerical results. Indeed we studied a large number of low lying perturbations and we found that at one-loop they all have  positive anomalous dimensions and they do not seem to ever cross the marginality line.

This result is important for two main reasons.  Firstly it is a check of the RG framework of \cite{paper2}. Indeed we gained more confidence that this framework is correct since gives results compatible with the RF $\phi^3$ expectations. Notice that it is crucial to test this framework since  it provides a new understanding of when and why the RFIM undergoes dimensional reduction, which is a question which was debated for almost fifty years. 
Secondly it explains in the same language used for the RFIM why dimensional reduction is not lost for the RF $\phi^3$ itself. The latter is also an important question.  Indeed the RF $\phi^3$ theory  captures the phase transition of a number of interesting statistical physics models like branched polymers and lattice animals. For these models dimensional reduction was observed to hold giving rise to the Lee-Yang universality class in $d-2$ dimensions. Our calculations are in perfect agreement with these results. 

Interestingly, for the problem of branched polymers Brydges and Imbrie proposed a model that has explicit Parisi-Sourlas supersymmetry at the microscopic level and which, by construction, undergoes dimensional reduction  \cite{zbMATH02068689}. On one hand the result of  Brydges and Imbrie is impressive because it explicitly shows how dimensional reduction works in the problem of branched polymers without even using field theory or RG. On the other hand \cite{zbMATH02068689} only shows dimensional reduction for a very fine tuned model where supersymmetry is taken as a starting point. 
 In particular in \cite{zbMATH02068689} it is missing an explanation of why supersymmetry should emerge in the IR for a generic non-supersymmetric model of branched polymers.\footnote{E.g. the model proposed in \cite{zbMATH02068689} is not sensitive to susy-null and non-susy-writable perturbations which were crucial to understand the destabilization of the SUSY fixed point in the RFIM.} Our work provides an answer to this question by showing that all SUSY-breaking deformations are indeed irrelevant.
\\

There are many open problems which deserve further investigation (see also the discussion section of \cite{paper2}) and our work should be viewed as a first step in this direction.
Some of the most interesting ones are listed below.
\begin{enumerate}
	
\item \textit{Higher loops and other operators} : All the anomalous dimensions computed in this paper are at one loop. It would be interesting to compute higher loop corrections of the low lying leader operators. 
This would allow one to verify that SUSY fixed point is indeed stable for larger values of $\e$.
Susy-null and non-susy-writable leaders like  $(\mathcal{N}_0)_L$ and $(\mathcal{F}_6)_L$ need to be studied in the $\mathcal{L}_L$ theory. This is slightly more complicated than the usual $\widehat  \phi^3$ theory (Lee-Yang) fixed point in $\widehat d=6-\e$, since  the $\mathcal{L}_L$ Lagrangian contains two interaction vertices and three different propagators. However using modern multi-loops technologies it should be fairly easy to extend the results of this work to higher orders in perturbation theory. 
For the susy-writable leaders like the box operators the computation can be directly done in the usual $\widehat  \phi^3$ theory (because of dimensional reduction), by computing higher loop anomalous dimensions of the $\widehat{B}$ operators. 

We also ignored other perturbations with higher classical dimensions, e.g. susy-null leaders not in the class $(\mathcal{N}_k)_L$ and non-susy-writable leaders that are not of the form $(\mathcal{F}_{6,k})_L$ or of the Feldman type. It would be interesting to compute their anomalous dimensions. This may be useful in order to understand nonperturbative mixing when $\e$ is large. It should be kept in mind that for operators with higher classical dimensions good accuracy is expected only at higher loops. 

	\item \textit{Conformal Bootstrap} : It would be interesting to set up a conformal bootstrap problem for the RF $\phi^3$ fixed point.\footnote{See \cite{Hikami:2017sbg} for an attempt by Hikami, reviewed in App.A.10 of \cite{paper2}.} 
	This would be a strong check of the stability of the SUSY fixed point and dimensional reduction. Choosing a 4-point function appropriately one can shed light on the various operators allowed in the theory in the nonperturbative regime $8-d=O(1)$. E.g. non-susy-writable operators like $(\mathcal{F}_k)_L$ can be exchanged in the conformal block decomposition of $\langle\chi_i(x_1)\chi_j(x_2)\chi_k(x_3)\chi_l(x_4)\rangle$, so  we could numerically estimate their dimensions by bootstrapping this correlator. 
	
	There are a number of challenges in setting up this bootstrap problem. Firstly, we do not have a CFT for positive integer values $n$ which we can analytically continue to $n\to 0$. This is because the fixed point only exists at $n=0$ (the fixed point arises as the limit $n\to 0$ of a sequence of approximate fixed points as we review in section \ref{replicas}). Second, since we are working in an $S_n$ invariant theory with $n\to 0$ we expect to find a logarithmic  CFT \cite{Cardy:2013rqg}. Indeed for the RFIM case we find logarithmic multiplets (see section 9.2 and app. H of \cite{paper2}), and one should expect the same for the RF $\phi^3$ model. In order to account for logarithmic multiplets one needs to use a different bootstrap algorithm which makes use of logarithmic conformal blocks \cite{Hogervorst:2016itc}. Finally the CFT is nonunitary as evident e.g. from the dimension of $\vf$. So one cannot impose unitarity bounds or positivity of OPE coefficients. For nonunitary CFTs one needs to resort to Gliozzi's bootstrap algorithm \cite{Gliozzi:2013ysa} or to variations thereof, which typically are less systematic. Some work is required to tackle these problems, but certainly it would be priceless to have at our disposal the bootstrap toolbox to study random field theories.

	

	\item \textit{Dimensional reduction $2 \to 0$}  : The  RF $\phi^3$ model can also be studied in $d=2$.  In this case, according to dimensional reduction, one would obtain a relation to the   pure $\widehat \phi^3$ theory in zero dimensions. This case is of course bound to be singular. E.g. dimensional reduction for correlation functions is obtained by localizing all operators to a $d-2$ dimensional hyperplane, however this prescription is not well defined for $d=2$, since all insertions would collapse to a single point. On the other hand one can compute the partition function (and related observables) in the RF $\phi^3$ model for $d=2$ and relate them to the zero-dimensional $\widehat \phi^3$ counterpart, where the zero dimensional path integral is just understood as an ordinary integral.
	Parisi and Sourlas in \cite{PhysRevLett.46.871} checked that critical exponents behave according to dimensional reduction even at $d=2$. The problem was later reconsidered in \cite{Miller:1992uv}, where it was suggested that the $d=2$ theory does not have the structure of a conformal theory. 
	Indeed for $d=2$ the Parisi-Sourlas supersymmetric theory is of a subtle type since it is not clear if it can possess a  traceless super-stress tensor. This problem can be explicitly seen in free theory by looking at formula (C.4) of \cite{paper1} where the improvement term which makes the stress tensor traceless is singular for $d=2$.
	We think that it would be worth to revisit this problem in a modern language to get a more comprehensive understanding of this two-dimensional theory and of its dimensional reduction.

\item \textit{Dimensional reduction $3 \to 1$} : A less singular case arises for the dimensional reduction $3 \to 1$. Here one can still consider correlation functions restricted to a line. On the other hand, again we cannot define a stress tensor in $\hat{d}=1$ since all one-dimensional theories are non-local. This can be again seen in Parisi-Sourlas supersymmetric free theory where one finds that the dimensionally reduced stress tensor ---see e.g. in  (C.6) of \cite{paper1}---  vanishes. It would be interesting to study in more detail the three-dimensional SUSY theories and their dimensional reduction.

\item \textit{Dimensional reduction $4 \to 2$} :  Finally we would like to mention that the dimensional reduction $4 \to 2$ is not singular but it is very interesting. Indeed the dimensionally reduced models have emergent Virasoro symmetry  (e.g. the RF $\phi^3$ theory in $d=4$ dimensionally reduces to the Yang-Lee minimal model with central charge $c=-22/5$). It would be very interesting to investigate how the Virasoro symmetry is embedded in the $d=4$ supersymmetric theories. This direction deserves further investigation \cite{progress}.
				
\item \textit{Applications to other models} : Beside RFIM and RF $\phi^3$ theory, there are other models which could be potentially studied in the RG framework used in this paper. Indeed in statistical physics one often uses the replica method and sometimes one lands on Lagrangians of the form \eqref{Sr} or generalizations thereof (see e.g. \cite{Franz_2013} for a model of the continuous phase transition of glassy materials).
It would be interesting to further apply the RG framework of \cite{paper2} to these cases both as a test of the method and to see whether it provides a deeper understanding of a broader class of phenomena.
	
\end{enumerate}

\section*{Acknowledgements}
We would like to thank  the participants of the workshop “Bootstat 2021”  which took place at Institut Pascal (Université Paris-Saclay) with the support of the program “Investissements d’avenir” ANR-11-IDEX-0003-01. We  especially  thank Silvio Franz for interesting discussions and Slava Rychkov for comments on the draft and collaboration in the early stages of the project. The work of A.K. is funded by the German Research
Foundation DFG under Germany’s Excellence Strategy -- EXC 2121 ``Quantum Universe" -- 390833306.
The work of E.T. is supported by the European Research Council (ERC) under the European Union’s Horizon 2020 research and innovation programme (grant agreement No 852386).

\appendix


\section{Review of branched polymers and lattice animals}\label{app:BPreview}

In this appendix we review how to define the problem of diluted branched polymers/lattice animals and see why it is captured by the RF $\phi^3$ theory. 
We will follow the construction of \cite{LubIsslong} and we aim at giving a pedagogical and explicit definition of all the ingredients of the model (see also chapter 9 of  \cite{Cardy-book}). 

\begin{figure}[h]
 \centering
     \begin{subfigure}[b]{0.3\textwidth}
         \centering
         \includegraphics[width=\textwidth]{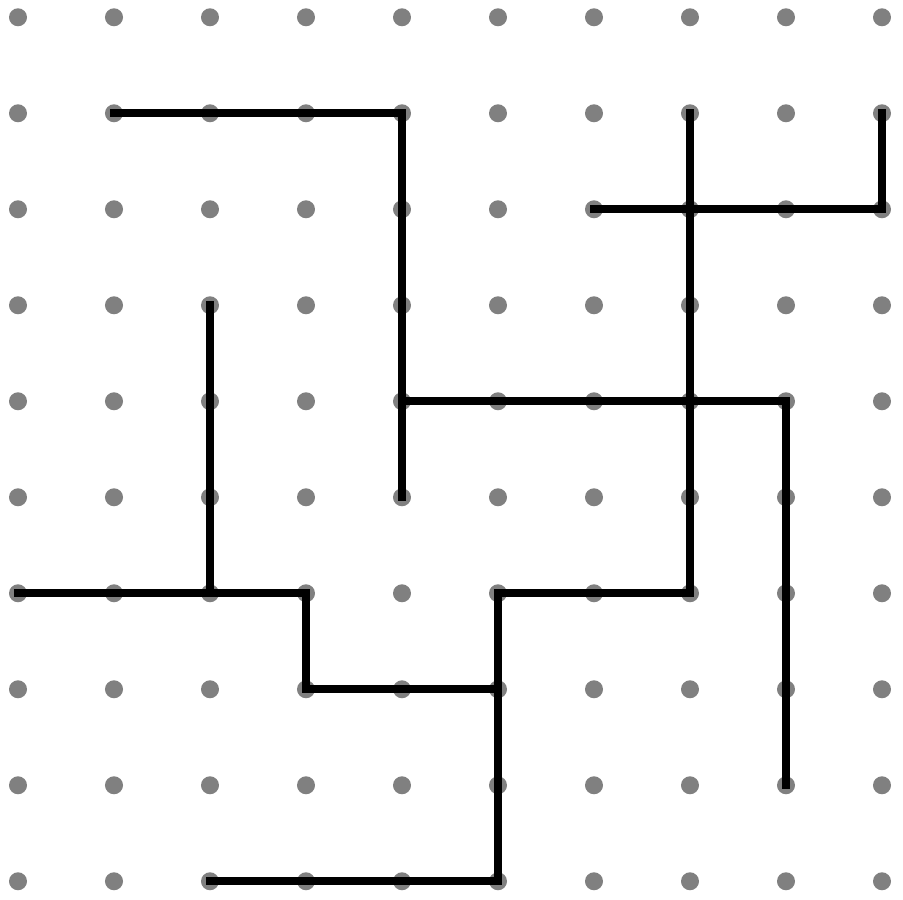}
         \caption{Branched polymer}
         \label{fig:Branched_polymer}
     \end{subfigure}
    \qquad 
        \qquad
        \qquad       
  \begin{subfigure}[b]{0.3\textwidth}
         \centering
         \includegraphics[width=\textwidth]{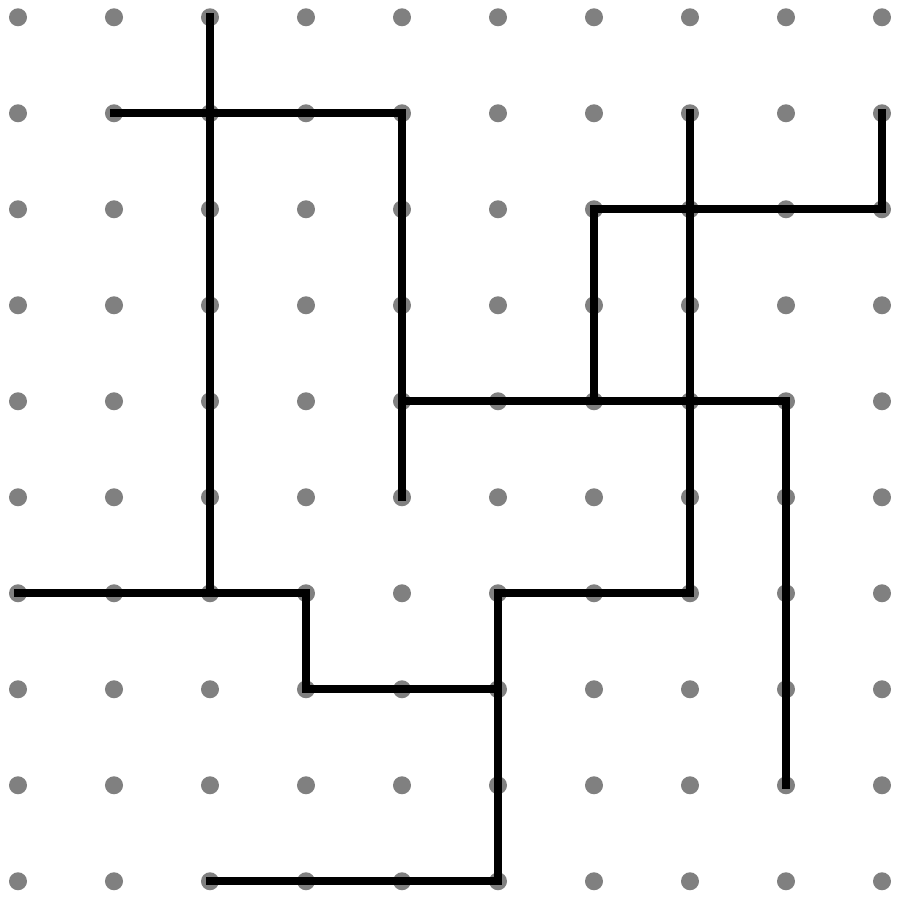}
         \caption{Lattice animal}
         \label{fig:Lattice_animals}
     \end{subfigure}
 	\caption{
	Different clusters on the lattice.
	\label{clusters} }
\end{figure}

Let us start with some basic definitions. A polymer is defined as a linear chain of units called monomers. A branched polymer is built by sewing together linear polymers.
It is convenient to define the polymers on a lattice so that for any fixed number of monomers, there is a finite number of configurations.
We define a polymer on a $d$ dimensional hypercubic lattice by placing the monomers at the sites of the lattice and by connecting neighbouring sites with lines.
Sets of connected sites on the lattice are called ``lattice animals'' or ``clusters''. When lattice animals have a tree-like topology they define branched polymers. General lattice animals also contain self intersections (\emph{i.e.} closed loops) as shown in figure \ref{clusters}.

We are interested in studying the statistics of branched polymers/lattice animals when the number $N$ of monomers is very large: in this limit the number of possible configurations $P(N)$ and the average size $R$ of a cluster scale as $P(N)\sim N^{- \theta} \l^{N}$ and $R \sim N ^\nu$ where $\theta, \nu$ are critical exponents, while $\l$ is a non-universal constant. 
The models of branched polymers and lattice animals  are known to be in the same universal class  (see discussion below). 
In the following we define a model which counts configurations of lattice animals, 
which allows us to choose how many links, loops, clusters and vertices are present in a given configuration.
Finally we will be interested in the so called diluted limit, which suppresses configurations with multiple clusters. 

We will introduce the model in steps. We will start by defining a toy-model that is used to build random walks (which are equivalently  understood as  linear polymers).
We then introduce vertices to allow for branching. Finally we explain how to modify the model to count the number of separate clusters. 
We conclude the appendix by showing that the IR behaviour of this model can be described in the continuum limit by an action with $S_n$ symmetry which is of the same form as the replica action of \eqref{Sr}.

To start we consider a simple $O(m)$ model with an $m$-state vector $s_i$ with partition function  
\be\label{ONmodel}
Z=
\tr \prod_{\langle x x'\rangle} (1+K \sum_{i=1}^m s_i(x) s_i(x'))\,.
\ee
where $\la x,x'\ra$ denotes nearest neighbour sites. 
We want to consider $s_i$ to be variables in $\mathbb{R}$ which are spatially uncorrelated with trace  
defined as
\be
\tr = \int_{-\infty}^{+ \infty} \prod_{x} (\prod_{i=1}^{m} d s_i(x)) P(\sum_{i=1}^m s_i^2(x)) \, ,
\ee
where $P$ is an $O(m)$ invariant distribution. By $O(m)$ invariance we must have that 
 \be
\label{si_corr}
\tr s_{i_1}(x) \cdots s_{i_k}(x) = 
C_{k} \; ( \d_{i_1 i_2} \dots \d_{i_{k-1} i_k} + \dots) \, , \qquad (k \mbox{ even})
\ee
and zero if $k$ is odd, for some coefficients $C_{k}$. The dots in \eqref{si_corr} represent products of Kronecker deltas for all possible inequivalent pairings of the indices $i_k$. 
In order to build the model we further require the coefficients $C_{k}$ to satisfy the following conditions
\be
\label{tr_prop}
\tr 1=1 \, ,
\qquad
 \tr s_{i_1}(x) s_{i_2}(x)=\delta_{i_1 i_2} \, ,
 \qquad
C_k=m^{\frac{k}{2}-1}  c_k \,, \quad  \mbox{for } k=4,6,\dots,4d \, ,
\ee
where the coefficients $ c_k$ must be finite in the limit $m\to0$.
Finally we also need to be able to tune the coefficients $c_k$ to the values we want.
Later we will explain why these conditions are useful.

These conditions can be recast as restrictions on the possible distributions $P$.
In other words it is possible to choose $P$ in such a way that all these conditions are satisfied. 
A simple way to do this is to choose a distribution written as a Gaussian $\exp (-\frac{\sum_i s_i^2}{2})$ times a polynomial in $\sum_i s_i^2$.
This distribution gives rise to computable traces (by Wick contractions)  and one can easily check that by tuning the coefficients of the polynomial one can fix the same number of coefficients $C_{k}$.\footnote{To be explicit, given the distribution 
 \be
P = \frac{e^{-\frac{\sum_i s_i^2}{2}}}{(2 \pi)^{m/2}}  \sum_{l=0}^{l_{max}} a_l (\sum_i s_i^2)^l \, ,
 \ee 
 one finds that $ C_{k}= \sum_{l=0}^{l_{max}} 2^l \left(\frac{k+m}{2}\right)_l a_l $, where $k$ are even integers. The linear map between $a_l$ with $l=0,1,\dots,l_{max}$ and $C_k$ with $k=0,2,\dots , 2l_{\max}$ is invertible and we can therefore use it  to fix this set of $C_k$ to any value by opportunely tuning the coefficients $a_l$. This means that by considering $l_{max}=2d$ it is possible to satisfy \eqref{tr_prop} and to even choose the values of $c_k$.}

Formula \eqref{ONmodel} corresponds to a Hamiltonian $\sum_{\langle x x'\rangle} \log (1+K \sum_{i=1}^m s_i(x) s_i(x') )$, which is a slight modification of the usual $O(m)$ model due to the presence of the logarithm. 
This formulation gives a simpler partition function which has at most a single bond between two sites and thus it can be conveniently described diagrammatically in terms of clusters.
Let us show how this is done. We first expand the product in \eqref{ONmodel} to get a sum of terms each of which is just a product of bonds $K \sum_{i=1}^m s_i(x) s_i(x') $ for a given set of $x, x'$. This product is then depicted by drawing a line between each two sites if a bond between them is present. A cluster is thus a set of connected bonds.
Now the important observation is that by tracing over $s_i$ the only clusters that survive are closed loops. Indeed for each open end we get $\tr s_i =0$.
Moreover for each separate loop we get a factor of $m$ due to the sum over $i$.
There are also configurations with intersecting loops which arise when more than two spins are present at the same site. Forgetting about the latter (which we will discuss below), the partition function can be expressed as
\be
Z=\sum_{\mathcal{C}} m^{N_L} K^{N_B} 
\ee
where $\mathcal{C}$ is a cluster configuration which only contains loops, $N_L$ counts the number of loops in $\mathcal{C}$ and $N_B$ the number of bonds.
Now we want to take the limit $m \to 0$. This has two important consequences: first it suppresses loop configurations, secondly because of \eqref{si_corr} it suppresses traces of more than two spins at a given vertex, thus eliminating all configurations with intersecting loops and leaving only linear configurations. 
E.g. at order $m$ the limit $m \to 0$  selects only configurations with a unique self avoiding loop. 
While this limit trivializes the partition function, it is typically used to study two point functions of  a spin variable $s_1$
\be
\label{2ptSAW}
 \langle s_1(x_1)s_1(x_2) \rangle = \frac{1}{Z} \tr s_1(x_1) s_1(x_2) \prod_{\langle x x'\rangle} (1+K \sum_{i=1}^m s_i(x) s_i(x')) \, .
\ee
 By expanding again the product we notice that presence of $s_1$ makes it possible to draw new diagrams which are lines that connect $x_1$ to $x_2$.  These configurations are not suppressed by the $m \to 0$ limit because the operator $s_1$ is selecting a single component of the possible  $m$. In practice  all configurations are written in terms of lines that connect $s_1$ at different sites (all the other spin components $s_{i>1}$ are not present in any configuration).
  Conversely, choosing the two point function of  $\sum_i s_i$ would have generated an extra factor of $m$ for the lines due to the sum over $i$. Equation \eqref{2ptSAW} is used to compute the statistics of self avoiding walks between two points. 
  
  For the problems of branched polymers and lattice animals we want to sew together many self avoiding walks.
 A first step to do so is to allow many endpoints, namely we can multiply the partition function \eqref{ONmodel} by a term $\prod_{x} (1+H  s_1(x))$. This is not enough since  as $m\to0$ it would only generate  configurations involving a number $N_c$ of linear self avoiding walks weighted by $H^{2 N_c}$. In order to obtain branched configurations we need to introduce vertices $v_k(x)$  which allow us to connect $k$ lines to the site $x$. 
 These can be defined by the requirement
 \be
 \label{Vksi}
\lim_{m \to 0} \tr v_k(x) s_{i_1}(x) \cdots s_{i_p}(x) = \delta_{p k} \delta_{1 i_1 \dots i_k } \, ,
 \ee
 where  $\delta_{a_1 \dots a_k } =1 $ if all $a_i$ are equal, otherwise it is zero.
 Since the index $k$ of $v_k(x)$ counts how many lines are attached to a given site, for a hypercubic lattice we need only to build a finite number of vertices with $3\leq k\leq 2d$. Also the $k$ label is restricted by the possible number of spins $s_{i}(x)$ at a given site $x$ in the nearest neighbour interaction \eqref{ONmodel}, which gives $0\leq p\leq 2d$.
 The operators $v_k(x)$ are written in terms of the spin variable $s_i(x)$. 
 We ask $v_k(x)$ to preserve the $O(m-1)$ symmetry that rotates the variables $s_{i}$ with $i=2,\dots,m$. In practice we can define  $v_k(x)$ as a polynomial of order $k$ in the spin variables which depends on $s_1$ and $\sum_{i=1}^{m} s^2_i$.  Let us exemplify this by considering the simplest vertex for $k=3$. We start by the  ansatz $v_3(x)=a_1 s_1+a_2 s_1^3 + a_3 s_1 \sum_{i=1}^m s_i^2 $ for some coefficients $a_i$.\footnote{We did not include even powers of the spin variable because \eqref{Vksi} gives rise to an independent set of linear equations for the correspondent coefficients which is homogeneous and thus  has trivial solution, namely we can set all these coefficient to zero.}
The coefficients $a_i$ of the ansatz are fixed by requiring that \eqref{Vksi} is satisfied for $p=1,3$.
Notice that from these two requirements we obtain three equations which arise from the different tensor structures. E.g. from $p=1$ we get a single equation that multiplies $\d_{1 i_1}$ which we want to set to zero. From $p=3$ we get two equations. One from the coefficient of $(\d_{1 i_1} \d_{i_2 i_3}+\d_{1 i_2} \d_{i_3 i_1}+\d_{1 i_3} \d_{i_1 i_2})$ which we want to set to zero and one from the coefficient of $\d_{1 i_1}\d_{1 i_2} \d_{1 i_3}$ which we want to set to one.
In particular using  \eqref{si_corr} and \eqref{tr_prop} we can solve the full set of equations obtaining
 \be
a_1 = \frac{c_4}{m \left(c_4^2 (m+2)-c_6 (m+4)\right)}
\, , \quad  
a_2= \frac{1}{6 c_6 m^2}
\, , \quad  
a_3= \frac{c_6-c_4^2}{2 c_6 m^2 \left( c_4^2 (m+2)-c_6 (m+4)\right)} \, .
 \ee
 By simple power counting this automatically satisfies  \eqref{Vksi} for all $p>k$ because of the scaling in $m$ of  \eqref{tr_prop}. 
 
 This example not only shows how to construct the vertices but also it clarifies  why we need to be able to tune the coefficient $c_k$. Indeed  we must ask that the denominators of the equations above are not vanishing, namely $c_4^2 (m+2)-c_6 (m+4)\neq 0$ and $c_6\neq 0$.\footnote{E.g. these requirements are not satisfied by the distribution $P(x)=\delta(x-m)$ which gives $c_k=\frac{2^{1-\frac{k}{2}}}{\left(\frac{n}{2}+1\right)_{\frac{k}{2}-1}}$ and thus $c_4^2 (m+2)-c_6 (m+4)=0$.}
%
 In other words, the existence of vertices with the property \eqref{Vksi}  is an extra requirement which is not trivially satisfied by all distributions $P$.
 We can go on by repeating this construction for vertices $v_k$ with higher $k$. 
The number of coefficients in the ansatz for the vertices matches the number of independent tensor structures appearing in the equations \eqref{Vksi} for $p\leq k$ (this is ensured by symmetry) and thus we are able to fix all the coefficients in terms of the variable $c_k$. 
We checked this algorithm and found explicit formulae for the vertices $v_k$ for $k=3,\dots 6$. We do not report  the results here since the expressions are lengthy.

Using the vertices we can write a new expression for the partition function
  \be\label{Zbranched1}
Z=
\tr \prod_{\langle x x'\rangle} (1+K \sum_{i=1}^m s_i(x) s_i(x')) \prod_{x} (1+H  s_1(x)+ \sum_{3\leq k \leq 2d} W_k v_k(x))\, ,
\ee
where the coefficients  $W_k$ are introduced to count the number of vertices with $k$ lines in a  configuration. It is important to notice that for a given site we can either have a simple point, an endpoint or a vertex $v_k$. This means that we never have to consider traces of two vertices $v_{k_1}(x)$  and $v_{k_2}(x)$ and therefore that \eqref{Vksi} is enough to compute all possible configurations.  It is also worth mentioning that in this model there cannot be more than $4d$ spin variables at a given site (this happens when a vertex $v_{2d}(x)$ is attached to $2d$ lines) which explains why we required the maximal value of $k$ in the conditions \eqref{tr_prop} to take this value.

Because of the vertices,  the partition function now also generates  loop configurations (thus generic lattice animals and not only branched polymers) as long as they are constructed by connecting $v_k$ and endpoints. E.g. the simplest loop configuration is obtained by attaching an endpoint to one of the lines of a $v_3$ vertex and by making a closed loop with the remaining two lines --- giving a weight $H W_3 $. This term will not scale as $m$ because all the spin variables in the diagram must be of the form $s_1$. 

\begin{figure}[h]
 \centering
     \begin{subfigure}[b]{0.3\textwidth}
         \centering
         \includegraphics[width=\textwidth]{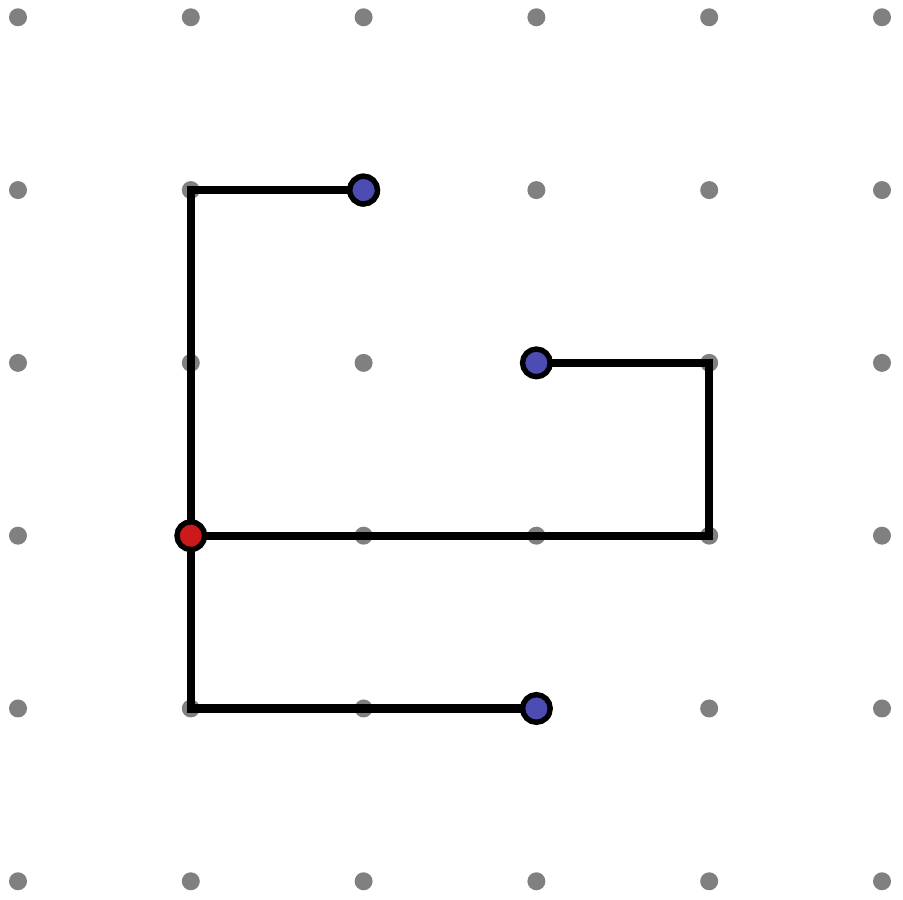}
     \end{subfigure}
    \qquad 
        \qquad
        \qquad       
  \begin{subfigure}[b]{0.3\textwidth}
         \centering
         \includegraphics[width=\textwidth]{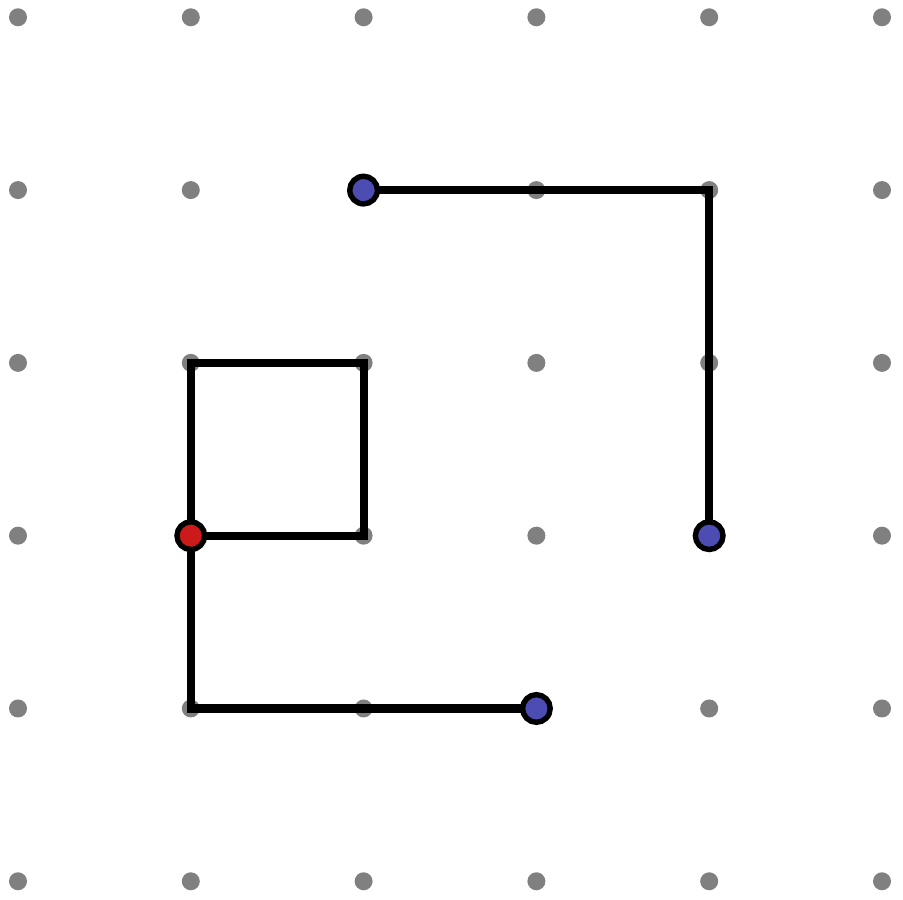}
     \end{subfigure}
 	\caption{
	Two configurations that scale like  $H^3 W_3$ (both made with eleven bonds which thus give an extra factor of $K^{11}$). The blue dots represent end points weighted by $H$, while the red dot represents a vertex $v_3$ weighted by $W_3$. The configuration on the left contains a single cluster while the configuration on the right contains two.
	\label{eg_clusters} }
\end{figure}

A problem of this model is that it does not give us any handle to count the number of separated clusters.
E.g. a diagram weighted by $H^3 W_3 $ (see figure \ref{eg_clusters}) could either be a vertex attached to three open ends or it could be made of two separate clusters, the first being like the one described above with weight $H W_3 $ and the second one being a line that connects two points with weight $H^2$.
This is not what we wanted, since we aimed at counting the number of configurations of a single lattice animal. In the following we explain how to modify the model in order to count the clusters. This will be crucial to take the dilute limit in which configurations with multiple clusters are suppressed.

There is a very simple way to count clusters. Indeed, as mentioned above, a linear cluster obtained by inserting at the open ends the operator $\sum_i s_i$ (instead of $s_1$)  is weighted by a factor of $m$ (similarly the weight of $m$ is kept for branched clusters if one uses vertices that satisfy $ \tr v_k(x) s_{i_1}(x) \cdots s_{i_p}(x) = \delta_{p k} \delta_{i_1 \dots i_k } $ instead of \eqref{Vksi}). 
Here however we do not want to use $m$ to count the number of clusters, since this variable  is already used to count the number of loops.
The idea is thus to introduce a new index $j$ in the spin variable obtaining $s_{i j}$ with $i=1, \dots m$ and $j=1,\dots n$, where the new parameter $n$ will be used to count clusters.
The label $j$ is independent of $i$ but works in a very similar way.
The trace is defined as 
\be
\tr = \int_{-\infty}^{+ \infty} \prod_{x} (\prod_{i=1}^{m} \prod_{j=1}^{n} d s_{ij}(x)) P(\sum_{i=1}^m\sum_{j=1}^n s_{ij}^2(x)) \, ,
\ee
where $P$ is an $O(m)\times O(n)$ invariant distribution which thus gives 
 \be
\label{si_corr1}
\tr s_{i_1 j_1}(x) \cdots s_{i_k j_k}(x) = 
D_{k} \; ( \d_{i_1 i_2}\d_{j_1 j_2} \dots \d_{i_{k-1} i_k} \d_{j_{k-1} j_k} + \dots) \, , \qquad (k \mbox{ even})
\ee
and vanishes when $k$ is odd. The dots contain products of Kronecker deltas for all pairings of the indices $(i_k,j_k)$ ---such that every $\d_{i_p i_q}$ always appears multiplied by $\d_{j_p j_q}$ with the same $p,q$.   
We require the coefficients $D_{k}$ to satisfy the conditions
\be
\label{tr_prop1}
\tr 1=1 \, ,
\qquad
 \tr s_{i_1 j_1}(x) s_{i_2 j_2}(x)=\delta_{i_1 i_2} \delta_{j_1 j_2} \, ,
 \qquad
D_k=m^{\frac{k}{2}-1}  d_k \,, \quad  \mbox{for } k=4,6,\dots,4d \, ,
\ee
where again we want to be able to tune the values of $d_k$ which must be finite in the limit $m \to 0$. As before all these conditions can be met by considering $P$ as a Gaussian distribution multiplied by a polynomial with opportunely chosen coefficients.
%
%

 We thus define the partition function
  \be\label{Zbranched}
Z=
\tr \prod_{\langle x x'\rangle} (1+K \sum_{i=1}^m \sum_{j=1}^n s_{ij}(x) s_{ij}(x')) \prod_{x} (1+H  \sum_{j=1}^n s_{1 j}(x)+ \sum_{3\leq k \leq 2d} W_k w_k(x))\, ,
\ee
where the new vertex  $w_k(x)$ satisfies
\be
\label{trw}
\lim_{m \to 0} \tr w_k(x) s_{i_1 j_1}(x) \cdots s_{i_p j_p}(x) = \delta_{p k} \delta_{1 i_1 \dots i_k } \delta_{ j_1 \dots j_k } \, ,
\ee
for $0\leq p \leq 2d$ and $3\leq k \leq 2d$. The vertices $w_k$ are built as polynomials of order $k$ in the spin variable $s_{ij}$ which preserved $S_n\times O(m-1)$ symmetry and whose terms are assembled from the building blocks $\sum_{j=1}^n s_{1 j}^p(x)$ and $\sum_{i=1}^m \sum_{j=1}^n s^2_{i j}(x)$.
It is easy to see that the same ideas used to construct $v_k$ can be generalized to define the vertices $w_k$. 
E.g. a vertex $w_3$ that satisfies \eqref{trw} can be defined as follows
\be
w_3(x)= a_1 \sum_{j=1}^n s_{1 j}(x)+ a_2 \sum_{j=1}^n s_{1 j}^3(x)+ a_3  \sum_{i=1}^m \sum_{j=1}^n s_{i j}^2(x)  \sum_{j'=1}^n s_{1 j'}(x) 
 \ee
 with 
 \be
 a_1=\frac{d_4}{m \left(d_4^2 (m n+2)-d_6 (m n+4)\right)}
  \, , \quad
 a_2=\frac{1}{6 d_6 m^2}
  \, , \quad
 a_3=\frac{d_6-d_4^2}{2 d_6 m^2 \left(d_4^2 (m n+2)-d_6 (m n+4)\right)} \, .
 \ee
 As before the coefficients $d_k$ are tuned to avoid singularities.
 
By expanding the partition function in the $m \to 0$ limit we obtain 
\be
Z = \sum_{\mathcal{C}} K^{N_B} H^{N_E} n^{N_C} \prod_ {3\leq k \leq 2d} W_k^{N_{V_k}} \, ,
\ee
where $\mathcal{C}$ is a configuration that contains $N_B$ bonds, $N_E$ endpoints, $N_C$ separated clusters and $N_{V_k}$ vertices with $k$ legs. The presence of all these parameters allows us to count the number of loops $N_L$ in a given configuration, which is obtained through the relation $  N_L =N_C-\frac{N_E}{2}+\sum_{k} (\frac{k}{2}-1)N_{V_k} $. We  thus obtain a partition function of the form
\be
\label{Z_lattice_animal}
Z = \sum_{\mathcal{C}} \Lambda_B^{N_B} \Lambda_L^{N_L} \Lambda_C^{N_C} \prod_ {3\leq k \leq 2d} \Lambda_{V_k}^{N_{V_k}} \, ,
\ee
where 
\be
\Lambda_B=K\, , \qquad
\Lambda_L=H^{-2}\, , \qquad
\Lambda_C=n H^{2} \, , \qquad
\Lambda_{V_k}=W_k H^{k-2}  \, . 
\ee
We further want to consider the diluted limit, namely the case in which there are few clusters. To do so we take the limit $n \to 0$. 
The resulting partition function generates lattice animals; in principle one could get branching polymers by forcing the number of loops to zero by taking $\Lambda_L \to 0$ (or equivalently $H \to \infty$) in \eqref{Z_lattice_animal} while keeping the rest of the $\Lambda$'s finite. 
We avoid this procedure and keep $\Lambda_L$ finite since lattice animals and branched polymers are known to be in the same universality class.  
While there is no full proof of the latter statement, this was checked  in various numerical works (see e.g. \cite{animals}) and also at low order in perturbation theory in \cite{LubIsslong}. 
Given the formulation \eqref{Zbranched}, the fact  the two models are in the same universality class can be rephrased by saying that the RG flows which start at infinite $H$ or finite $H$ in the UV should end up in the same IR fixed point. 
 As a final comment, we would like to mention that in physical systems where branched polymers arise one cannot typically exclude the presence of closed loops \cite{PhysRevLett.41.829} so in fact lattice animals define themselves a physically relevant system.

Now that we constructed the model \eqref{Zbranched} which counts diluted lattice animals, we want to show how to reformulate it  in terms of a field theory. 
We start by rewriting the partition function as follows
\be
Z=\int (\prod_{x} \prod_{i=1}^m\prod_{j=1}^n d s_{ij}(x)) e^{-\Hcal[s_{ij}]}
\ee
where the Hamiltonian takes the form
\begin{align}
\begin{split}
\label{final_H_lattice}
&\Hcal=-\sum_{\langle x x'\rangle} \log (1+K \sum_{i=1}^m \sum_{j=1}^n s_{ij}(x) s_{ij}(x')) 
\\
&\qquad \qquad\qquad 
-\sum_{x} [\log(1+H  \sum_{j=1}^n s_{1 j}(x)+ \sum_{k\geq 3} W_k w_k(x))+ \log P(\sum_{ij} s_{ij}^2(x))] \, .
 \end{split}
\end{align}
The model has a continuous symmetry $O(m-1)$ which rotates the $i=2,\dots m$ components of $s_{ij}$. This arises because the original $O(m)$ symmetry is broken by the end points and vertices which are built in terms of powers of spins in the direction $i=1$. 
The  $O(m-1)$ symmetry is not going to play any role in the final field theory since the directions $i\neq 1$ are not going to be critical. This may not be so counterintuitive since also in the cluster representation only the spin $s_{i=1,j}$ was contributing to the configurations. 
There is an extra $S_n$ symmetry  which acts by permuting the components $j$. This arises because the vertices $w_k$ contain powers $k$ of  $\sum_j s^k_{1j}$ with arbitrary $k$.
We thus expect to obtain a model in the continuum limit which has $S_n$ symmetry in the limit $n \to 0$.
In the following we shall explain how this arises   in more detail.

We first consider the nearest neighbour interaction term of \eqref{final_H_lattice}. Each term in the expansion of the logarithm can be further Taylor expanded for small lattice spacing $a$ between neighbouring sites. The result is written as a polynomial in $s\cdot s\equiv \sum_{ij}s_{ij}s_{ij}$ dressed with derivatives. E.g. let us consider the term $k=1$ in the expansion $\log (1+y) =\sum_{k=1}^{\infty} \frac{(-1)^k}{k} y^k$ of the logarithm and let us work for simplicity in $d=1$; in this case a site $x$ has two nearest neighbours at position $x+a$ and $x-a$, so $s(x) \cdot s(x+a) +s(x) \cdot s(x-a)=2 s(x)\cdot s(x) +a^2 s(x) \cdot \partial^2 s(x)+\frac{1}{12}a^4 s(x) \cdot \partial^4 s(x)+\dots $. Similarly the $k=2$ term give rise to $ [s(x) \cdot s(x+a)]^2  +[s(x) \cdot s(x-a)]^2=2 [s(x)\cdot s(x)]^2+2 [( \partial s(x)\cdot  s(x))^2 + s(x)\cdot s(x) \; s(x)\cdot \partial^2 s(x)]+ \dots$ and so on. Of course it is trivial to generalize this to higher dimensions by using derivatives $\partial^{\mu}$ along all space directions. The terms in the second sum in \eqref{final_H_lattice} are defined at a single site $x$ so they do not change by taking the continuum limit (however we still have to include in the final action all possible terms in the expansion of the logarithm). The result takes the form
\begin{align}
\begin{split}
\label{cont_lim}
\int d^d x & \bigg[  p_0 s(x)\cdot \partial^2  s(x)  +p_2 s(x) \cdot s(x)+ p_4 (s(x)\cdot s(x))^2 + \dots  +
\\
& \qquad +
 \sum_{j=1}^n  q_{1}  s_{1 j}(x)+ \sum_{k\geq 3} q_k   \sum_{j=1}^n s_{1 j}^{k}(x)+ \sum_{k\geq 1} q'_k   \sum_{j=1}^n s_{1 j}^{k}(x) (s(x)\cdot s(x)) + \dots  \bigg] \, ,
 \end{split}
\end{align}
where the first dots correspond to terms with higher powers of $s(x)\cdot s(x)$ possibly dressed with derivatives and the second dots take into account higher powers of $s(x)\cdot s(x)$ which multiply $\sum_{j=1}^n s_{1 j}^{k}$.
The coefficients $p_i$ are written in terms of $K$ and the lattice spacing $a$ 
while the  coefficients $q_i, q'_i, \dots$ are written in terms of polynomials in the variables $H$ and $W_i$.  
In order to have an action written in a canonical form we first to rescale the fields $s_{ij}(x)$ to get a kinetic term normalized as $\frac{1}{2} \partial^\m s(x) \cdot \partial_\m  s(x) $. Secondly we get rid of linear terms proportional to $q_1$ in the action by shifting $s_{1 j}(x)$. Because of this shift, the coefficients multiplying the monomials in $s_{1 j}(x)$ and in $s_{i\neq1\, j}(x)$ become different. In particular the mass terms $s^2_{1j}(x)$ and $\sum_{i>1} s^2_{ij}(x)$ appear with a different coefficient. When we study this action in the IR we tune the mass for $i=1$ in order to reach the critical point. All other fields are massive and are integrated out. The resulting action takes the form of an $S_n$ model
\be
\label{S_LI}
\int d^d x  \bigg[\frac{1}{2} \sum_j  \partial^\mu s_{1 j} \partial_\mu   s_{1 j} + c_1  \sum_j  s_{1 j}^2 + c_2  (\sum_j s_{1 j})^2 + c_3 \sum_j  s_{1 j}^3 + \dots
\bigg] \, ,
\ee
where the dots stand for any $S_n$ invariant interaction defined by powers of $s_{1 j} $ possibly dressed with derivatives and the coefficients $c_i$ are written in terms of the $p_i$ and $q_i$ of \eqref{cont_lim}. 
After mapping $s_{1 j} \to \phi_j$, this action is exactly of the form \eqref{Sr} (with $V(\phi)=c_1 \phi^2+ c_3\phi^3$ and $c_2=-H/2$)  which is the starting point for analyzing the RF $\phi^3$ model in terms of replica variables.\footnote{Notice that in the dots of \eqref{S_LI} there is also the term $\sum_{j_1, j_2}  s_{1 j_1} s_{1 j_2}^2$. 
This term corresponds to what we call $\sigma_1\sigma_2 $ in the main text, which is an irrelevant deformation.
}

\section{Classification of leaders in RF $\phi^3$ theory}\label{class}
The classification of leaders of the RF $\phi^3$ theory is very similar to the one of the RFIM described in \cite{paper2}. Indeed, the form of the leaders does not depend on the potential $V(\phi)$, nor on the space dimensions. 
The only differences are the following. First the classical dimensions of the fields depend on the spacetime dimension $d$ and the latter is different in the two models (since we work close to the respective upper critical dimension). Secondly  the RFIM has an extra $\mathbb{Z}_2$ symmetry, and thus one is allowed to focus on  leaders which are singlet under  $\mathbb{Z}_2$ (which are all composite operators made out of an even number of elementary fields). In the RF $\phi^3$ theory the symmetry $\mathbb{Z}_2$ is absent, we will thus need to consider also  operators made out of an odd number of elementary fields. 

Let us call $N_{\phi}$ the number of elementary fields and $N_{\tmop{der}}$ the number of derivatives that are contained in the composite operators. As an example of even-$N_{\phi}$ leaders, let us consider the case of $N_{\phi} = 2$. As for the RFIM, the possible singlets without derivatives are $\sigma_2$, $\sigma_1^2$. All leaders with $N_{\phi} = 2$ must be susy-writable as it was discussed in \cite{paper2}. Here the only difference is that their dimensions is computed in $d=8$, as shown in table \ref{Nf2Nd0}.  
\begin{table}[h]\centering
    \begin{tabular}{@{}lll@{}}
      \toprule
      Singlet & Leader & Leader type\\
      \midrule
      $\sigma_2$ & $\left[ 2 \omega \vf + \chi_i^2  \right]_{\Delta = 6}$ &
      susy-writable\\
      $\sigma_1^2$ & $[\omega^2]_{\Delta = 8}$ & susy-writable\\
      \bottomrule
    \end{tabular}
  \caption{\label{Nf2Nd0}Scalar leaders with $N_{\phi} =
  2$, $N_{\tmop{der}} = 0$ in $d=8$.}
\end{table}
Similarly operators with derivatives will have the same exact form as the one described in \cite{paper2}. E.g.  $\sigma_{2 (\mu)
(\mu)} = \left[ 2 \partial \omega \partial \vf + (\partial \chi_i)^2 
\right]_{\Delta = 8}$.
This should make it clear that we do not need to repeat the classification for all even $N_{\phi}$, which is given in \cite{paper2} and still holds, up to the scaling dimension of the fields which can be very easily recomputed following the $d=8$ rules that $\D_\varphi=2$, $\D_{\chi_i}=3$ and $\D_\omega=4$.
 It is easy to see that this amounts to shift the $d=6$ dimension in all tables of \cite{paper2} by $N_{\phi}$, namely 
\begin{equation}
\Delta_{\textrm{leader}}^{(d=8)}=\Delta_{\textrm{leader}}^{(d=6)}+N_{\phi} \, .
\end{equation}
Applying the equation above to the cases studied in \cite{paper2} we find the lowest dimensional leaders in the even-$N_{\phi}$ spectrum of $\phi^3$ theory in $d=8$:  the first susy-null leader is $(\chi_i^2)^2$ with dimension $\D=12$ while the first non-susy-writable leader is $(\mathcal{F}_6)_L$ and only appears  at dimension $\D= 18$. In  \cite{paper2} other susy-null leaders appear at $\D \leq 18$, the second one is $\varphi^2(\chi_i^2)^2$ with dimension  $\D= 16$  in $d=8$.
In the following we want to check if in the odd-$N_{\phi}$ spectrum of $\phi^3$ theory there are new low lying leaders  in the  susy-null and non-susy-writable sectors. 
We will see that there are no new non-susy-writable leaders at $\D\leq 18$, therefore the lowest dimensional leader of this type is still $\mathcal{F}_6$.\footnote{Another non-susy-writable leader is $\mathcal{F}_{6,1}$ and can be found at dimension $20$ as explained in the main text. In this appendix we will not show this leader because we decided to focus on $\D\leq 18$.  } 
 Similarly we find that the first susy-null leader is still $(\chi_i^2)^2$, however we also find other susy-null leaders at $\D\leq 18$ with odd $N_{\phi}$.  The first new one is $\varphi(\chi_i^2)^2$ which appears at $N_{\phi}=5$ and has dimension $14$ in $d=8$.  In equation \eqref{def:NkSingl} in the main text we show that this operator is a member of an infinite class of susy-null operators named $\mathcal{N}_k$, which have leaders equal to $\varphi^k(\chi_i^2)^2$, which also describe the cases $(\chi_i^2)^2$ and $\varphi^2(\chi_i^2)^2$ discussed above. 
\subsection{$N_{\phi} = 1$}
A single operator with $N_{\phi} = 1$ is present,
\begin{equation}
\sigma_1 =  \omega \, ,
\end{equation}
which is susy-writable.
When we add this operator to the Lagrangian, the physics does not change since $\omega$ can be reabsorbed by shifting the fields. In SUSY language it corresponds to adding $\Phi$ to the SUSY Lagrangian. Equivalently this corresponds to adding $\hat \phi$ to the dimensionally reduced Lagrangian.
 In both cases this perturbation can simply be undone by a shift in the fields (respectively $\Phi$ and $\hat \phi$).
\subsection{$N_{\phi} = 3$}
There are only three leaders with $N_{\phi} = 3$ and no derivatives. They are all of the susy-writable type as shown in table \ref{Nf3Nd0}. 
\begin{table}[h]\centering
    \begin{tabular}{@{}lll@{}}
      \toprule
      Singlet & Leader & Leader type\\
      \midrule
      $\sigma_3$ & $\left[ \omega \vf^2 + \chi_i^2  \vf \right]_{\Delta = 8}$ &
      susy-writable\\
      $\sigma_1 \sigma_2$ & $[2 \varphi  \omega^2 +\chi_i^2 \omega]_{\Delta = 10}$ & susy-writable\\
      $\sigma_1^3 $ & $[\omega^3]_{\Delta = 12}$ & susy-writable\\
      \bottomrule
    \end{tabular}
  \caption{\label{Nf3Nd0}Scalar leaders with $N_{\phi} =
  3$, $N_{\tmop{der}} = 0$ in $d=8$.}
\end{table}
We can also consider leaders with $N_{\phi} = 3$ and two derivatives. The result is still a list of susy-writable operators as shown in table \ref{Nf3Nd2}. \\
\begin{table}[]
\centering
    \begin{tabular}{@{}lll@{}}
      \toprule
      Singlet & Leader & Leader type\\
      \midrule
      $\sigma_{3 (\m) (\m)}$ & $\left[ 2 \chi _i \varphi_{,\m}  \chi_{i,\m}+\varphi   \chi _{i,\m}^2+\omega  \varphi_{,\m}^2+2 \varphi   \varphi_{,\m}   \omega_{,\m} \right]_{\Delta = 10}$ &
      susy-writable\\
      $\sigma_{1(\m)} \sigma_{2(\m)}$ & $[\omega _{\mu } \left(\chi _i \chi _{i,\mu }+\omega  \varphi _{\mu }+\varphi  \omega _{\mu }\right)]_{\Delta = 12}$ & susy-writable\\
      $\sigma_1  \sigma_{2(\m)(\m)}$ & $[\omega  \left(\chi _{i,\mu }^2+2 \varphi _{\mu } \omega _{\mu }\right)]_{\Delta = 12}$ & susy-writable\\
   $\sigma_1  \sigma_{1(\m)}\sigma_{1(\m)}$ & $[\omega  \omega _{\mu }^2]_{\Delta = 14}$ & susy-writable\\
      \bottomrule
    \end{tabular}
  \caption{\label{Nf3Nd2}Scalar  leaders with $N_{\phi} =
  3$, $N_{\tmop{der}} = 2$ in $d=8$.}
\end{table}
By increasing the number of derivatives we do not find new interesting operators. Indeed all $\s_k$ with $k\leq2$ have a susy writable leader with no followers. So, any combination of these with any number of derivatives will be susy-writable. One can thus focus on $\s_3$ ---which is the only operator with $N_{\phi} =
  3$ that has followers (which are not always susy-writable)--- and try to dress it with derivatives. But one can see that these do not produce new  non-susy-writable or susy-null operators. Indeed we first notice that there exists a single independent\footnote{All other dressings are equal to $\s^{(k)}_3$ plus total derivatives plus terms  built by dressing $\s_1$ and $\s_2$ with derivatives (which we already considered above). 
  }  way to write a Lorentz scalar by dressing $\s_3$ with derivatives:  $\s^{(k)}_3 \equiv \sum_i \phi_i (\partial_{\m_1}\dots \partial_{\m_k}\phi_i)^2$. 
  The associated leaders are susy-writable. 
  One may still worry that another singlet could have the same susy-writable leader and that their difference should be considered as a new leader. However the leader of $\s^{(k)}_3$ has lower dimension than any other leader built by composing products of $\s_1$ and $\s_2$ dressed by derivatives, as one can see for the example of two derivatives in table \ref{Nf3Nd2}. This ensures that there is no other singlet with the same leader and therefore that the followers of  $\s^{(k)}_3$ can always be neglected. In summary we find that no matter how many derivatives we consider, any operator built out of $N_{\phi}=3$ fields must have a susy-writable leader. 
    
\subsection{$N_{\phi} = 5$}
We now focus on the operators built out of five fields. The list of $N_{\phi} = 5$ leaders with $\D\leq 18$ and $N_{\tmop{der}} = 0$ is shown in table \ref{Nf5Nd0}. Here we find the first odd-$N_{\phi} $ susy-null operator which we call $(\mathcal{N}_5)_L$ according to the definition \eqref{def:NkSingl} in the main text. This operator has dimension $\Delta=14$ and leader equal to $\varphi(\chi_i^2)^2$. 
At dimensions $\Delta=16$  we also find the new susy-null leader  $\omega(\chi_i^2)^2$.
\begin{table}[h]\centering
    \begin{tabular}{@{}lll@{}}
      \toprule
      Singlet & Leader & Leader type\\
      \midrule
      $\sigma_5$ & $\left[ \varphi ^3 (\varphi  \omega +2 \chi_i^2) \right]_{\Delta = 12}$ &
      susy-writable\\
      $\sigma_1 \sigma_4$ & $[\varphi ^2 \omega  (2 \varphi  \omega +3 \chi_i^2)]_{\Delta = 14}$ & susy-writable\\
      $\mathcal{N}_1$ & $[\varphi(\chi_i^2)^2]_{\Delta = 14}$ & susy-null\\
 $\sigma_3 \sigma_1^2$ & $[\varphi  \omega ^2 (\varphi  \omega +\chi_i^2)]_{\Delta = 16}$ & susy-writable\\
 $4\sigma_3 \sigma_1^2-3 \sigma_2^2 \sigma_1$ & $[\omega(\chi_i^2)^2]_{\Delta = 16}$ & susy-null\\
 $\sigma_2 \sigma_1^3$ & $[\omega ^3 (2 \varphi  \omega +\chi_i^2)]_{\Delta = 18}$ & susy-writable\\
      \bottomrule
    \end{tabular}
  \caption{\label{Nf5Nd0}Scalar leaders with $N_{\phi} =
  5$, $N_{\tmop{der}} = 0$ in $d=8$.}
\end{table}

By considering $N_{\tmop{der}} = 2$ we must take into account the following $S_n$ singlets 
\begin{align}
\begin{split}
\s_{5, (\m)(\m)} \ , \quad 
 \s_{4, (\m)(\m)} \s_{1} \ , \quad 
  \s_{3, (\m) (\m)} \s_{2} \ , \quad 
  \s_{3} \s_{2, (\m) (\m)} \ , \quad 
  \s_{3, (\m)(\m)} \s^2_1 \ , \quad 
  \s_{2, (\m) (\m)}  \s_{2}\s_{1} \ , \quad 
  \s_{2, (\m) (\m)}  \s_{1}^3 \ , 
  \\
  \s_{4, (\m)} \s_{1, (\m)} \ , \quad 
 \s_{3, (\m)} \s_{2, (\m)} \ , \quad 
\s_{3, (\m)} \s_{1, (\m)} \s_1 \ , \quad 
\s_{2, (\m)} \s_{2, (\m)} \s_1 \ , \quad
\s_{2, (\m)} \s_{1, (\m)} \s_1^2 \ , \quad 
\s_{1, (\m)}^2 \s_1^3 \ . 
\end{split}
\end{align}
Combinations of these singlets give either susy-writable or susy-null leaders. 
At dimension $16$ there are five leaders: three susy-writable and two susy-null.
Since the two susy-null leaders have dimension greater than $(\mathcal{N}_5)_L$, they are not very important for our classification. However it is instructive to show their explicit form:
\begin{align}
\Ocal^{(1)}_{\textrm{null}} &\equiv 2\s_{3} \s_{2, (\m) (\m)} - 6 \s_{2} \s_{3, (\m) (\m)} + 6 \s_{1} \s_{4, (\m) (\m)} + 3 \s_{2}^2 \s_{1} -4 \s_3 \s_1^2  \, , \\
\Ocal^{(2)}_{\textrm{null}}&\equiv -2 \sigma_1 \sigma_{4(\m)(\m)}+\frac{2 \sigma_{2(\m)(\m)} \sigma_3}{3}-2 \sigma_{1 (\m)}\sigma_{4 (\m)}+2 \sigma_{2 (\m)}\sigma_{3 (\m)} \, ,
\end{align}
which gives rise to the susy-null leaders
\begin{align}
(\Ocal^{(1)}_{\textrm{null}})_L =  -3 \partial _{\mu }[\varphi_{, \m}(\chi_i^2)^2]_{\D=16}\, ,
\qquad 
(\Ocal^{(2)}_{\textrm{null}})_L= [\partial _{\mu } \left(\varphi  \chi_i^2 \partial_{\mu }(\chi_j^2)\right) ]_{\D=16}\, .
\end{align}
It is interesting to notice that the leaders of $\Ocal^{(1)}$ and $\Ocal^{(2)}$ are just total derivatives, indeed these operators can be written as derivative of spin-one $S_n$-singlets,
\begin{align}
\Ocal^{(1)}_{\textrm{null}} &=\partial_{\mu}\left(2 \s_3 \s_{2 (\m)}-3 \s_2 \s_{3 (\m)}-\frac{1}{2} \s_4 \s_{1 (\m)}+2 \s_1 \s_{4 (\m)}\right) \, , \\
\Ocal^{(2)}_{\textrm{null}}&=\partial_{\mu} \left(-\frac{2}{3} \sigma_{1} \sigma_{4 (\m)}+\frac{2}{3} \s_3 \sigma_{2 (\m)}-\frac{1}{3}\s_4\sigma_{1 (\m)}\right)\, .
\end{align}
 Since the operators are total derivatives they do not affect the RG and we can discard them for the purpose of the present classification. Analogous considerations are valid for the next null operators with $N_{\textrm{der}}=2,N_{\phi}=5$ which have dimensions $18$ and are built by dressing $\omega(\chi_i^2)^2$ with derivatives.

By studying higher number of derivatives with $N_{\phi} = 5$ we do not find any non-susy-writable leader with $\Delta\leq 18$. E.g. by considering $N_{\tmop{der}} = 4$  the lowest dimensional operator is  $\s_5$ dressed by derivatives which has a susy-writable leader with dimension $16$, while the first susy-null operator has already dimensions $\D=18$.

To summarize for $N_{\phi} = 5$  we found that the lowest dimensional susy-null leader is $\mathcal{N}_1$ and that there are no non-susy-writable leaders up to $\Delta\leq 18$ .


\subsection{$N_{\phi} = 7$}
The $N_{\phi} = 7$ leaders with $N_{\tmop{der}} = 0$ and $\D\leq 18$ are considered in table \ref{Nf7Nd0}
We notice that the only susy-null operator belongs to the family $\mathcal{N}_k$ for $k=7$ and it has dimension $18$ in $d=8$.
\begin{table}[!h]\centering
    \begin{tabular}{@{}lll@{}}
      \toprule
      Singlet & Leader & Leader type\\
      \midrule
      $\sigma_7$ & $\left[ \varphi ^5 (\varphi  \omega +3 \chi_i^2) \right]_{\Delta = 16}$ &
      susy-writable\\
      $\sigma_1 \sigma_6$ & $[ \varphi ^4 \omega  (2 \varphi  \omega +5 \chi_i^2)]_{\Delta = 18}$ & susy-writable\\
      $\mathcal{N}_3$ & $[\varphi ^3 (\chi_i^2)^2]_{\Delta = 18}$ & susy-null\\
      \bottomrule
    \end{tabular}
  \caption{\label{Nf7Nd0}Scalar leaders with $N_{\phi} =
  7$, $N_{\tmop{der}} = 0$ in $d=8$.}
\end{table}
It is easy to see that all  leaders with $N_{\tmop{der}} > 0$ and  $\D\leq 18$ are susy-writable.
\section{Box operators}
\label{App:Susy_Writable}
The only susy-writable operators which may destabilize the SUSY fixed point are those which transform in the $(2,2)$ irrep of OSp$(d|2)$. Since these are susy-writable their anomalous dimensions are determined by computing anomalous dimensions of the corresponding operators in $\hat \phi^3$ theory in $d-2$ dimensions. In this appendix we show the form of these operators both in $\hat \phi^3$ theory and in the SUSY theory.

In order to define operators which have good scaling dimension at one loop, we consider the primary operator built out of free scalars $ \widehat{\phi}$ in $\widehat d=d-2$ dimension.
We are interested in primaries that transform in the $(2,2)$ representation of $SO(d)$. It is easy to show that they take the form
\begin{equation}
 \widehat{B}^{(k)}_{\mu \nu , \rho \sigma} \equiv \widehat{\phi}^{k-3} \left( \widehat{\phi}_{, \mu \nu} \widehat{\phi}_{,
  \rho \sigma} \widehat{\phi} - \frac{2 \widehat{d}}{\widehat{d} - 2} \widehat{\phi}_{, \mu}
  \widehat{\phi}_{, \nu} \widehat{\phi}_{, \rho \sigma}  \right)^Y \, ,
  \label{Boxn}
\end{equation}
where $k\geq 3$. To unpack \eqref{Boxn}, we contract it with polarization vectors $z_1^{\m}z_1^{\n}z_2^{\rho}z_2^{\s}$ such that $z_1^2=z_2^2=z_1\cdot z_2=0$. We thus obtain
\begin{align}\label{Bncontracted}
\! \! \! \! \! \! \! 
\begin{split}
\widehat{B}^{(k)}(z_1,z_2) \propto &\widehat{\phi} ^{k-3} \bigg[
d \left(z_1\cdot \partial \right)^2 \widehat{\phi}  \left(\left(z_2\cdot \partial \right)\cdot \widehat{\phi} \right){}^2
-2 d \left(z_1\cdot \partial \right) \widehat{\phi}  \left(z_1\cdot \partial \right) \left(z_2\cdot \partial \right) \widehat{\phi}  \left(z_2\cdot \partial \right) \widehat{\phi} 
\\
&
+(d-2) \widehat{\phi}  \left(\left(z_1\cdot \partial \right) \left(z_2\cdot \partial \right) \widehat{\phi} \right){}^2
+  \left[d \left(\left(z_1\cdot \partial \right) \widehat{\phi} \right){}^2-(d-2) \widehat{\phi}  \left(z_1\cdot \partial \right)^2 \widehat{\phi} \right]
\left(z_2\cdot \partial \right)^2\widehat{\phi}
\bigg]\, ,
\end{split}
\end{align}
where each derivative is meant to act only on the closest $\widehat{\phi}$ on its right.
The two lowest dimensional operators of this family are thus $\widehat{B}^{(3)}$ and $\widehat{B}^{(4)}$.\footnote{To study the RFIM in \cite{paper2} we considered the operator $\widehat{B}^{(4)}$, since it is the lowest dimensional operator in this representation which is $\mathbb{Z}_2$-even. Here the  $\mathbb{Z}_2$ symmetry is not present thus we are allowed to consider operators built out of an odd number of elementary fields like $\widehat{B}^{(3)}$.
}

We now want to write down the correspondent SUSY superprimary $\mathcal{B}^{(k),a b, c d}$.
We are especially interested  in the highest component $\mathcal{B}^{(k), \theta \bar \theta,\theta \bar \theta}_{\theta \bar \theta}$ with all indices in directions $\theta$ and $\bar \theta$. 
The operator $\mathcal{B}^{(4)}$ was explicitly written in \cite{paper2} (in $d=6$). In the following we thus focus on $\mathcal{B}^{(3)}$, which can be written in components as follows
\begin{align}
\label{B3SUSYtop}
\begin{split}
\mathcal{B}^{(3), \theta \bar \theta,\theta \bar \theta}_{\theta \bar \theta}\propto&\;
 \omega  \left(-86 \psi _{,\mu } \bar{\psi }_{,\mu }-\varphi _{,\mu \nu }^2-72 \varphi _{,\mu } \omega _{,\mu }\right)
 +6 \varphi _{,\mu } (\psi _{,\nu } \bar{\psi }_{,\mu \nu }+ \psi _{,\mu \nu } \bar{\psi }_{,\nu })
 -2 \varphi _{,\mu \nu } (\psi  \bar{\psi }_{,\mu \nu }+ \psi _{,\mu \nu }\bar{\psi })
\\
& +6 \varphi _{,\mu \nu } \psi _{,\mu } \bar{\psi }_{,\nu }
+14 \omega _{,\mu } (\psi  \bar{\psi }_{,\mu }+ \psi _{,\mu }\bar{\psi })
+6 \varphi _{,\mu } \omega _{,\nu } \varphi _{,\mu \nu }
+3 \varphi _{,\mu } \varphi _{,\nu } \omega _{,\mu \nu }
+158 \omega ^3
\\
&
+2 \varphi  \left(-\psi _{,\mu \nu } \bar{\psi }_{,\mu \nu }-\varphi _{,\mu \nu } \omega _{,\mu \nu }+7 \omega _{,\mu }^2\right) \ ,
\end{split}
\end{align}
where here, for simplicity, we set $d=8$. To compute anomalous dimensions of this operator, one can alternatively consider the lowest component of the same supermultiplet, which has a considerably simpler form
\be
\label{B3SUSY0}
\mathcal{B}^{(3), \theta \bar \theta,\theta \bar \theta}_{0}\propto
\varphi  
\left(64 \omega ^2-2 \left(14 \psi _{\mu } \bar{\psi }_{\mu }+\varphi _{\mu ,\nu }^2\right)\right)+6 \varphi _{\mu } \left(7 \psi  \bar{\psi }_{\mu }+7  \psi _{\mu }\bar{\psi }+\varphi _{\nu } \varphi _{\mu ,\nu }\right)-6 \omega  \left(42 \psi  \bar{\psi }+5 \varphi _{\mu }^2\right) \, .
\ee
In App.  \ref{app:susywrit}  we will compute the anomalous dimensions of $\mathcal{B}^{(3)}$ using the formulation \eqref{Bncontracted} in terms of the $\hat \phi^3$ theory. It would be nice to check this computation using the SUSY descriptions  \eqref{B3SUSYtop} or \eqref{B3SUSY0}. 

\section{RG computations}\label{2loop}\label{app:RG}

In this appendix we give the details of the results shown in section
\ref{sec:anomalous_dim}. All our computations will be at one loop order. For our
theory this typically involves two vertices. It is suitable to use Feynman
diagrams for such cases, as opposed to the OPE method we used for one-loop computations in \cite{paper2}. We will use dimensional regularization scheme in our
computations for which we refer to {\cite{srednicki_2007,Kleinert:2001ax}}  for a good review.

We will work with the $\chi_i$ fields instead of $\psi,\psib$ fields. This is suitable for computing anomalous dimensions of general susy-writable, susy-null and non-susy-writable operators.

\subsection{Wavefunction renormalization}\label{app:wvefn}

Our theory is given by the Lagrangian {\eqref{L0}} in $d = 8 - \e$ dimension,
written in Cardy variables and mass term set to zero:
\begin{equation}
\Lcal_L = \partial \varphi_B \partial \omega_B - \frac{H}{2} \omega_B^2 +
\frac{1}{2} (\partial \chi_B)^2 + \frac{g_B}{6} \left( \left. 3 \omega_B 
\vf_B^2 + 3 \chi_B^2  \vf_B \right) \right. . \label{lagL0}
\end{equation}
The subscript `$B$' denotes bare quantities that are related to the
renormalized ones as follows: \footnote{Although we could consider separate
	renormalizations for the couplings corresponding to $\vf^2 \omega$ and $\chi^2
	\vf$, it is sufficient to focus on just one. We comment later that the two
	vertices lead to the same renormalization constant and beta function. }
\begin{equation}
\vf_B = Z_{\vf} \vf, \hspace{0.5cm} \omega_B = Z_{\omega} \omega,
\hspace{0.5cm} \chi_B = Z_{\chi} \chi, \hspace{0.5cm} g_B = Z_g Z_{\vf}^{-
	2} Z_{\omega}^{- 1} \mu^{\e / 2} g. \label{counterterms}
\end{equation}
Here $\mu$ is an arbitrary mass scale. The $Z$-s denote renormalization
constants which can be expanded in powers of $1 / \epsilon$ and $g^2$ as
follows:
\begin{equation}
\quad Z_A = 1 + \sum_{p \geqslant 1} \sum_{1 \leqslant q \leqslant p} z^{(p,
	q)}_A g^{2 p} \epsilon^{- q}, \label{Zform}
\end{equation}
where the subscript $A = g, \vf, \omega, \chi$. The above is very much similar
to the usual $\phi^3$-theory fixed point in $d = 6 - \e$ dimension.
Correlation functions of the bare quantities contain poles. Since those of the
renormalized ones must be finite we should choose $z^{(p, q)}_{}$ accordingly.
\ \

\begin{figure}[h]
	\centering \includegraphics[width=400pt]{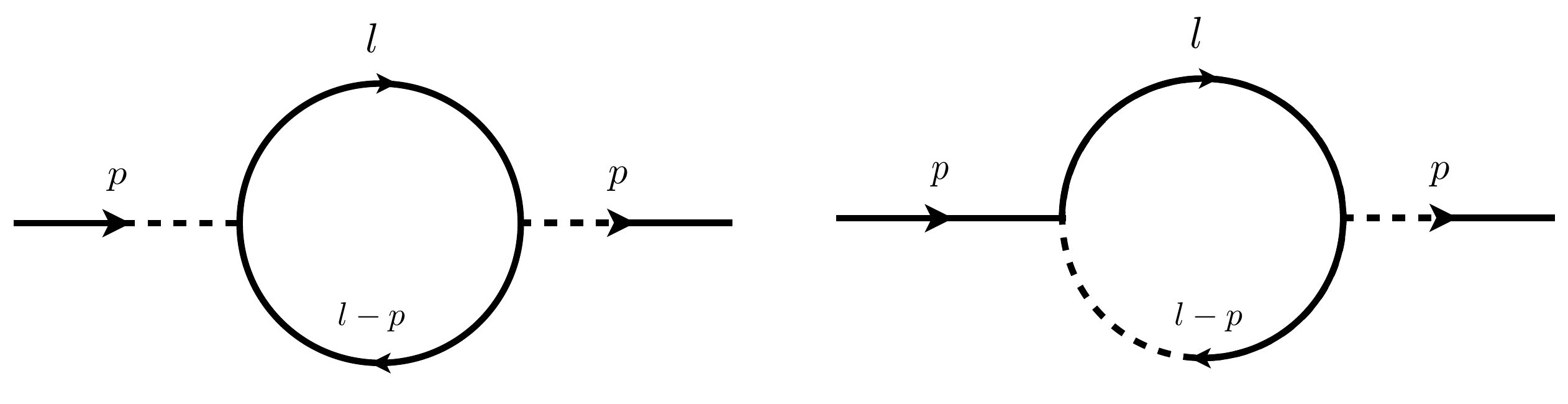}
	\
	\caption{The Feynman diagrams for the one-loop
		renormalization of $\langle \vf (p) \vf
		(-p)   \rangle$.\label{wvfnfig1} }
\end{figure}

\begin{figure}[h]
	\centering \includegraphics[width=400pt]{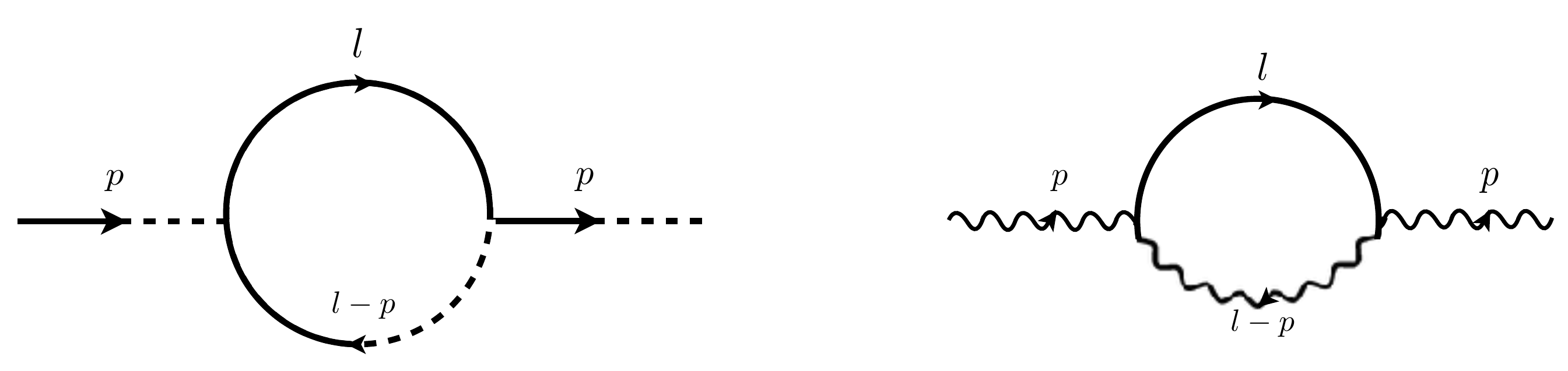}
	\
	\caption{\label{wvfnfig2}The Feynman diagrams for the one-loop
		renormalization of $\langle \vf (p) \omega(-p)   \rangle$ (left) and $\langle\chi_i \vf (p) \chi_j(-p) \rangle$ (right). We have suppressed the $i,j$ indices.}
\end{figure}

We begin with the most basic computation which is to obtain the wavefunction
renormalization constants $Z_{\vf}, Z_{\omega}, Z_{\chi}$. For this we
consider the correlators $\langle \vf (p) \omega (- p) \rangle$, $\langle \vf
(p) \vf (- p) \rangle$ and $\langle \chi_i (p) \chi_j (- p) \rangle$ and
demand that they should be finite. Their corrections are shown in Fig.
\ref{wvfnfig1} and \ref{wvfnfig2}. The corrections to $\langle \vf (p) \omega (- p) \rangle$ and
$\langle \chi_i (p) \chi_j (- p) \rangle$ are equal and given by the integral:
\begin{equation}
I_{\vf \omega} (p^2) = \frac{g^2 H^{}}{(2 \pi)^d} \int \frac{d^d
	l_{}}{(l^2)^2  (p + l)^2} = \; - \frac{g^2 H p^2}{6 (4 \pi)^4  \e} + O
\left( \e^0 \right) . \label{phiomegaint1}
\end{equation}

The correction to $\langle \vf (p) \vf (- p) \rangle$ contains the above and
also the additional integral :
\begin{equation}
I_{\vf \vf} (p^2) = \frac{g^2 H^{}}{(2 \pi)^d} \int \frac{d^d l_{}}{(l^2)^2 
	((p + l)^2)^2} = \; \frac{g^2 H}{3 (4 \pi)^4  \e} + O \left( \e^0 \right) .
\label{phiomegaint2}
\end{equation}

Taking into account the respective symmetry factors the one-loop corrected
2-point functions are given by \ :
\begin{eqnarray}
\left\langle \vf (p) \omega (- p) \right\rangle & = & \frac{1}{p^2} \left[
\frac{1}{Z_{\vf} Z_{\omega}} - \frac{g^2 H}{6 (4 \pi)^4  \e} + O \left( g^2
\e^0 \right) \right], \nonumber\\
\langle \chi_i (p) \chi_j (- p) \rangle & = & \frac{1}{p^2} \left[
\frac{1}{Z_{\chi}^2} - \frac{g^2 H}{6 (4 \pi)^4  \e} + O \left( g^2 \e^0
\right) \right], \\
\langle \vf (p) \vf (- p) \rangle & = & \frac{H}{(p^2)^2} \left[
\frac{1}{Z_{\vf}^2} - \frac{g^2 H}{6 (4 \pi)^4  \e} + O \left( g^2 \e^0
\right) \right] . \nonumber
\end{eqnarray}
This leads to
\begin{equation}\label{phiomega}
 Z_{\vf} = Z_{\omega} =
Z_{\chi}^{} = 1 - \frac{g^2 H}{12 (4 \pi)^4  \e} . 
\end{equation}

\subsection{Beta function}\label{app:betafn}

\begin{figure}[h]
	\centering \includegraphics[width=400pt]{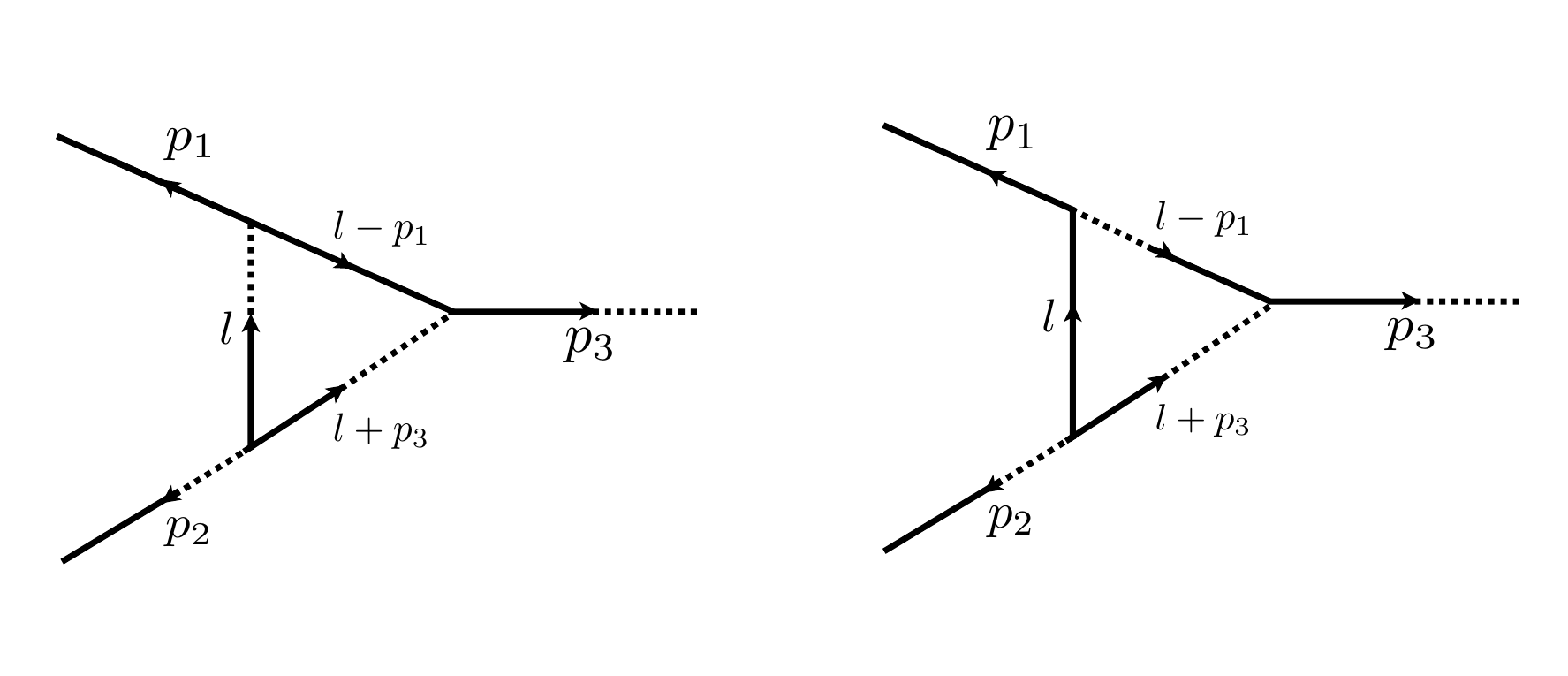}
	\
	\caption{\label{betafnfig}The Feynman diagrams for the one-loop
		renormalization of $g$. The shown diagrams correct the correlator $\langle \vf (p_1) \vf
		(p_2) \omega (p_3)  \rangle$ at one loop.}
\end{figure}

We will now compute the one-loop beta function. For this we consider the
3-point function $\langle \vf (p_1) \vf (p_2) \vf (p_3) \rangle$ and demand
that it is free of $\e$ poles. The tree level diagram involves the $\omega
\vf^2$ vertex. The one-loop diagrams are of the type shown in
Figure \ref{betafnfig} . \ The amputated form of the 3-point
function is given by:
\begin{equation}
\langle \vf (p_1) \vf (p_2) \omega (p_3) \rangle_{\tmop{amp}} \, = g_B +
\frac{3}{2} g_B^3 [I (p_1, p_2) + I (p_3, p_1)], \label{4ptamp} 
\end{equation}
The renormalization constants appear through $g_B$ via {\eqref{counterterms}}.
\ $I_{\omega \vf^3}$ is given by the integral
\begin{equation}
I (p_1, p_2) =  \frac{H}{(2 \pi)^d} \int \frac{d^d l}{(l^2) 
	(l + p_2)^2 ((l - p_1)^2)^2 }, \label{loopint}
\end{equation}
This is a triangular loop integral that can be evaluated using standard
methods (see e.g. {\cite{srednicki_2007,Kleinert:2001ax}}) , and gives the following $\e$ pole:
\begin{equation}
I (p_1, p_2) = \frac{H}{3 (4 \pi)^4 \e} + O \left( \e^0 \right) .
\label{Ipole}
\end{equation}

\

Demanding that the pole cancels, and using the wave function renormalization
constants from {\eqref{phiomega}} we obtain
\begin{equation}
Z_g = 1 - \frac{g^2 H}{(4 \pi)^4 \e}, \label{Zl} 
 .
\end{equation}
We thus obtain the beta function $\left( \beta_g \equiv \mu \frac{\partial
	g}{\partial \mu} \right)$ as follows:
\begin{eqnarray*}
	&  & 0 \: = \frac{\partial \log g_B }{\partial \log \mu} =
	\frac{\partial}{\partial \log \mu} \log \left[ Z_g Z_{\vf}^{- 2}
	Z_{\omega}^{- 1} \mu^{\e / 2} g \right] = \frac{\e}{2} -
	\frac{{3 g H}}{2 (4 \pi)^3 \e} \beta_g + \frac{1}{g}
	\beta_g,\\
	&  & \Rightarrow \: \beta_g = {- \frac{\epsilon g}{2} -
		\frac{3 g^3 H}{4 (4 \pi)^4}}
\end{eqnarray*}
This gives a fixed point at
\begin{equation}
g_{\text{${\star}$}}^2 = - \frac{2  (4 \pi)^4 \e}{3 H} \label{fp} 
 .
\end{equation}
The two solutions to $g_{\star}$ are equivalent as they are related by transforming $\vf\to -\vf$, $\omega\to - \omega$\,.

The same result can be obtained by considering the 3-point function $\langle
\chi_j (p_1) \chi_j (p_2) \vf (p_3) \rangle$. It involves the vertex $\chi_i^2
\vf^2$ at tree level. At one loop we have two diagrams which
give the same $Z_g$. This also follows from the equivalence of the Lagrangian
{\eqref{lagL0}} with the SUSY theory {\eqref{Lsusy}}.

\subsection{Renormalization of local operators}\label{localOp}

For RG analysis of local operators we consider bare operators
$\mathcal{O}_i^B$ (composed of bare fields), that are related to renormalized
operators $\mathcal{O}_j^B$ by a mixing (renormalization) matrix $Z_{i j}$:
\begin{equation}
\mathcal{O}^B_i = Z_{i j} \mathcal{O}_j . \label{Zij}
\end{equation}
Similar to the renormalization constants of previous subsection, the matrix
$Z_{i j}$ can be expanded in powers of $1 / \epsilon$ and $g$ as follows
\begin{equation}
Z_{i j} = \tmd_{i j} + \frac{g^2}{\epsilon} z_{i j} + \ldots \hspace{0.17em}
. \label{ZijForm}
\end{equation}
The matrix $z_{i j}$ has to be fixed by computing one-loop corrections to
correlation functions of $O_j$ and requiring them to be free of $1 / \epsilon$
poles. In general, this matrix is of block diagonal form. A certain block in
$z_{i j}$ corresponds to operators with equal number of fields \ and equal
classical dimension $\D^0_i = \D^0_j$.

The matrix $Z$ gives the anomalous dimension matrix $\Gamma$ as follows:
\begin{equation}
\begin{array}{ll}
\Gamma (g) & \equiv \: Z^{- 1} . \frac{d}{d \log \mu} Z \,
\end{array} . \hspace{0.17em}  \label{anomdimmat}
\end{equation}
Writing $\frac{d}{d \log \mu} Z = \frac{d Z}{d g^2}  \frac{d (g^2)}{d \log
	\mu}$ and using the definition of beta function it is then simple to obtain:
\begin{equation}
\G_{i j} (g) = - g^2 z_{i j} + O (g^4) . \label{ztoGamma}
\end{equation}
This has to be evaluated at the fixed point given by {\eqref{fp}}.

Diagonalization of $\G$ gives the renormalized operators, whose dimensions
are given by the eigenvalues. For an eigenvector $e^{(m)}$ satisfying $\sum_i
e^{(m)}_i \G_{i j} = \g_m e^{(m)}_i$, we get a renormalized operator
$\mathcal{O}^R_m = \left( 1 + \frac{\g_m}{\e} \right) \sum_j e^{(m)}_j
\mathcal{O}^B_j$ having an anomalous dimension $\D_m-\D^0_m=\g_m$. 

\subsubsection{Anomalous dimension of $\vf, \omega, \chi$}\label{app:fieldanomdim}

Let us first compute the anomalous dimension of the fundamental fields $\vf, \omega,\chi_i$. Their renormalization constants we computed in \eqref{phiomega}\,. Using \eqref{ztoGamma} we immediately get the anomalous dimension:
\begin{equation}
\g_\vf=\g_{\omega}=\g_{\chi} =-\frac{\e}{18}\,.
\end{equation}
As expected this is same as the well-known field anomalous dimension for Lee-Yang fixed point in $d=6-\e$ dimensions \cite{srednicki_2007}.

\subsubsection{Susy-null leaders}\label{app:susynull}

\begin{figure}[h]
	\centering \includegraphics[width=400pt]{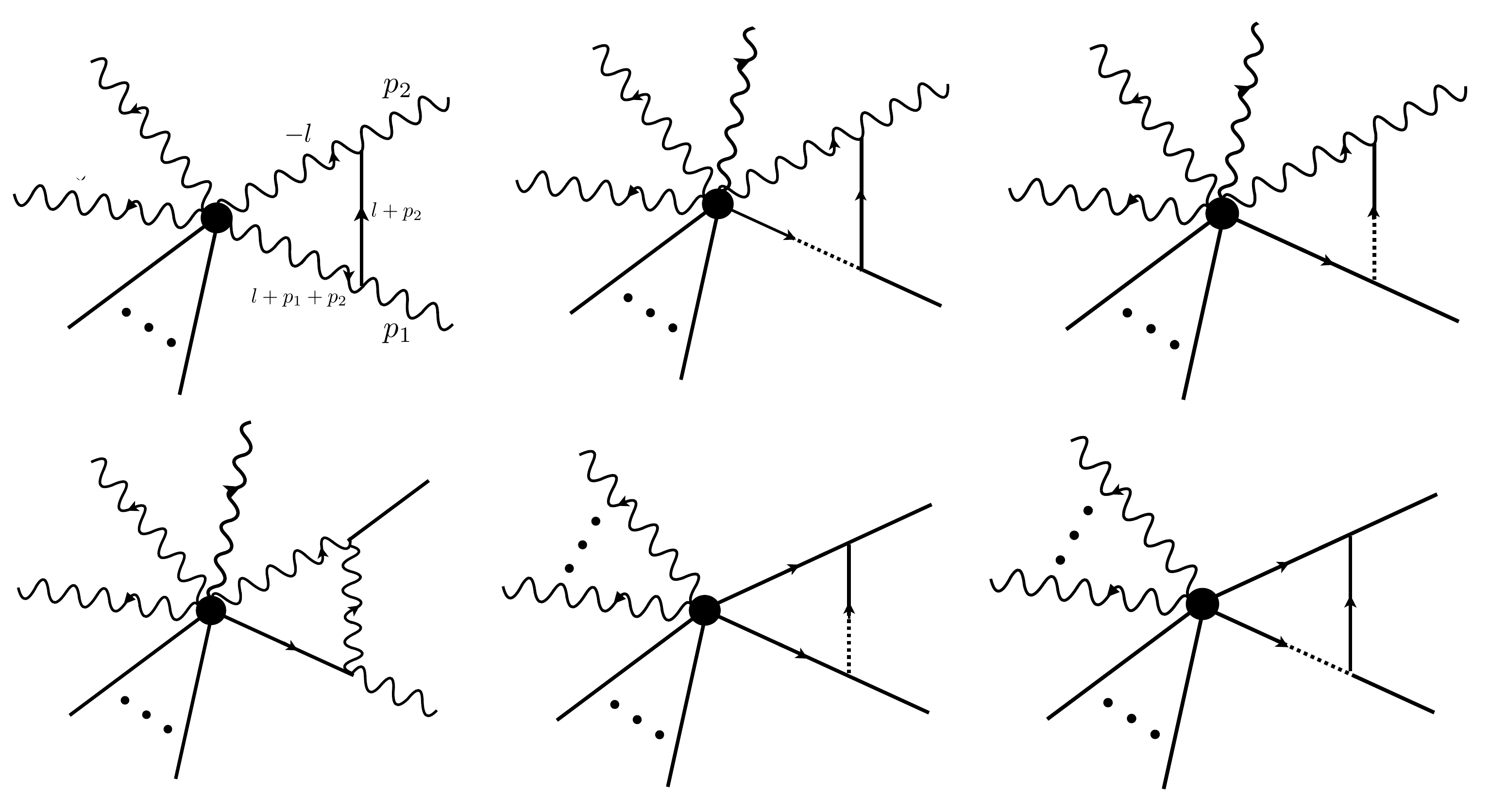}
	\
	\caption{Diagrams that contribute to the one-loop correction of $(\mathcal{N}_k)_L$. \label{susynull}}
\end{figure}

In this section we will compute the anomalous dimensions of some susy-null
leader operators. The ones we consider correspond to the class of singlet operators $\mathcal{N}_k$ we introduced in
section \ref{sec:susynull}. For $k = 4, 5, 6$ their leaders $(\mathcal{N}_k)_L=(\chi_i^2)^2\vf^{k-4}$ give susy-null operators
with the three lowest classical dimensions. These are given by $\;
(\chi_i^2)^2, \; \vf (\chi_i^2)^2, \; \vf^2 (\chi_i^2)^2$ respectively. Since each
$(\mathcal{N}_k)_L$ has the lowest classical dimension among operators with $k$ fields we may simply represent their respective renormalization constants
as follows:
\begin{eqnarray*}
	\; [(\mathcal{N}_k)_L]_B & = & Z_k  (\mathcal{N}_k)_L\,.
\end{eqnarray*}
To compute the renormalization constant $Z_k$ we consider the correlator
$$\langle (\mathcal{N}_k)_L (p=0) \chi_j (p_1) \chi_k (p_2) \chi_l (p_3)
\chi_m (p_4) \vf (p_5) \cdots \vf (p_{k}) \rangle .$$ 
The one-loop triangle diagram corrections are shown in Figure {{\ref{susynull}}}. All the
diagrams involve the same loop integral that we had earlier in
{\eqref{loopint}}.

\

It is straghtforward to compute the symmetry factors associated to each
diagram. Demanding that the $\frac{1}{\e}$ term vanishes from the correlator
above we get the following condition:
\begin{equation}
Z_{k\,}^{- 1} = - \frac{3}{2} (k^2-k-8) [I (p_1, p_1 + p_2)]_{\e^{-
		1}} - {\frac{1}{2}}  \sum_{i = 1}^{k + 4} \left[ \frac{I_{\vf
		\omega} (p_i^2)}{p_i^2} \right]_{\e^{- 1}} \quad . \label{correlchi}
\end{equation}
Here \ $I_{\vf \omega}$ is to account for external leg corrections and is given
by {\eqref{phiomegaint1}}. With $I (p_1, p_1 + p_2)$ computed previously in
{\eqref{Ipole}} we get the following condition for cancellation of
$\frac{1}{\e}$ singlarities:

\begin{equation}
Z_{k\,}^{- 1} = 1 - (6 k^2-7 k-48) \frac{H^{} g^2}{12 (4 \pi)^4  \e} 
. \label{ZH21}
\end{equation}
Using the general formula {\eqref{ztoGamma}} we get the anomalous dimensions
of leaders of $\mathcal{N}_k$ :
\begin{equation}
\gamma_{(\mathcal{N}_k)_L} = \frac{1}{18}  (6 k^2-7 k-48) \epsilon
 . \label{anomchi22}
\end{equation}

\subsubsection{Non-susy-writable leaders}\label{app:nonsusy}

\begin{figure}[h]
	\centering \includegraphics[width=200pt]{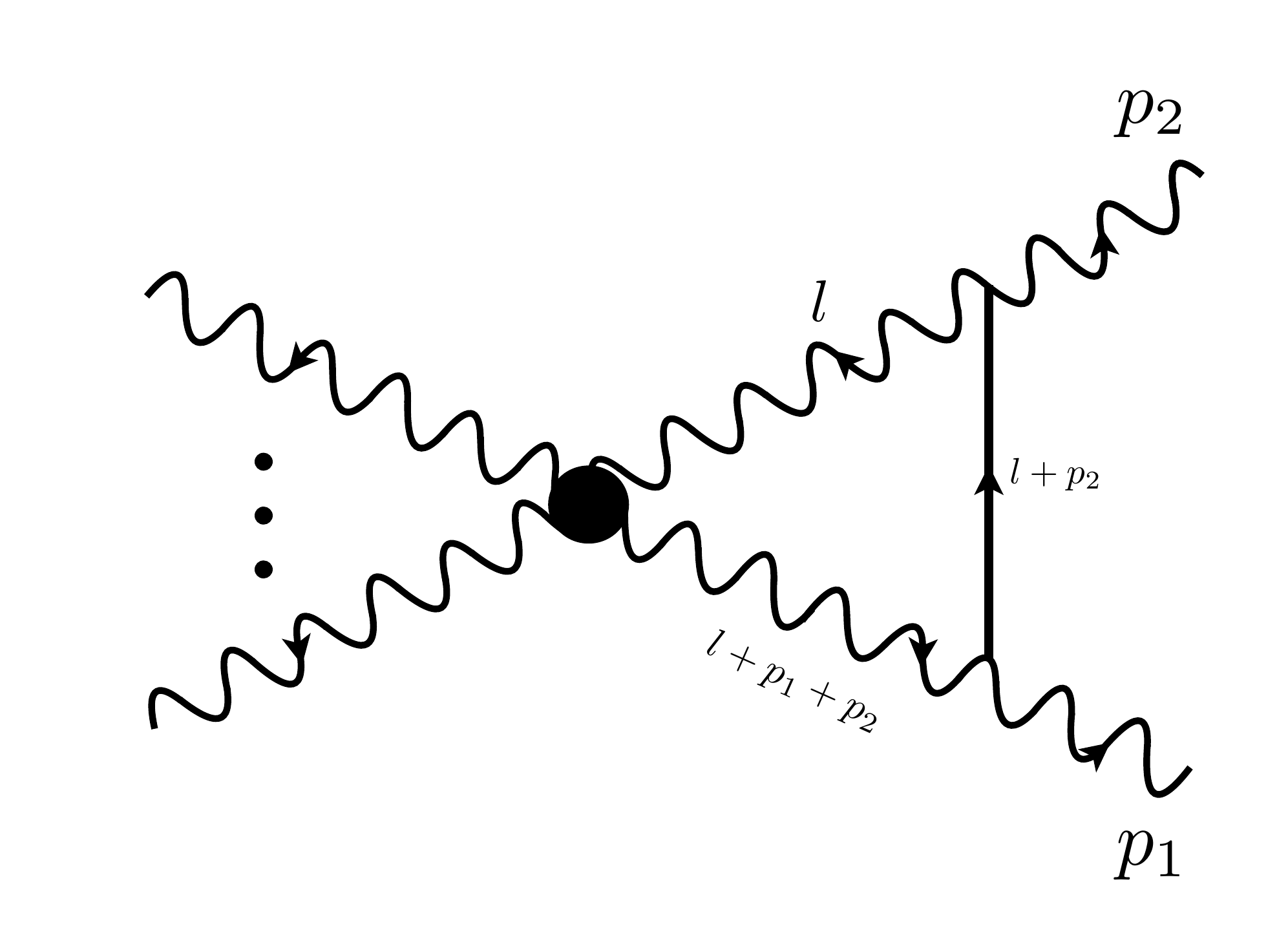}
	\
	\caption{One-loop corrections to $(\mathcal{F}_k)_L$. \label{Feldman}}
\end{figure}

Next we consider non-susy-wriable leader operators. We will focus on leaders
of the particular class of singlets, call Feldman operators. We defined them
in equation \eqref{Fk} which we repeat for convenience:
\begin{eqnarray}
(\mathcal{F}_k)_L & = & \sum_{l = 2}^{k - 2} (- 1)^l \binom{k}{l} \left(
\sum' \chi^l_i \right)  \left( \sum' \chi^{k - l}_j \right) . 
\label{FkCardy1}
\end{eqnarray}
To compute the anomalous dimension of this we should, in principle, consider
all opoerators $\mathcal{O}_i$ with $k$ fields and the same classical
dimension as $(F_k)_L$. Let us write their renormalization matrix as
$\mathcal{O}^B_i = Z_{i j} \mathcal{O}_j$. We take $\mathcal{O}_1 =
(\mathcal{F}_k)_L$. We then have a renormalization matrix that has the following form:
\begin{equation}
Z = \left(\begin{array}{cccc}
Z_{11} & 0 & 0 & \ldots\\
0 & \ast & \ast & \ldots\\
0 & \ast & \ast & \ldots\\
\vdots & \vdots & \vdots & \ddots
\end{array}\right), \label{ZF}
\end{equation}
To show this consider the correlator $\langle (\mathcal{F}_k)_L (p = 0)
\chi_{i_1} (p_1) \ldots \chi_{i_k} (p_k) \rangle$. It has a one loop
correction as shown in Figure \ref{Feldman}. From the
$\chi$-$\chi$ legs and the associated contractions of $K$ matrices it is clear
that we only get a correction to $Z_{11}$ and no other component $Z_{1 j}$. The fact that the components $Z_{i 1}$ ($i > 1$) would be  0 was argued in app H.3 of {\cite{paper2}}, and this also applies for the cubic interactions considered in this paper.  Hence for computation of anomalous dimension of $(\mathcal{F}_k)_L$ we have
only to compute the renormalization constant $Z_{11} = Z_{(\mathcal{F}_k)_L}$
and can neglect the other operators $\mathcal{O}_{i \neq 1}$. 

The loop diagram is again that of {\eqref{Ipole}} and comes with the symmetry
factor $\frac{1}{2} k (k - 1)$. Considering the correction due to $k$ external
legs we get:
\begin{equation}
Z_{k\,}^{- 1} = - \frac{1}{2} k (k - 1) [I (p_1, p_1 + p_2)]_{\e^{- 1}} -
{\frac{1}{2}}  \sum_{i = 1}^k \left[ \frac{I_{\vf \omega}
	(p_i^2)}{p_i^2} \right]_{\e^{- 1}}  . \label{Zfeld}
\end{equation}
This gives:
\begin{equation}
Z_{k\,}^{- 1} = 1 - \frac{g^2 H}{12 (4 \pi)^4  \e} k (2 k - 3) .
\label{Zfeld1}
\end{equation}
Using {\eqref{ztoGamma}} we get the anomalous dimension:
\begin{equation}
\gamma_{(F_k)_L} = \frac{k (2 k - 3)}{18} \epsilon
 . \label{anomFk}
\end{equation}

\subsubsection{Susy-writable leaders}\label{app:susywrit}

In this section we will compute the anomalous dimension of a
class of susy-writable leaders, specifically the box operators discussed in section \ref{sec:susywritable} and App. \ref{App:Susy_Writable}.
For convenience we will compute anomalous dimension of the box operator in the $\widehat\phi^3$ theory
in $\widehat d= 6 - \e$ dimension. Via dimensional reduction that should give the same anomalous dimenision as the  operator we want. 

The $\widehat\phi^3$ theory has the Lagrangian shown in {\eqref{hphi3}} which we rewrite below for
convenience (here we canonically normalize the action since the anomalous dimensions are independent of this normalization):
\be
\mathcal{S}=
\int d^{\widehat d}x \Big[ \frac{1}{2}(\partial\widehat{\phi})^2+m^2\widehat{\phi}^2+\frac{g}{6}\widehat{\phi}^3 \Big]\,.
\ee
The arguments are similar to what we described above in section \ref{localOp}.
The standard details of this theory (beta function, critical point value) are given in section \ref{sec:RG}.

We want to use a form of the box operator that is simple for computation. In \cite{paper2} App. F we used the idea of equivalence class $\{\mathcal{O}\}$ which contains an operator $\mathcal{O}$ and other operators that are related to it by a total derivative. Hence all operators in $\{\mathcal{O}\}$ are   equivalent for perturbative RG computations. The box operator equivalence class $\{\widehat{B}^{(k)} (z_1, z_2)\}$ has the following operator:
\begin{equation}
\widehat{B}^{(k)} (z_1, z_2) \equiv \widehat\phi^{k - 2} (((z_1 . \partial) (z_2 . \partial)
\widehat\phi)^2 - (z_1 . \partial)^2 \widehat\phi (z_2 . \partial)^2 \widehat\phi) . \label{box}
\end{equation}
which we defined as $\widehat{B}^{(k)}$ for this section. It is easy to show that this is related to the form \eqref{Bncontracted} by a total derivative.

We define its renormalization as follows:
\begin{equation}
[\widehat{B}^{(k)} (z_1, z_2)]_B = Z_{\widehat B^{(k)}}  \widehat{B}^{(k)} (z_1, z_2). \label{boxZ}
\end{equation}
We consider the correlator $\langle
\widehat{B}^{(k)} (z_1, z_2) (p = 0) \widehat\phi (p_1) \cdots \widehat\phi (p_n) \rangle$ for computing $Z_{\widehat B^{(k)}}$. It can be shown that the box operator as shown in \eqref{Boxn} is a primary in free theory. Therefore for computing the one-loop anomalous dimension we can ignore mixing with other operators. 

\begin{figure}[h]
	\centering \includegraphics[width=200pt]{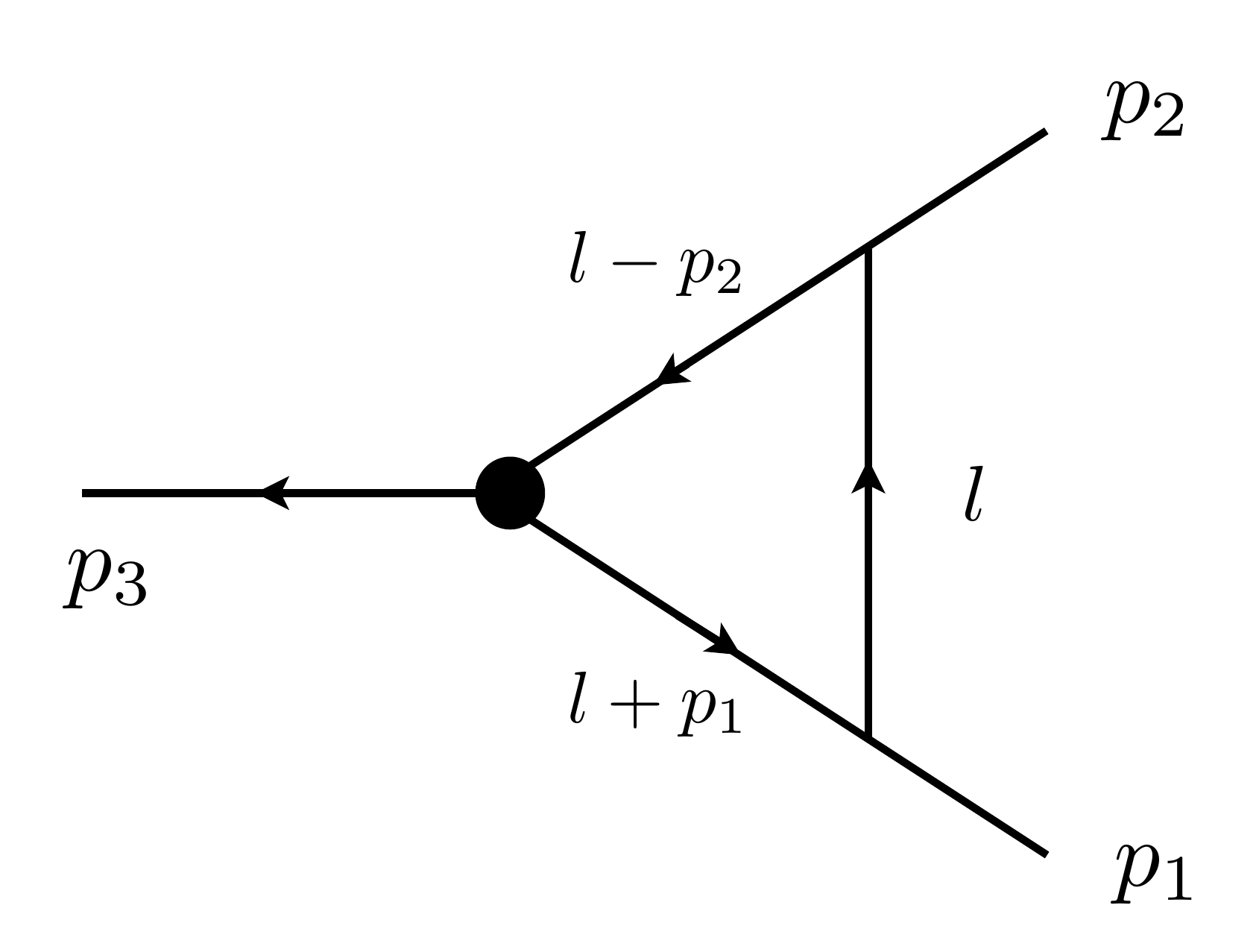}
	\
	\caption{One-loop correction to $\widehat{B}^{(3)}$ in $\widehat\phi^3$ theory in $d=6-\e$. \label{boxdiag}}
\end{figure}

We will focus on the case $k = 3$. The operator $\widehat{B}^{(3)} $ may be written as follows:
\begin{equation}
\widehat{B}^{(3)} (z_1, z_2) = (z_1^{\mu} z_2^{\nu} z_1^{\rho} z_2^{\sigma} -
z_1^{\mu} z_1^{\nu} z_2^{\rho} z_2^{\sigma}) \widehat\phi (\partial^{\mu}
\partial^{\nu} \widehat\phi) (\partial^{\rho} \partial^{\sigma} \widehat\phi) .
\label{boxrep}
\end{equation}
At tree level the correlator is proportional to:
\be\label{treebox}
p_1^{\mu}\, p_1^{\nu}\,p_2^{\r}\,p_2^{\s}+p_2^{\mu}\,p_2^{\nu}\,p_3^{\r}\,p_3^{\s}+p_3^{\mu}\,p_3^{\nu}\,p_1^{\r}\,p_1^{\s} \ +\ (\mu\leftrightarrow\r,\nu\leftrightarrow\s)\,.
\ee
The correction to the correlator is given by the triangle diagram as
shown in Figure \ref{boxdiag}. 

Then we have integrals of the following forms depending on which leg(s) the
derivatives act:
\begin{eqnarray*}
	I_1 & = & p_3^{\g} p_3^{\d} \int d^d l \frac{(l + p_1)^{\a} (l + p_1)^{\b}
		}{(l^2)  (l + p_1)^2 (l - p_2)^2 },\\
	I_2 & = & p_3^{\g} p_3^{\d} \int d^d l\frac{(l - p_2)^{\a} (l - p_2)^{\b}
		}{(l^2)  (l + p_1)^2 (l - p_2)^2 },\\
	I_3 & = & \int d^d l\frac{(l + p_1)^{\a} (l + p_1)^{\b} (l - p_2)^{\g} (l -
		p_2)^{\d} }{(l^2)  (l + p_1)^2 (l - p_2)^2 }, \label{loopintbox}
\end{eqnarray*}
where the indices $\a, \b, \g, \rho$ have to be chosen as appropriate.

Let us focus on $I_1$ to show how to evaluate these integrals. We may write
the denominator in $I_1$ as \
\begin{equation}
\frac{1}{(l^2)  (l + p_1)^2 (l - p_2)^2 } = 2 \int d x_1 d x_2 d x_3
\frac{\tmd (x_1 + x_2 + x_3 - 1)}{[x_1 (l + p_1)^2 + x_2 (l - p_2)^2 + x_3
	l^2]^3} .
\end{equation}
Now using the standard change of variable $q = l + x_1 p_1 - x_2 p_2$ the
above becomes
\begin{equation}
2 \int d x_1 d x_2 d x_3 \frac{\tmd (x_1 + x_2 + x_3 - 1)}{[q^2 + F]^3},
\qquad [F = x_1 (1 - x_1) p_1^2 + x_2 (1 - x_2) p_2^2 + 2 x_1 x_2 p_1 .p_2]
.
\end{equation}
From the numerator of $I_1$ one gets:
\begin{eqnarray}
	(l + p_1)^{\a} (l + p_1)^{\b} & = & q^{\a} q^{\b} + \left[
	q^{\alpha}  (x_2 p_2^{\beta} - (x_1 - 1) p_1^{\beta}) + \left( \a
	\leftrightarrow \b \right) \right] \label{numerator}\nonumber\\
	&  & + (p_1^{\alpha} (x_1 - 1) - x_2 p_2^{\alpha})  (p_1^{\beta} (x_1 - 1)
	- x_2 p_2^{\beta}) .
\end{eqnarray}
The $q^{\a} q^{\b}$ gives a term proportional to $\eta^{\a \b}$ upon
integration. Since the multiplicative factors of $z_1$ and $z_2 $tensors in
{\eqref{boxrep}} eliminates all traces, we can ignore it. The terms
proportional to $q^{\a}$ or $q^{\b}$ are also simultaneously zero. \ So we
are left with the last term of {\eqref{numerator}} which gives:
\begin{equation}
I_1 = \frac{p_3^{\g} p_3^{\tmd}}{12 (4 \pi)^3 \e}  (6 p_1^{\alpha}
p_1^{\beta} + 3 p_1^{\alpha} p_2^{\beta} + 3 p_2^{\alpha} p_1^{\beta} + 2
p_2^{\alpha} p_2^{\beta}) + O \left( \e \right) .
\end{equation}
The integral $I_2$ is obtained by simply replacing $p_1 \leftrightarrow -
p_2$. The $I_3$ is slightly more tedious as it is quartic in the numerator.
But it follows exactly the same method, and gives:
\begin{align}
I_3=&  \frac{1}{(4\pi)^3\e}\Big[\frac{1}{30} p_1^\mu p_1^\nu p_1^\rho  p_1^\sigma +\frac{1}{60} p_2^\mu  p_1^\nu p_1^\rho  p_1^\sigma +\frac{1}{60} p_1^\mu  p_2^\nu  p_1^\rho  p_1^\sigma  +\frac{1}{30} p_2^\mu  p_2^\nu  p_2^\rho  p_2^\sigma  +\frac{1}{90} p_2^\mu  p_2^\nu  p_1^\rho  p_1^\sigma +\frac{1}{15} p_1^\mu  p_1^\nu  p_2^\rho  p_1^\sigma \nonumber \\ 
&  +\frac{1}{36} p_2^\mu  p_1^\nu p_2^\rho  p_1^\sigma +\frac{1}{36} p_1^\mu  p_2^\nu  p_2^\rho  p_1^\sigma +\frac{1}{60} p_2^\mu  p_2^\nu  p_2^\rho  p_1^\sigma  +\frac{1}{15} p_1^\mu  p_1^\nu  p_1^\rho  p_2^\sigma +\frac{1}{36} p_2^\mu  p_1^\nu  p_1^\rho p_2^\sigma  +\frac{1}{36} p_1^\mu  p_2^\nu  p_1^\rho  p_2^\sigma   \nonumber \\ 
& +\frac{1}{60} p_2^\mu  p_2^\nu  p_1^\rho  p_2^\sigma  +\frac{19}{90} p_1^\mu  p_1^\nu  p_2^\rho  p_2^\sigma  +\frac{1}{15} p_2^\mu  p_1^\nu  p_2^\rho  p_2^\sigma    +\frac{1}{15} p_1^\mu  p_2^\nu  p_2^\rho  p_2^\sigma \Big]\,.
\end{align}

The final answer should be written in a form that matches the tree level momentum
dependence \eqref{treebox}. The result from the loop integral
can be brought to this form by using momentum conservation $p_1 + p_2 + p_3 =
0$.

Demanding that the correlator is free of $\e^{- 1}$ poles allows us to
compute $Z_{\widehat B^{(3)}}$ as follows:
\begin{equation}
Z_{\widehat B^{(3)}}^{- 1} = 1 - \frac{g^2}{2 (4 \pi)^3 \epsilon} + \frac{3 g^2}{12 (4
	\pi)^3 \epsilon} . \label{ZBn}
\end{equation}
The last part is the standard $\phi^3$ theory  correction to the
3 external legs (see \cite{srednicki_2007}). This gives the following
anomalous dimension of $\widehat{B}^{(3)} (z_1, z_2)$:
\begin{equation}
\g_{\widehat B^{(3)}} = \frac{\e}{6} . \label{anomB3}
\end{equation}
Now let us discuss the case of $\widehat B^{(k)}(z_1,z_2)$ for $k>3$. The steps are  similar to above. But there is one extra loop integral where the derivatives of $\widehat B^{(k)}$ are not distributed over the internal legs. It is given by:
\be
	I_4 =  \int \frac{d^d l}{l^2  (l + p_1)^2 (l - p_2)^2 }=\frac{1}{(4\pi)^3\e}\,.
\ee
We explicitly computed the anomalous dimension of $\widehat B^{(k)}(z_1,z_2)$ for $k=3, \dots, 9$. By extrapolation we expect the result for generic  $k$ to be given by
\be
\g_{\widehat B^{(k)}} =\frac{1}{6} \left(2 k^2-5 k-2\right) \epsilon . \label{anomB4}
\ee


\small
\bibliographystyle{utphys}
\bibliography{mybib}

\end{document}